\documentclass[iop,apj,twocolumn]{emulateapj}
\usepackage{times}
\usepackage{epsfig}
\usepackage{amsmath, amsthm, amssymb}
\usepackage[plainpages=false, colorlinks=true, anchorcolor=blue, linkcolor=blue, citecolor=blue, bookmarks=false]
{hyperref}
\usepackage{color}
\usepackage{microtype}
\usepackage{float}
\usepackage[caption = false]{subfig}
\usepackage{graphicx}

\newcommand{\beq}{\begin{equation}}
\newcommand{\eeq}{\end{equation}}
\newcommand{\bea}{\begin{eqnarray}}
\newcommand{\eea}{\end{eqnarray}}

\newcommand{\supernu}{{\tt SuperNu}}
\newcommand{\phoenix}{{\tt PHOENIX}}

\begin{document}

\title{Radiation Transport for Explosive Outflows: Opacity Regrouping}
\author{Ryan T. Wollaeger$^{1,2}$ and Daniel R. van~Rossum$^{2}$}

\affil{$^{1}$Department of Nuclear Engineering \& Engineering Physics, University of Wisconsin, Madison
1500 Engineering Drive, 410 ERB, Madison, WI, 53706; wollaeger@wisc.edu}
\affil{$^{2}$Flash Center for Computational Science, Department of Astronomy
  \& Astrophysics, University of Chicago, Chicago,
IL, 
60637; daan@flash.uchicago.edu}

\begin{abstract}

Implicit Monte Carlo (IMC) and Discrete Diffusion Monte Carlo (DDMC) are methods used to stochastically solve the radiative transport and diffusion equations, respectively.
These methods combine into a hybrid transport-diffusion method we refer to as IMC-DDMC.
We explore a multigroup IMC-DDMC scheme that, in DDMC, combines frequency groups with sufficient optical thickness.
We term this procedure ``opacity regrouping''.
Opacity regrouping has previously been applied to IMC-DDMC calculations for problems in which the dependence of the opacity on frequency is monotonic.
We generalize opacity regrouping to non-contiguous groups and implement this in \supernu, a code designed to do radiation transport in high-velocity outflows with non-monotonic opacities.
We find that regrouping of non-contiguous opacity groups generally improves the speed of IMC-DDMC radiation transport.
We present an asymptotic analysis that informs the nature of the Doppler shift in DDMC groups and summarize the derivation of the Gentile-Fleck factor for modified IMC-DDMC.
We test \supernu\ using numerical experiments including a quasi-manufactured analytic solution, a simple ten-group problem, and the W7 problem for Type Ia supernovae.
We find that the opacity regrouping is necessary to make our IMC-DDMC implementation feasible for the W7 problem and possibly Type Ia supernova simulations in general.
We compare the bolometric light curves and spectra produced by the \supernu\ and \phoenix\ radiation transport codes for the W7 problem.
The overall shape of the bolometric light curves are in good agreement, as are the spectra and their evolution with time.
However, for the numerical specifications we considered, we find that the peak luminosity of the light curve calculated using \supernu\ is $\sim$10\% less than that calculated using \phoenix.


\end{abstract}

\section{Introduction}
\label{sec:Intro}

Type Ia supernovae (SNe Ia) are the explosions of
Carbon-Oxygen (C-O) white dwarf stars.
In the most widely studied model of SNe Ia, a C-O white dwarf
approaching the Chandrasekhar mass releases energy from nuclear
fusion that exceeds gravitational binding energy of the star,
causing the star to explode~\citep{branch1995}.
The resulting high-velocity outflow becomes ballistic in a matter
of minutes, and thereafter expands homologously.
During this expansion, gamma rays from the radioactive decay
of $^{56}$Ni heat the outflow, causing it to radiate, with a
peak luminosity that can exceed the host galaxy of the
supernova.

The majority of observed SNe Ia have similar peak luminosities
and spectra~\citep{hillebrandt2000}.
The light curves of most SNe Ia obey a peak luminosity-width
relationship~\citep{phillips1993}.
As a result, the light curve data for SNe Ia may be fit to
a template, enabling its peak luminosity, and therefore its
relative distance, to be determined.
Consequently, SNe Ia are important ``standard candles'' for
measuring cosmic distances and the expansion rate of the
universe, and their use for these purposes led to the
discovery of dark energy [see, e.g.,~\citet{riess1998,perlmutter1999}].

Given the significance of SNe Ia in galaxy formation and evolution
\citep{scannapieco2008} and in nucleosynthesis, as well as in
cosmology, much research has been done to understand how model parameters
affect the observable properties of these events; for example, the connection
between explosion asymmetry and anomalies in luminosity~\citep{calder2004,kromer2009}.
Other research efforts have focused on generating methods, algorithms,
and codes that can adequately treat the physics of SNe Ia, along with
other hydrodynamic and radiative events in astrophysics.
\citet[pp.~128,144,160]{mihalas1984} derived the equations of relativistic
fluid flow.
\citet[pp.~41,49]{castor2004} describes standard Lagrangian and Eulerian
methods to solving hydrodynamic problems.
The {\tt FLASH} code~\citep{fryxell2000,calder2002} provides a means of
solving the Euler equations for compressive, reactive hydrodynamics with
nuclear reactions.

Radiation transport in Type Ia SNe is a complex problem
both theoretically and practically.
From the theoretical perspective, photons may interact
with millions of spectral lines in a heterogeneous material
that has multiple ionization states [see, e.g.~\citet{vanrossum2012}].
A photon may see an optically thin environment in one location
of the outflow and subsequently redshift into resonance with a
line opacity elsewhere.
Such situations provide a challenge to Local Thermodynamic
Equilibrium (LTE) calculations, and especially, Nonlocal Thermodynamic
Equilibrium (NLTE) calculations.
There is also the question of the leading-order behavior of the
radiation at different time scales in the presence of material fluid.
\cite{lowrie2001} make the distinction between the radiation time scale
and the fluid time scale as a means of preserving correct
relativistic principles in first-order comoving transport.

From the practical perspective, high-fidelity Type Ia SNe simulations
are generally seen to be demanding in memory and algorithm efficiency
\citep{baron2007}.
For an end-to-end simulation, one needs to couple a progenitor
explosion-phase hydrodynamic simulation to the beginning of the
homologous-expansion phase, and then appropriately treat
radiation transport in the latter~\citep{seitenzahl2012,long2014}.
Numerical simulations of the full evolution of the supernova, regardless
of the particular explosion model, involve a large range of densities,
temperatures, length scales, time scales, and physical phenomena.

Codes can apply transport theory to the homologous-expansion
phase of Type Ia supernovae to synthesize light curves and spectra.
Broadly speaking, transport calculations may be performed
deterministically with some subset of matrix-solution
techniques or stochastically with random-number sampling.
The stochastic approach gives terms in the transport equation
a probabilistic interpretation; this gives rise to ``particles''
with sampled properties that can be manipulated and
tallied to solve the transport equation.
Common methods of computational transport described by
\cite{lewis1993} include: discrete ordinates~\citep[pp.~116,156]
{lewis1993}, integral transport~\citep[p.~208]{lewis1993},
multigroup~\citep[p.~61]{lewis1993}, and finite elements~\citep{adams2001}.
The listed methods may be implemented in or in conjunction with Monte Carlo
(MC) or deterministic schemes~\citep{urbatsch1999}; the resulting
scheme might be deemed a composite method.

Several radiation transport codes have been developed and
applied to the W7 model of~\cite{nomoto1984} and to SN Ia
models generally.
Deterministic codes include \phoenix, a code
based on the iterative, short characteristic method
\citep{hauschildt1999,baron2007,olson1987}.
Recently,~\cite{vanrossum2012} extended \phoenix\ to be able to
calculate self-consistently the temporal evolution of the SN Ia outflow.
\cite{hauschildt1991} investigate a discrete ordinates method that
incorporates relativistic effects to be able to treat explosive outflow.
The MC codes {\tt SEDONA} of~\cite{kasen2006}, the code of~\cite{lucy2005},
and the {\tt ARTIS} code of~\cite{kromer2009} solve multi-dimensional,
time-dependent radiation transport in homologous outflow.
\cite{kasen2006} and~\cite{kromer2009} solve multifrequency transport
by applying the Solobev approximation~\citep[p.~122]{castor2004} to line transport.

Monte Carlo in the context of a velocity field has the favorable property that
particles (which are also referred to as packets) may be tracked
in one inertial (lab) frame and interact with the fluid in the comoving
frame.
A particle may have its properties converted to the comoving frame,
updated according to the interaction, and converted back to the lab
frame if the particle history is not discontinued.
\cite{kasen2006} applies MC iteratively within a time step to
obtain converged electron temperatures while~\cite{kromer2009} find
the contribution of MC iteration to be insignificant if small time steps are chosen.

Instead of treating the temperature structure iteratively or explicitly,
there exist transport methods that are made fully implicit
\citep{nkaoua1991,brooks1989} or semi-implicit~\citep{fleck1971,carter1973}
through time discretization of the material equation(s) and adjustment
of Monte Carlo interpretations~\citep{densmore2004}.
To our knowledge, these methods have not been extensively examined for
application in the SN Ia problem.

Implicit Monte Carlo (IMC) is a stochastic method that may be
applied to solve the time-dependent, nonlinear radiation transport
equations~\citep{fleck1971,fleck1984}.
Of the implicit methods referenced towards the end of the preceding
paragraph, IMC is quite possibly the simplest to implement.
The IMC method is made semi-implicit through a non-dimensional
quantity, referred to as the Fleck factor, that converts a portion
of absorption and reemission to instantaneous ``effective scattering.''
\footnote{Note the Fleck factor is not a
directly tunable parameter but follows naturally from
linearizing the thermal transport equations within each
time step.}
By introducing effective scattering, the Fleck factor stabilizes
large-time-step
\footnote{Roughly speaking, time steps that result in the
deposition of a radiation energy density that is greater than or of
order the material energy density may cause an IMC simulation to
become unstable; hence the pathology depends on the
evolution of the radiation field~\citep{gentile2011}.}
radiation transport calculations that might otherwise
suffer significant non-physical temperature fluctuations
\citep{fleck1971}.
However~\cite{larsen1987} demonstrate that IMC may still be prone to
spurious temperature fluctuation for large time steps and derive a
sufficient but not necessary constraint on time step size to prevent
non-physical behavior, which they call the ``Maximum Principle'' (MP).
Recent extensions have been made to IMC that mitigate the pathologies
associated with the MP [see, e.g.,~\citet{mcclarren2009,gentile2011,
mcclarren2012}].

IMC may suffer in performance when effective scattering dominates
over other particle processes.
Performance may be improved for calculations having
significant physical or effective scattering by combining IMC
with either a deterministic or stochastic diffusion method.
Stochastic methods include Random Walk (RW)~\citep{fleck1984},
Implicit Monte Carlo Diffusion (IMD)~\citep{gentile2001,cleveland2010},
and Discrete Diffusion Monte Carlo (DDMC)
[see, e.g.,~\citet{densmore2007,densmore2008,densmore2012}].
The methods listed have been hybridized with IMC and applied to both
grey and multifrequency or multigroup problems.
Additionally, each method may benefit IMC by replacing small-mean-free-path
particle processes with large diffusion processes.
The larger diffusion steps of the RW method developed by
\cite{fleck1984} place a diffusive particle isotropically on the
surface of a sphere of several mean free paths in radius centered
at the particle's initial position.
This sphere must be bounded by the spatial grid that stores the
material properties~\citep{fleck1984}.
Hence, histories in diffusive domains near cell boundaries will
not have sufficiently large displacement spheres; this is found
to limit the increase in IMC efficiency~\citep{densmore2008}.

DDMC and IMD differ from RW by discretizing the diffusion equation
in space; after some algebra, the resulting terms are given a Monte
Carlo interpretation~\citep{densmore2007,gentile2001}.
The discretization implies that a DDMC particle position within a spatial
cell is ambiguous~\citep{wollaeger2013}.
IMD discretizes the diffusion equation in time while DDMC keeps
particle time continuous.
Continuous particle time precludes causal ambiguity for each
particle~\citep{densmore2007}.

The hybridization of IMC and DDMC, referred to as IMC-DDMC,
has been investigated in multigroup problems
\citep{densmore2012,abdikamalov2012,wollaeger2013}.
In each of the IMC-DDMC implementations, there is a mean-free-path
threshold that dictates whether or not a cell and group of the
spatial and wavelength grids is amenable to diffusion theory.
\cite{densmore2012} investigate a hybrid for monotonic opacity
dependence on frequency that applies grey DDMC in a ``large'' lower
group below a frequency threshold and multifrequency or multigroup IMC
above the frequency threshold.
\cite{abdikamalov2012} describe a general multigroup IMC-DDMC scheme
for application to neutrino transport in the presence of a fluid;
this makes the method velocity dependent.
\cite{wollaeger2013} delineate a velocity-dependent method for
photons that reconciles IMC-DDMC to high-velocity, homologous
Lagrangian grids.

Here, we present some extensions to the
particular IMC-DDMC method described by~\cite{wollaeger2013}.
The extensions are opacity regrouping~\citep{densmore2012} and the
Gentile-Fleck factor~\citep{gentile2011}.
We implement these features in the IMC-DDMC radiation transport code,
\supernu~\citep{wollaeger2013}.
We first briefly discuss the thermal radiation transport equations.
Then we apply an asymptotic analysis to the continuous, comoving
transport equation on an interior of a frequency domain and in a
boundary layer of a frequency domain; this clarifies where the
DDMC redshift scheme is generally applicable.
We summarize standard IMC, the Gentile-Fleck factor
modified IMC scheme~\citep{gentile2011}, and the hybrid IMC-DDMC
equations.
Next, we discuss IMC-DDMC processes and a scheme for combining
groups that have DDMC into larger groups to increase computational
efficiency.
The groups belong to the same spatial cell and must all have opacities
that make the cell sufficiently optically thick; this is an
optimization since effective scattering for particles in either
of the original groups is reduced~\citep{densmore2012}.
We term this optimization ``opacity regrouping.''
Opacity regrouping was first implied by~\cite{densmore2012} with a
low-frequency DDMC group adaptively adding or subtracting adjacent IMC
groups based on the mean free path threshold.
Moreover, the extension of the optimization to strongly non-monotonic
opacity was anticipated by~\cite{densmore2012}.
Recently, an opacity regrouping procedure for non-contiguous
groups was implemented by~\cite{cleveland2014} for
Hybrid Implicit Monte Carlo Diffusion (HIMCD); in addition
to improving code performance, their approach addresses the
effects of teleportation error~\citep{fleck1984} with new
method coupling criteria.
In addition to the IMC-DDMC mean free path threshold, $\tau_{D}$, we
introduce an additional mean free path threshold, $\tau_{L}$, that
determines regroupable DDMC groups.
We investigate the effect of changing regrouping parameters on
a simple ten-group problem and the one-dimensional W7 problem presented
by~\cite{nomoto1984}.
Additionally, we explore the effect of a modified Fleck factor,
presented by~\cite{gentile2011}, on mitigating erroneous
fluctuations in the temperature profile in the W7 test problem.

This article is organized as follows.
In Section~\ref{sec:trans}, we discuss the approximations to the
radiation transport and fluid equations assumed in our code.
In Section~\ref{sec:dopge}, we perform an asymptotic analysis which
indicates a potential source of discrepancy between full multigroup
IMC with a discretized Doppler shift correction and continuous-frequency
IMC in a multigroup material setting.
In Section~\ref{sec:mceqs}, we describe the Gentile-Fleck factor used
in some numerical results and we summarize the IMC-DDMC equations.
Additionally, we write the equations for opacity regrouping.
In Section~\ref{sec:gregroup}, we write the formulae used to
regroup subsets of groups.
In Section~\ref{sec:procs}, we describe IMC-DDMC particle processes
including the opacity regrouping and DDMC redshift schemes.
In Section~\ref{sec:NumRes}, we present some calculations that
highlight the advantages of the Gentile-Fleck factor and opacity
regrouping and demonstrate the application of \supernu\ to SNe Ia.
In Section~\ref{sec:manu}, combining the techniques of~\cite{oberkampf2010}
and~\cite{gentile2011}, we use a simple quasi-manufactured transport solution
for high-velocity outflow to verify the Gentile-Fleck factor's
ability to mitigate spurious overheating.
In Section~\ref{sec:heav}, we demonstrate the improved performance that
using DDMC opacity regrouping produces for the multigroup outflow problems
presented by~\cite{wollaeger2013}.
Finally, in Section~\ref{sec:W7}, we explore the application of
IMC-DDMC with opacity regrouping and the Gentile-Fleck factor to the W7
problem.
We also investigate the effects of group opacities that are a composite of
Rosseland-like and Planck-like opacities.

\section{Radiation and Fluid Equations}
\label{sec:trans}

We review the underlying theory of the IMC-DDMC scheme tested.
Following~\cite{pomraning1973} and~\cite{castor2004}, terms
in the comoving fluid frame are subscripted with 0.
The thermal equation of radiation transport in the lab frame is
\citep{szoke2005,abdikamalov2012}
\begin{multline}
\label{eq1}
\frac{1}{c}\frac{\partial I_{\nu}}{\partial t}+\hat{\Omega}\cdot
\nabla I_{\nu}+\sigma_{\nu,a}I_{\nu}=\sigma_{\nu,a}B_{\nu}-
\sigma_{\nu,s}I_{\nu}+\\\int_{4\pi}\int_{0}^{\infty}\frac{\nu}{\nu'}
\sigma_{s}(\vec{r},\nu'\rightarrow\nu,\hat{\Omega}'\rightarrow
\hat{\Omega})I_{\nu'}(\vec{r},\hat{\Omega}',t)d\nu'd\Omega'
\end{multline}
where $c$ is the speed of light, $t$ is time, $\vec{r}$ is the
spatial coordinate, $\hat{\Omega}$ is unit direction, $\nu$ is
frequency, $\sigma_{a,\nu}$ is absorption opacity, $\sigma_{s,\nu}$
is scattering opacity, $\sigma_{s}(\vec{r},\nu'\rightarrow\nu,
\hat{\Omega}'\rightarrow\hat{\Omega})$ is differential scattering
opacity, $I_{\nu}$ is the radiation intensity, and $B_{\nu}$ is the
thermal emission source.
The first order comoving form of Eq.~\eqref{eq1} is
\citep[p.~111]{castor2004}
\begin{multline}
\label{eq2}
\left(1+\hat{\Omega}_{0}\cdot\frac{\vec{U}}{c}\right)\frac{1}{c}
\frac{DI_{0,\nu_{0}}}{Dt}+\hat{\Omega}_{0}\cdot\nabla I_{0,\nu_{0}}-
\frac{\nu_{0}}{c}\hat{\Omega}_{0}\cdot\nabla\vec{U}\cdot
\nabla_{\nu_{0}\hat{\Omega}_{0}}I_{0,\nu_{0}}\\+
\frac{3}{c}\hat{\Omega}_{0}\cdot\nabla\vec{U}\cdot\hat{\Omega}_{0}
I_{0,\nu_{0}} = \sigma_{0,\nu_{0},a}(B_{0,\nu_{0}}-I_{0,\nu_{0}})
-\sigma_{0,\nu_{0},s}I_{0,\nu_{0}}\\+
\int_{4\pi}\int_{0}^{\infty}\frac{\nu_{0}}{\nu_{0}'}\sigma_{0,s}
(\vec{r},\nu_{0}'\rightarrow\nu_{0},
\hat{\Omega}_{0}'\cdot\hat{\Omega}_{0})I_{0,\nu_{0}'}d\nu_{0}'d\Omega_{0}'
\;\;,
\end{multline}
where $\vec{r}$ is an Eulerian spatial coordinate, $\vec{U}$ is
the velocity field, and we have used Castor's notation to denote
the photon comoving momentum derivative with $\nabla_{\nu_{0}\hat{\Omega}_{0}}$.
The homologous flow equation is~\citep{kasen2006}
\begin{equation}
\label{eq3}
\vec{r}=\vec{U}t \;\;,
\end{equation}
Equation~\eqref{eq3} allows for some simplification to material-
radiation coupling.
The Lagrangian momentum and energy equations, respectively, are
\begin{equation}
\label{eq4}
\rho\frac{D\vec{U}}{Dt}+\nabla P = -\vec{g} \;\;,
\end{equation}
and
\begin{equation}
\label{eq5}
C_{v}\frac{DT}{Dt}+P\nabla\cdot\vec{U} = -g^{(0)} \;\;,
\end{equation}
where $\rho$ is density, $P$ is fluid pressure, $T$ is
fluid temperature, $C_{v}$ is heat capacity per unit volume,
and $(g^{(0)},\vec{g})$ is a radiation energy-momentum
coupling 4-vector.
Following the justification provided by~\cite{kasen2006} and
\cite{vanrossum2012}, we neglect $P$.
For the time scales and physical specifications of interest,
much more energy is in the radiation field than the material.
Incorporating Eq.~\eqref{eq3} and $P=0$ into Eqs.~\eqref{eq4}
and~\eqref{eq5} yields
\begin{multline}
\label{eq6}
C_{v}\frac{DT}{Dt}=\int_{4\pi}\int_{0}^{\infty}\sigma_{0,\nu_{0},a}
(I_{0,\nu_{0}}-B_{0,\nu_{0}})d\nu_{0}d\Omega_{0}\\
+\int_{4\pi}\int_{0}^{\infty}\sigma_{0,\nu_{0},s}
I_{0,\nu_{0}}d\nu_{0}d\Omega_{0}-
\int_{4\pi}\int_{0}^{\infty}\int_{4\pi}\int_{0}^{\infty}\\
\frac{\nu_{0}}{\nu_{0}'}\sigma_{0,s}
(\vec{r},\nu_{0}'\rightarrow\nu_{0},\hat{\Omega}_{0}'\cdot\hat{\Omega}_{0})
I_{0,\nu_{0}'}d\nu_{0}'d\Omega_{0}'d\nu_{0}d\Omega_{0}\\
=-g_{0,a}^{(0)}-g_{0,s}^{(0)}\;\;,
\end{multline}
where $g_{0,a}^{(0)}$ and $g_{0,s}^{(0)}$ are absorption and
scattering contributions to the comoving radiation-material
coupling, respectively.
Equation~\eqref{eq6} is similar in form to the material equation
presented by~\cite{szoke2005} but with a Lagrangian temporal
derivative.

\section{Doppler Shift Group Edge Analysis}
\label{sec:dopge}

Monte Carlo particles may be tracked by either discrete
groups or continuous values in frequency space.
In the context of relativistic velocity, Doppler shift
has an important effect on the radiation intensity's
interaction with a group structure.
When considering how to track particles through phase
space, it is informative to consider approaches to sustaining
consistency between multigroup transport and multigroup
diffusion.
Specifically IMC may have particle frequency tracked
and updated continuously in a multigroup setting through
explicit changes in reference frame.
In contrast, a DDMC particle wavelength is essentially
unknown within a group since a DDMC particle step in
theory replaces multiple corresponding IMC collision steps.
Hence, each time a continuous frequency value is needed
from a DDMC particle, it must be sampled from a subgroup
distribution~\citep{densmore2012}.
DDMC particles may be tracked with continuous
frequencies or wavelengths but the values then merely serve as a
label for the surrounding group.
Consequently, multigroup IMC may simulate the frequency derivative in
Eq.~\eqref{eq2} exactly while the DDMC scheme described
by~\cite{wollaeger2013} can not exactly simulate the frequency
derivative.
We perform an asymptotic analysis for frequency-dependent,
semi-relativistic, comoving transport with the simplification
of homologous outflow before considering a group grid
that is constant in the comoving frame along with the upwind
redshift approximation~\citep[p.~475]{mihalas1984}.
A group edge of an optically thick region of frequency is
treated in a manner analogous to spatial boundary layers
\citep{habetler1975,malvagi1991}.
Incorporating Eq.~\eqref{eq3} in Eq.~\eqref{eq2},
\begin{multline}
\label{eq7}
\frac{1}{c}\frac{\partial I_{0,\nu_{0}}}{\partial t}+
\hat{\Omega}_{0}\cdot\nabla I_{0,\nu_{0}}+\sigma_{0,\nu_{0}}I_{0,\nu_{0}}\\
-\frac{\nu_{0}}{ct}\frac{\partial I_{0,\nu_{0}}}{\partial\nu_{0}}
+\frac{\vec{r}}{ct}\cdot\nabla I_{0,\nu_{0}}+\frac{3}{ct}I_{0,\nu_{0}}
= j_{0,\nu_{0}} \;\;,
\end{multline}
where $\sigma_{0,\nu_{0}}=\sigma_{0,\nu_{0},a}+\sigma_{0,\nu_{0},s}$
is isotropic, $j_{0,\nu_{0}}$ is the total source due to scattering
and external sources, and the $\hat{\Omega}_{0}\cdot\vec{U}/c$ term
multiplying the Lagrangian derivative has been neglected.
Following prior authors~\citep{habetler1975,malvagi1991},
we introduce a parameter, $\varepsilon\ll 1$, and make the
following scalings: $c\rightarrow c/\varepsilon$, 
$\sigma_{0,\nu_{0}}\rightarrow\sigma_{0,\nu_{0}}/\varepsilon$,
$\sigma_{0,\nu_{0},a}\rightarrow\varepsilon\sigma_{0,\nu_{0},a}$,
$\omega=(\nu-\nu_{b})/\varepsilon^{m}$,
$q\rightarrow\varepsilon q$, where $\nu_{b}$ is
a frequency at boundary $b$ in frequency space and
$q$ is the external or thermal source in $j_{0,\nu_{0}}$.
The value $m$ is a number introduced to control the
amount of variation in intensity with respect to frequency.
If $\partial I_{0,\nu_{0}}/\partial\omega$ is O(1), then
$\partial I_{0,\nu_{0}}/\partial\nu$ is O($1/\varepsilon^{m}$).
Incorporating the scalings into Eq.~\eqref{eq7},
\begin{multline}
\label{eq8}
\frac{\varepsilon^{2}}{c}\frac{\partial I_{0,\nu_{0}}}{\partial t}+
\varepsilon\hat{\Omega}_{0}\cdot\nabla I_{0,\nu_{0}}+
\sigma_{0,\nu_{0}}I_{0,\nu_{0}}-\\
\frac{\varepsilon^{2-m}}{ct}\nu_{0}
\frac{\partial I_{0,\nu_{0}}}{\partial\omega}+
\frac{\varepsilon^{2}}{ct}\vec{r}\cdot\nabla I_{0,\nu_{0}}
+\frac{3\varepsilon^{2}}{ct}I_{0,\nu_{0}}=\varepsilon j_{0,\nu_{0}}\;\;,
\end{multline}
and assuming isotropic elastic scattering,
\begin{equation}
\label{eq9}
\varepsilon j_{0,\nu_{0}} = \varepsilon^{2}\frac{q}{4\pi}+
\left(\sigma_{0,\nu_{0}}-\varepsilon^{2}\sigma_{0,\nu_{0},a}\right)
\frac{1}{4\pi}\int_{4\pi}I_{0,\nu_{0}}d\Omega_{0}' \;\;.
\end{equation}
For our purposes, we need only consider $m\in\{0,1\}$
for an interior group solution ($m=0$) and a frequency
boundary layer solution ($m=1$).
The intensity may then be decomposed as $I_{0,\nu_{0}}=I_{i}+I_{b}$
\citep{malvagi1991} where $I_{i}$ is the interior frequency
solution and $I_{b}$ is the boundary layer frequency solution.
Moreover, all solutions may be expanded as a power
series in $\varepsilon$,
$I_{(i,b)}=\sum_{k=0}^{\infty}I_{(i,b)}^{(k)}\varepsilon^{k}$.
Additionally, we constrain $\lim_{\omega\rightarrow\infty}I_{b}=0$;
this constraint is analogous to the spatial boundary layer
constraint of~\cite{malvagi1991} where the value $\omega$
would instead correspond to distance away from a surface
along a normal vector.

To ensure validity of the stated scalings, we demonstrate
the resulting interior solution is the diffusion approximation
to the semi-relativistic moment equations presented by
\citet[p.~113]{castor2004}.
The interior intensity is subsequently used along with
the boundary layer to obtain the desired result.
Setting $m=0$ and incorporating the power series in $\varepsilon$,
Eq.~\eqref{eq8} may be separated into O($\varepsilon^{0}$),
O($\varepsilon^{1}$), and O($\varepsilon^{2}$) equations:
\begin{equation}
\label{eq10}
I_{i}^{(0)}=\frac{\phi_{i}^{(0)}}{4\pi}
\end{equation}
for O($\varepsilon^{0}$),
\begin{equation}
\label{eq11}
I_{i}^{(1)}=\frac{\phi_{i}^{(1)}}{4\pi}-
\frac{1}{4\pi}\frac{\hat{\Omega}_{0}}{\sigma_{0,\nu_{0}}}\cdot
\nabla\phi_{i}^{(0)}
\end{equation}
for O($\varepsilon^{1}$), and
\begin{multline}
\label{eq12}
I_{i}^{(2)}=\frac{1}{4\pi}\left[\frac{q}{\sigma_{0,\nu_{0}}}+
\phi_{i}^{(2)}-\frac{\sigma_{0,\nu_{0},a}}{\sigma_{0,\nu_{0}}}
\phi_{i}^{(0)}+\right.\\
\frac{\nu_{0}}{ct\sigma_{0,\nu_{0}}}
\frac{\partial\phi_{i}^{(0)}}{\partial\nu_{0}}
-\frac{\vec{r}}{ct\sigma_{0,\nu_{0}}}\cdot\nabla\phi_{i}^{(0)}
-\frac{3}{ct\sigma_{0,\nu_{0}}}\phi_{i}^{(0)}\\
\left.
-\frac{1}{c\sigma_{0,\nu_{0}}}\frac{\partial\phi_{i}^{(0)}}
{\partial t}-\frac{\hat{\Omega}_{0}}{\sigma_{0,\nu_{0}}}\cdot\nabla
\left(\phi_{i}^{(1)}-\frac{\hat{\Omega}_{0}}{\sigma_{0,\nu_{0}}}
\cdot\nabla\phi_{i}^{(0)}\right)
\right]
\end{multline}
for O($\varepsilon^{2}$), where Eq.~\eqref{eq10}
has been used in Eq.~\eqref{eq11} and Eqs.~\eqref{eq10}
and~\eqref{eq11} have been used in Eq.~\eqref{eq12}.
The values $\phi_{i}^{(k)}=\int_{4\pi}I_{i}^{(m)}d\Omega_{0}$
are the $\varepsilon$ power series coefficients for
scalar intensity.
Integrating Eq.~\eqref{eq12} over comoving solid angle,
\begin{multline}
\label{eq13}
\frac{1}{c}\frac{\partial\phi_{i}^{(0)}}{\partial t}-
\nabla\cdot\left(\frac{1}{3\sigma_{0,\nu_{0}}}\nabla\phi_{i}^{(0)}
\right)+\sigma_{0,\nu_{0},a}\phi_{i}^{(0)}\\
-\frac{\nu_{0}}{ct}\frac{\partial\phi_{i}^{(0)}}{\partial\nu_{0}}
+\frac{\vec{r}}{ct}\cdot\nabla\phi_{i}^{(0)}+\frac{3}{ct}\phi_{i}^{(0)}
= q \;\;.
\end{multline}
With some manipulation (by reverting $\vec{r}/t$ to $\vec{U}$
and $3/t$ to $\nabla\cdot\vec{U}$), Eq.~\eqref{eq13} can be
seen to be the diffusion approximation to the zeroth-moment,
frequency-dependent transport equation presented by
\citet[p.~113]{castor2004} under the assumptions of isotropic,
elastic scattering in the comoving frame and homologous flow.

Next we set $m=1$ and asymptotically analyze the frequency
boundary.
In the domain examined, the optically thick region will
be at higher frequency, or $\omega>0$.
Applying the $\varepsilon$ power series again, the
O($\varepsilon^{0}$), O($\varepsilon^{1}$), and O($\varepsilon^{2}$)
equations for $I_{b}$ are
\begin{equation}
\label{eq14}
I_{b}^{(0)}=\frac{\phi_{b}^{(0)}}{4\pi}
\end{equation}
for O($\varepsilon^{0}$),
\begin{equation}
\label{eq15}
I_{b}^{(1)}=\frac{1}{4\pi}\left(\phi_{b}^{(1)}-\frac{\hat{\Omega}_{0}}
{\sigma_{0,\nu_{0}}}\cdot\nabla\phi_{b}^{(0)}+
\frac{\nu_{b}}{ct\sigma_{0,\nu_{0}}}\frac{\partial\phi_{b}^{(0)}}
{\partial\omega}\right)
\end{equation}
for O($\varepsilon^{1}$), and
\begin{multline}
\label{eq16}
I_{b}^{(2)}=\frac{1}{4\pi}\left[
\phi_{b}^{(2)}-\frac{\sigma_{0,\nu_{0},a}}{\sigma_{0,\nu_{0}}}
\phi_{b}^{(0)}+\right.\\
\frac{\nu_{b}}{ct\sigma_{0,\nu_{0}}}
\frac{\partial\phi_{b}^{(1)}}{\partial\omega}
-\frac{\vec{r}}{ct\sigma_{0,\nu_{0}}}\cdot\nabla\phi_{b}^{(0)}
-\frac{3}{ct\sigma_{0,\nu_{0}}}\phi_{b}^{(0)}\\
\left.
-\frac{1}{c\sigma_{0,\nu_{0}}}\frac{\partial\phi_{b}^{(0)}}
{\partial t}-\frac{\hat{\Omega}_{0}}{\sigma_{0,\nu_{0}}}\cdot\nabla
\left(\phi_{b}^{(1)}-\frac{\hat{\Omega}_{0}}{\sigma_{0,\nu_{0}}}
\cdot\nabla\phi_{b}^{(0)}\right)
\right]
\end{multline}
for O($\varepsilon^{2}$), where
$\phi_{b}^{(k)}=\int_{4\pi}I_{b}^{(m)}d\Omega_{0}$.
The term $\partial\phi_{b}^{(0)}/\partial\omega=0$ from
integration of Eq.~\eqref{eq16}; this is an important
result for the remainder of the derivation and has
been used in Eqs~\eqref{eq15} and~\eqref{eq16}.
If Eq.~\eqref{eq16} is integrated, closure for
$\phi_{b}^{(0)}$ is not obtained.
In particular, $\partial\phi_{b}^{(1)}/\partial\omega$ persists.
The O($\varepsilon^{3}$) solution in terms of $I_{b}^{(1,2,3)}$
is
\begin{multline}
\label{eq17}
\frac{1}{c}\frac{\partial I_{b}^{(1)}}{\partial t}+
\hat{\Omega}_{0}\cdot\nabla I_{b}^{(2)}+\sigma_{0,\nu_{0}}I_{b}^{(3)}
-\frac{\nu_{b}}{ct}\frac{\partial I_{b}^{(2)}}{\partial\omega}\\
+\frac{\vec{r}}{ct}\cdot\nabla I_{b}^{(1)}+\frac{3}{ct}I_{b}^{(1)}
=\frac{\sigma_{0,\nu_{0}}}{4\pi}\phi_{b}^{(3)}-
\frac{\sigma_{0,\nu_{0},a}}{4\pi}\phi_{b}^{(1)} \;\;.
\end{multline}
To obtain an equation for $\phi_{b}^{(1)}$, Eq.~\eqref{eq16}
may be incorporated into the second and fourth terms on
the left hand side of Eq.~\eqref{eq17} and the overall
result may be integrated in $\Omega_{0}$.
Upon integration of $\hat{\Omega}_{0}\cdot\nabla I_{b}^{(2)}$,
values in Eq.~\eqref{eq16} that are even in $\hat{\Omega}_{0}$
vanish.
Upon integration of $\partial I_{b}^{(2)}/\partial\omega$,
values in Eq.~\eqref{eq16} that are odd in $\hat{\Omega}_{0}$
vanish.
Fortunately, any terms with $\partial\phi_{b}^{(0)}/\partial\omega$
vanish as well.
The result is
\begin{multline}
\label{eq18}
\frac{1}{c}\frac{\partial\phi_{b}^{(1)}}{\partial t}-
\nabla\cdot\left(\frac{1}{3\sigma_{0,\nu_{0}}}\nabla\phi_{b}^{(1)}
\right)+\sigma_{0,\nu_{0},a}\phi_{b}^{(1)}\\
-\frac{\nu_{b}}{ct}\frac{\partial}{\partial\omega}\left(
\frac{\nu_{b}}{ct\sigma_{0,\nu_{0}}}\frac{\partial\phi_{b}^{(1)}}
{\partial\omega}\right)+\frac{3}{ct}\phi_{b}^{(1)}\\
-\frac{\nu_{b}}{ct}\frac{\partial\phi_{b}^{(2)}}
{\partial\omega}+\frac{\vec{r}}{ct}\cdot\nabla\phi_{b}^{(1)}
 = 0 \;\;.
\end{multline}
The first and fourth terms in Eq.~\eqref{eq18} together
resemble a diffusion equation in frequency space.
The system of equations is still not closed, but Eq.~\eqref{eq17}
along with Eq.~\eqref{eq18} imply
\begin{equation}
\label{eq19}
\frac{\partial}{\partial\omega}\left(
\frac{\nu_{b}}{ct\sigma_{0,\nu_{0}}}\frac{\partial\phi_{b}^{(1)}}
{\partial\omega}\right) = 0 \;\;.
\end{equation}
Taking $\sigma_{0,\nu_{0}}=\sigma_{0,\nu_{b}}$, Eq.~\eqref{eq19}
solves to
\begin{equation}
\label{eq20}
\phi_{b}^{(1)}= \frac{ct\sigma_{0,\nu_{0}}}{\nu_{b}}A_{1}\omega + A_{2}\;\;,
\end{equation}
where $A_{1}$ and $A_{2}$ are constant in $\omega$.
But $\lim_{\omega\rightarrow\infty}\phi_{b}^{(1)}=0$, so
$\phi_{b}^{(1)}=A_{1}=A_{2}=0$.
With $\partial\phi_{b}^{(1)}/\partial\omega=0$, integration
of Eq.~\eqref{eq16} yields
\begin{multline}
\label{eq21}
\frac{1}{c}\frac{\partial\phi_{b}^{(0)}}{\partial t}-
\nabla\cdot\left(\frac{1}{3\sigma_{0,\nu_{0}}}\nabla\phi_{b}^{(0)}
\right)+\sigma_{0,\nu_{0},a}\phi_{b}^{(0)}\\
+\frac{\vec{r}}{ct}\cdot\nabla\phi_{b}^{(0)}+\frac{3}{ct}\phi_{b}^{(0)}
= 0 \;\;.
\end{multline}
Equation~\eqref{eq21} indicates the leading-order boundary
layer solution has no Doppler correction term when
$\partial I_{b}/\partial\nu$ varies strongly (or $m=1$).
Summing Eqs.~\eqref{eq13} and~\eqref{eq21},
\begin{multline}
\label{eq22}
\frac{1}{c}\frac{\partial\phi_{0,\nu_{0}}^{(0)}}{\partial t}-
\nabla\cdot\left(\frac{1}{3\sigma_{0,\nu_{0}}}\nabla\phi_{0,\nu_{0}}^{(0)}
\right)+\sigma_{0,\nu_{0},a}\phi_{0,\nu_{0}}^{(0)}\\
-\frac{\nu_{0}}{ct}\frac{\partial\phi_{i}^{(0)}}{\partial\nu_{0}}
+\frac{\vec{r}}{ct}\cdot\nabla\phi_{0,\nu_{0}}^{(0)}+\frac{3}{ct}
\phi_{0,\nu_{0}}^{(0)} = q \;\;.
\end{multline}
where $\phi_{0,\nu_{0}}=\phi_{i}^{(0)}+\phi_{b}^{(0)}$ is
the uniformly valid leading-order solution.
If the interior solution of the upper frequency range
is constant in frequency, then
\begin{multline}
\label{eq23}
\frac{1}{c}\frac{\partial\phi_{0,\nu_{0}}^{(0)}}{\partial t}-
\nabla\cdot\left(\frac{1}{3\sigma_{0,\nu_{0}}}\nabla\phi_{0,\nu_{0}}^{(0)}
\right)+\sigma_{0,\nu_{0},a}\phi_{0,\nu_{0}}^{(0)}\\
+\frac{\vec{r}}{ct}\cdot\nabla\phi_{0,\nu_{0}}^{(0)}+\frac{3}{ct}
\phi_{0,\nu_{0}}^{(0)} = q \;\;.
\end{multline}
The Doppler correction is removed from the leading-order
scalar intensity equation in the range of frequencies
$\nu_{0}>\nu_{b}$ when the leading-order interior solution
is constant in frequency.
In a piecewise-constant multigroup setting with high-contrast
opacities, the intensity can vary significantly between
groups and might be treated as constant within groups.
Integration of Eq.~\eqref{eq23} over a group interval does
not produce coupling between groups.

We now extend the analysis to problems with an inelastic
scattering component.
The extension is a model that serves to provide theoretical
evidence that group discretization may have a nontrivial
effect on problems with real or effective inelastic
scattering (such as those solved with IMC).
\cite{densmore2011} asymptotically analyzes the effect of
treating some absorption and re-emission as instantaneous
effective scattering while treating the remainder
explicitly with a linear spatial sampling distribution.
We draw an analogy here between elastic scattering, which
preserves $\nu_{0}$, and IMC effective scattering, which
preserves $\vec{r}$.
To complete the analogy, inelastic scattering redistributes
$\nu_{0}$ while IMC effective absorption/emission
redistributes $\vec{r}$.
We now generalize Eq.~\eqref{eq7} to include a pedagogical
model of inelastic scattering in the diffusive upper
frequency range.
This inelastic scattering component is meant to emulate
effective scattering in IMC within one group.
We rewrite Eq.~\eqref{eq7} as
\begin{multline}
\label{eq24}
\frac{1}{c}\frac{\partial I_{0,\nu_{0}}}{\partial t}+
\hat{\Omega}_{0}\cdot\nabla I_{0,\nu_{0}}+\sigma_{0,\nu_{0}}I_{0,\nu_{0}}\\
-\frac{\nu_{0}}{ct}\frac{\partial I_{0,\nu_{0}}}{\partial\nu_{0}}
+\frac{\vec{r}}{ct}\cdot\nabla I_{0,\nu_{0}}+\frac{3}{ct}I_{0,\nu_{0}}
=\frac{q}{4\pi}+\\
\frac{1}{4\pi}(1-\chi)\sigma_{s}\phi_{0,\nu_{0}}+
\frac{1}{4\pi}\chi\sigma_{s}p_{s}(\nu_{0})\phi_{0,g}
\end{multline}
where $\chi\in[0,1]$ is a elastic/inelastic splitting parameter,
$p_{s}(\nu_{0})$ is a probability density function, $\sigma_{s}$
is a frequency independent scattering opacity coefficient, and
$\phi_{0,g}=\int_{\nu_{b}}^{\nu_{t}}\phi_{0,\nu_{0}}d\nu_{0}$.
The value $\nu_{t}$ is the upper bound of the diffusive region.
Constraining $\int_{\nu_{b}}^{\nu_{t}}p_{s}(\nu_{0})d\nu_{0}=1$,
the integral of the total scattering source term over
frequency is $\sigma_{s}\phi_{0,g}$.
Considering Eq.~\eqref{eq24} implies
\begin{multline}
\label{eq25}
\int_{4\pi}\int_{\nu_{b}}^{\nu_{t}}\frac{\nu_{0}}{\nu_{0}'}\sigma_{0,s}
(\vec{r},\nu_{0}'\rightarrow\nu_{0},
\hat{\Omega}_{0}'\cdot\hat{\Omega}_{0})I_{0,\nu_{0}'}
d\nu_{0}'d\Omega_{0}' = \\\frac{1}{4\pi}(1-\chi)\sigma_{s}
\phi_{0,\nu_{0}}+\frac{1}{4\pi}\chi\sigma_{s}
p_{s}(\nu_{0})\phi_{0,g} \;\;,
\end{multline}
a consistent differential scattering opacity is
\begin{multline}
\label{eq26}
\sigma_{0,s}(\vec{r},\nu_{0}'\rightarrow
\nu_{0},\hat{\Omega}_{0}'\cdot\hat{\Omega}_{0}) = \\
\frac{\sigma_{s}}{4\pi}\left[(1-\chi)\delta(\nu_{0}-\nu_{0}')+
\chi\frac{\nu_{0}'}{\nu_{0}}p_{s}(\nu_{0})\right] \;\;,
\end{multline}
where $\delta(\nu_{0}-\nu_{0}')$ is the Dirac distribution.
Thus the total scattering opacity is
\begin{equation}
\label{eq27}
\sigma_{0,\nu_{0},s}=\sigma_{s}\left[(1-\chi)+
\chi\nu_{0}\int_{\nu_{b}}^{\nu_{t}}\frac{p_{s}(\nu_{0}')}{\nu_{0}'}
d\nu_{0}'\right] \;\;.
\end{equation}
Furthermore, we define a secondary distribution,
\begin{equation}
\label{eq28}
\tilde{p}_{s}(\nu_{0})=
\left(\int_{\nu_{b}}^{\nu_{t}}\frac{p_{s}(\nu_{0}')}{\nu_{0}'}d\nu_{0}'
\right)^{-1}\frac{p_{s}(\nu_{0})}{\nu_{0}} \;\;,
\end{equation}
which is shown below to be the O($\varepsilon^{0}$) and
O($\varepsilon^{1}$) frequency dependence of scalar intensity.
We define $\phi_{i,g}$ and $\phi_{b,g}$ as the interior and
boundary scalar intensity group integrated contributions to
to the diffusive range.
Applying the scalings with $m=0$, considering the interior
solution, and setting $\phi_{0,g}=\sum_{k=0}^{\infty}\phi_{0,g}^{(k)}
\varepsilon^{k}=\sum_{k=0}^{\infty}(\phi_{i,g}^{(k)}+\phi_{b,g}^{(k)})
\varepsilon^{k}$, the O($\varepsilon^{0}$), O($\varepsilon^{1}$),
and O($\varepsilon^{2}$) equations for intensity are
\begin{equation}
\label{eq29}
I_{i}^{(0)}=\frac{1}{4\pi}\tilde{p}_{s}(\nu_{0})\phi_{i,g}^{(0)}\;\;,
\end{equation}
\begin{equation}
\label{eq30}
I_{i}^{(1)}=\frac{1}{4\pi}\tilde{p}_{s}(\nu_{0})
\left(\phi_{i,g}^{(1)}-\frac{\hat{\Omega}_{0}}{\sigma_{0,\nu_{0},s}}
\cdot\nabla\phi_{i,g}^{(0)}\right)\;\;,
\end{equation}
and
\begin{multline}
\label{eq31}
I_{i}^{(2)}=\frac{1}{4\pi}\left[\frac{q}{\sigma_{0,\nu_{0},s}}+
\frac{\sigma_{s}}{\sigma_{0,\nu_{0},s}}
[(1-\chi)\phi_{i}^{(2)}+\chi p_{s}(\nu_{0})\phi_{i,g}^{(2)}]
\right.\\
-\frac{\sigma_{0,\nu_{0},a}}{\sigma_{0,\nu_{0},s}}
\tilde{p}_{s}(\nu_{0})\phi_{i,g}^{(0)}+
\frac{\nu_{0}\phi_{i,g}^{(0)}}{ct\sigma_{0,\nu_{0},s}}
\frac{\partial\tilde{p}_{s}}{\partial\nu_{0}}
-\frac{\tilde{p}_{s}(\nu_{0})}{ct\sigma_{0,\nu_{0},s}}
\vec{r}\cdot\nabla\phi_{i,g}^{(0)}
\\
-\frac{3\tilde{p}_{s}(\nu_{0})}{ct\sigma_{0,\nu_{0},s}}
\phi_{i,g}^{(0)}
-\frac{1}{c\sigma_{0,\nu_{0},s}}
\frac{\partial(\tilde{p}_{s}(\nu_{0})\phi_{i,g}^{(0)})}
{\partial t}-\\
\left.
\frac{\tilde{p}_{s}(\nu_{0})}{\sigma_{0,\nu_{0},s}}
\hat{\Omega}_{0}\cdot\nabla\left(\phi_{i,g}^{(1)}-
\frac{\hat{\Omega}_{0}}{\sigma_{0,\nu_{0},s}}
\cdot\nabla\phi_{i,g}^{(0)}\right)
\right]
\end{multline}
respectively.
Integration of Eq.~\eqref{eq31} gives a correct form of
the comoving diffusion equation.
Additionally, Eq.~\eqref{eq31} indicates the Doppler
coupling in the diffusion region is dependent on the
inelastic scattering profile.
The scattering profile determines the leading interior
solution.
For $m=1$, the O($\varepsilon^{0}$) and 
O($\varepsilon^{1}$) equations are
\begin{equation}
\label{eq32}
I_{b}^{(0)}=\frac{1}{4\pi}\tilde{p}_{s}(\nu_{b})\phi_{b,g}^{(0)}
\;\;,
\end{equation}
\begin{multline}
\label{eq33}
I_{b}^{(1)}=\frac{1}{4\pi}\left(
\frac{\sigma_{s}}{\sigma_{0,\nu_{0},s}}
[(1-\chi)\phi_{b}^{(1)}+\chi p_{s}(\nu_{b})\phi_{b,g}^{(1)}]
\right.
\\
\left.
-\frac{\hat{\Omega}_{0}}{\sigma_{0,\nu_{0},s}}\cdot\nabla\phi_{b}^{(0)}+
\frac{\nu_{b}}{ct\sigma_{0,\nu_{0},s}}\frac{\partial\phi_{b}^{(0)}}
{\partial\omega}\right)
\;\;,
\end{multline}
respectively, where it is assumed the inelastic
probability density does not vary strongly in the
boundary layer.
This assumption may be more clearly expressed
as an Taylor expansion of $p_{s}(\nu)$ around $\nu_{b}$
at a point in the boundary layer:
$p_{s}(\nu)=p_{s}(\nu_{b})+\varepsilon\omega
\partial p_{s}(\nu_{b})/\partial\nu$.
Equation~\eqref{eq32} is frequency independent; so
$\partial\phi_{b}^{(0)}/\partial\omega = 0$.
Integration of Eq.~\eqref{eq33} over solid angle yields
\begin{equation}
\label{eq34}
\phi_{b}^{(1)}=\tilde{p}_{s}(\nu_{b})\phi_{b,g}^{(1)}\;\;.
\end{equation}
Equation~\eqref{eq34} implies
$\partial\phi_{b}^{(1)}/\partial\omega = 0$.
Invocation of $\partial\phi_{b}^{(2)}/\partial\omega$
equation was not needed to obtain Eq.~\eqref{eq34}.
The O($\varepsilon^{2}$) boundary layer equation is
\begin{multline}
\label{eq35}
I_{b}^{(2)}=\frac{1}{4\pi}\left[\frac{\sigma_{s}}{\sigma_{0,\nu_{0},s}}
((1-\chi)\phi_{b}^{(2)}+\chi p_{s}(\nu_{b})\phi_{b,g}^{(2)})-
\right.\\
\frac{\sigma_{0,\nu_{0},a}}{\sigma_{0,\nu_{0},s}}
\phi_{b}^{(0)}
-\frac{\vec{r}}{ct\sigma_{0,\nu_{0},s}}\cdot\nabla\phi_{b}^{(0)}
-\frac{3}{ct\sigma_{0,\nu_{0},s}}\phi_{b}^{(0)}-\\
\left.
\frac{1}{c\sigma_{0,\nu_{0},s}}\frac{\partial\phi_{b}^{(0)}}
{\partial t}-\frac{\hat{\Omega}_{0}}{\sigma_{0,\nu_{0},s}}\cdot\nabla
\left(\phi_{b}^{(1)}-\frac{\hat{\Omega}_{0}}{\sigma_{0,\nu_{0},s}}
\cdot\nabla\phi_{b}^{(0)}\right)
\right] \;\;.
\end{multline}
Equation~\eqref{eq35} gives a diffusion equation,
\begin{multline}
\label{eq36}
\frac{1}{c}\frac{\partial\phi_{b}^{(0)}}{\partial t}-
\nabla\cdot\left(\frac{1}{3\sigma_{0,\nu_{0},s}}\nabla\phi_{b}^{(0)}
\right)+\sigma_{0,\nu_{0},a}\phi_{b}^{(0)}\\
+\frac{\vec{r}}{ct}\cdot\nabla\phi_{b}^{(0)}+\frac{3}{ct}\phi_{b}^{(0)}
= \chi\sigma_{s}\left(p_{s}(\nu_{b})\phi_{b,g}^{(2)}-\frac{p_{s}(\nu_{b})}
{\tilde{p}_{s}(\nu_{b})}\phi_{b}^{(2)}\right) \;\;,
\end{multline}
which has an inelastic scattering source from the O($\varepsilon^{2}$)
scalar flux.
Finally, integrating Eq.~\eqref{eq35} over $\Omega_{0}$,
differentiating the result with respect to $\omega$, and
using $\partial\phi_{b}^{(0)}/\partial\omega=\partial
\phi_{b}^{(1)}/\partial\omega=0$ yields
\begin{equation}
\label{eq37}
\frac{\partial\phi_{b}^{(2)}}{\partial\omega} =
\frac{(1-\chi)}{(1-\chi)+\chi\nu_{b}\int_{\nu_{b}}^{\nu_{t}}
\frac{p_{s}(\nu_{0}')}{\nu_{0}'}d\nu_{0}'}
\frac{\partial\phi_{b}^{(2)}}{\partial\omega} \;\;.
\end{equation}
If $\chi = 0$, scattering is entirely elastic and
Eq.~\eqref{eq37} is self-consistent.
Otherwise, Eq.~\eqref{eq37} is solved with
$\partial\phi_{b}^{(2)}/\partial\omega = 0$ (this
may be seen from differentiation of Eq.~\eqref{eq36}
with respect to $\omega$ as well).
The uniformly valid diffusion equation is
\begin{multline}
\label{eq38}
\frac{1}{c}\frac{\partial\phi_{0,\nu_{0}}^{(0)}}{\partial t}-
\nabla\cdot\left(\frac{1}{3\sigma_{0,\nu_{0}}}\nabla\phi_{0,\nu_{0}}^{(0)}
\right)+\sigma_{0,\nu_{0},a}\phi_{0,\nu_{0}}^{(0)}\\
-\frac{\nu_{0}\phi_{i,g}^{(0)}}{ct}\frac{\partial\tilde{p}_{s}}
{\partial\nu_{0}}
+\frac{\vec{r}}{ct}\cdot\nabla\phi_{0,\nu_{0}}^{(0)}+\frac{3}{ct}
\phi_{0,\nu_{0}}^{(0)} = q +\\
\chi\sigma_{s}\left(p_{s}(\nu_{0})\phi_{0,g}^{(2)}-\frac{p_{s}(\nu_{0})}
{\tilde{p}_{s}(\nu_{0})}\phi_{0,\nu_{0}}^{(2)}\right)\;\;.
\end{multline}
where we have made use of $\sigma_{0,\nu_{0}}=\sigma_{0,\nu_{0},s}+$
O($\varepsilon^{2}$).
Photon number density is proportional to $\phi_{0,\nu_{0}}/\nu_{0}$.
Setting $\tilde{\phi}=\phi_{0,\nu_{0}}^{(0)}/\nu_{0}$
gives an equation for number density in the comoving frame:
\begin{multline}
\label{eq39}
\frac{1}{c}\frac{\partial\tilde{\phi}}{\partial t}-
\nabla\cdot\left(\frac{1}{3\sigma_{0,\nu_{0}}}\nabla\tilde{\phi}
\right)+\sigma_{0,\nu_{0},a}\tilde{\phi}\\
-\frac{\phi_{i,g}^{(0)}}{ct}\frac{\partial\tilde{p}_{s}}
{\partial\nu_{0}}
+\frac{\vec{r}}{ct}\cdot\nabla\tilde{\phi}+\frac{3}{ct}
\tilde{\phi} = \frac{q}{\nu_{0}} +\\
\chi\sigma_{s}\left(\frac{p_{s}(\nu_{0})}{\nu_{0}}\phi_{0,g}^{(2)}
-\int_{\nu_{b}}^{\nu_{t}}\frac{p_{s}(\nu_{0}')}{\nu_{0}'}d\nu_{0}'
\phi_{0,\nu_{0}}^{(2)}\right)\;\;.
\end{multline}
Integration of Eq.~\eqref{eq39} causes the inelastic scattering
term on the right hand side to vanish.
Consequently, the Doppler correction is again dependent on
the interior solution but now also on the scattering distribution,
$\tilde{p}_{s}$.
If $\tilde{p}_{s}=1/(\nu_{t}-\nu_{b})$, then the comoving
photon number density diffusion equation has no Doppler
correction term.

The boundary layer solutions do not provide Doppler
corrections in the sense described by
\citet[p.~112]{castor2004}.
We thus focus on the Doppler correction that the interior
solution provides at the group boundary.
Additionally, sufficient inelasticity in collisions, or
$\chi\sim$ O(1) in Eq.~\eqref{eq24}, makes the Doppler
correction dependent on the redistribution profile.

To obtain the upwind approximation for Doppler shift in all
groups, the transport equation may first be group integrated.
We define a frequency grid in the comoving frame with $G$
groups: $\nu_{G+1/2}<\ldots<\nu_{1/2}$.
Integrating Eq.~\eqref{eq7} over a comoving group, $g$,
yields
\begin{multline}
\label{eq40}
\frac{1}{c}\frac{\partial I_{0,g}}{\partial t}+\hat{\Omega}_{0}
\cdot\nabla I_{0,g}+\sigma_{0,g}I_{0,g}+
\frac{4}{ct}I_{0,g}-\\
\frac{1}{ct}
(\nu_{g-1/2}I_{0,\nu_{g-1/2}}-\nu_{g+1/2}I_{0,\nu_{g+1/2}})+
\frac{\vec{r}}{ct}\cdot\nabla I_{0,g} = j_{0,g} \;\;,
\end{multline}
where $I_{0,g}=\int_{\nu_{g+1/2}}^{\nu_{g-1/2}}I_{0,\nu_{0}}d\nu_{0}$,
$\sigma_{0,g}=\int_{\nu_{g+1/2}}^{\nu_{g-1/2}}\sigma_{0,\nu_{0}}
I_{0,\nu_{0}}d\nu_{0}/\int_{\nu_{g+1/2}}^{\nu_{g-1/2}}I_{0,\nu_{0}}d\nu_{0}$,
and $j_{0,g}=\int_{\nu_{g+1/2}}^{\nu_{g-1/2}}j_{0,\nu_{0}}d\nu_{0}$.
In practice, $\sigma_{0,g}$, might be computed with an approximation
since the exact value is dependent on the solution.
Alternatively, one could define the opacity as piecewise constant
in frequency.
Applying the upwind approximation to the edge frequency-dependent
intensity terms yields~\citep[p.~475]{mihalas1984}
\begin{multline}
\label{eq41}
\frac{1}{c}\frac{\partial I_{0,g}}{\partial t}+\hat{\Omega}_{0}\cdot
\nabla I_{0,g}+\sigma_{0,g}I_{0,g}+
\frac{4}{ct}I_{0,g}+\\
\frac{\nu_{g+1/2}}{ct\Delta\nu_{g}}I_{0,g}+
\frac{\vec{r}}{ct}\cdot\nabla I_{0,g} = j_{0,g}+\frac{\nu_{g-1/2}}{ct}
\frac{I_{0,g-1}}{\Delta\nu_{g-1}} \;\;,
\end{multline}
where $\Delta\nu_{g}=\nu_{g-1/2}-\nu_{g+1/2}$.
The upwind approximation may be extended trivially to
find the multigroup form of Eq.~\eqref{eq2}.
The fifth term on the left hand side and the second term
on the right hand side of Eq.~\eqref{eq41} are responsible
for coupling groups through Doppler shifting.
If the group coupling terms in Eq.~\eqref{eq41} are removed,
then the result describes grey multigroup transport in the
context of homologous outflow.
If Eq.~\eqref{eq41} is solved with a grey MC transport
scheme that includes expansion effects (through frame
transformations and spatial grid expansion), then 
a stochastic interpretation must be given to the
Doppler shift group coupling terms.
The diffusion equation corresponding to Eq.~\eqref{eq41}
may be found by integrating Eq.~\eqref{eq41} over
comoving angle and applying Fick's Law,
\begin{multline}
\label{eq41b}
\frac{1}{c}\frac{\partial\phi_{0,g}}{\partial t}-\nabla\cdot
\left(\frac{1}{3\sigma_{0,g}}\nabla\phi_{0,g}\right)+\sigma_{0,g}
\phi_{0,g}\\+\frac{4}{ct}\phi_{0,g}+\frac{\nu_{g+1/2}}{ct\Delta\nu_{g}}
\phi_{0,g}+\frac{\vec{r}}{ct}\cdot\nabla\phi_{0,g} = 4\pi j_{0,g}+\\
\frac{\nu_{g-1/2}}{ct\Delta\nu_{g-1}}\phi_{0,g-1}\;\;,
\end{multline}
where opacities have been assumed piecewise constant in
frequency.
The Doppler correction terms in Eqs.~\eqref{eq41} and~\eqref{eq41b}
can be interpreted as ``Doppler shift opacities'', where
sampling the value $\nu_{g+1/2}/ct\Delta\nu_{g}$ would
induce a particle to transition from group $g$ to group $g+1$.
If an IMC particle samples a Doppler shift event, the
particle's frequency will be updated to an adjacent
group.

Instead of assuming a fully grouped approach, we
implement a Doppler shift scheme in IMC-DDMC that more
closely emulates continuous frequency transport in the
presence of piecewise constant opacities.
We make the constraint in our code that inelastic
redistribution at the subgroup level is uniform,
or
\begin{equation}
\label{eq41c}
p_{s}(\nu_{0}) = \frac{1}{\Delta\nu_{g}} \;\;.
\end{equation}
Considering Eqs.~\eqref{eq28}, and~\eqref{eq29}:
$\tilde{p}_{s}\sim 1/\nu \sim \phi_{i}^{(0)}$, and the
Doppler correction in Eq.~\eqref{eq38} and~\eqref{eq39}
satisfies
\begin{equation}
\label{eq41d}
-\frac{\nu_{0}\phi_{i,g}^{(0)}}{ct}\frac{\partial\tilde{p}_{s}}
{\partial\nu_{0}} = \frac{1}{ct}\phi_{i}^{(0)} \;\;.
\end{equation}
Since the equations for scalar flux in the frequency boundary
layer have no Doppler correction, we assume $I_{b}=0$; the
interior radiation field thus account for all radiation in
the diffusive frequency region.
Then the entire radiation field has the Doppler correction.
Consequently, incorporating Eq.~\eqref{eq41d} into
Eq.~\eqref{eq38}, neglecting higher order scattering terms,
assuming piecewise constant opacities and integrating over
the group range yields
\begin{multline}
\label{eq41e}
\frac{1}{c}\frac{\partial\phi_{0,g}^{(0)}}{\partial t}-
\nabla\cdot\left(\frac{1}{3\sigma_{0,g}}\nabla\phi_{0,g}^{(0)}
\right)+\sigma_{a,g}\phi_{0,g}^{(0)}\\
+\frac{\vec{r}}{ct}\cdot\nabla\phi_{0,g}^{(0)}+
\frac{4}{ct}\phi_{0,g}^{(0)} = q_{g}\;\;.
\end{multline}
Equation~\eqref{eq41e} is Eq.~\eqref{eq41b} without
upwind Doppler shift terms.
We infer that the degree of elasticity (in our model $\chi$)
is important to how DDMC groups redshift to other groups,
particularly when DDMC emulates continuous frequency
transport.
In order to have Eq.~\eqref{eq41e} represent grey diffusion
for the case of one group, we limit Doppler shift of
particles to adjacent groups for problems with
inelastic-dominant collisions, or $\chi\sim$ O(1).
Such a constraint should emulate IMC for problems with
inelastic-dominant collisions.
Assuming a non-zero velocity field exists and inelastic
opacity is large with respect to $\nu_{g+1/2}/ct\Delta\nu_{g}$,
IMC particles would have their frequencies redistributed
many times before streaming to the edge of a group;
this may greatly reduce the occurrence of Doppler shift
between groups in IMC.
In Section~\ref{sec:procs}, we describe a DDMC Doppler
shift scheme that takes into account the degree of
inelasticity in collisions.

\section{Multigroup IMC-DDMC Equations}
\label{sec:mceqs}

Equation~\eqref{eq6} is amenable to the semi-implicit
time difference described by~\cite{fleck1971}.
Moreover, the semi-implicit discretization procedure
may be applied on Eqs.~\eqref{eq2} and~\eqref{eq6} to obtain
IMC equations for the comoving frame.
The multigroup form of Eq.~\eqref{eq6} is
\begin{equation}
\label{eq42}
C_{v}\frac{DT}{Dt}=\sum_{g=1}^{G}\int_{4\pi}\sigma_{a,g}
I_{0,g}d\Omega_{0} - c\sigma_{P}aT^{4}-g_{0,s}^{(0)}
\end{equation}
where $\sigma_{a,g}$ is comoving grouped absorption opacity,
$\sigma_{P}$ is comoving Planck opacity, and we have
compressed the notation of the inelastic scattering
contribution since it is a material source with a treatment
described by~\cite{fleck1971}.
Introducing a parameter $\beta = 4aT^{3}/C_{v}$ and
integrating Eq.~\eqref{eq42} over a time step gives
\begin{multline}
\label{eq43}
(aT^{4})_{n+1}-(aT^{4})_{n}=\\
\int_{t_{n}}^{t_{n+1}}\beta\left(\sum_{g=1}^{G}\int_{4\pi}
\sigma_{a,g}I_{0,g}d\Omega_{0} - c\sigma_{P}aT^{4}-
g_{0,s}^{(0)}\right)dt \;\;,
\end{multline}
where a value subscripted with $n$ implies evaluation
at the beginning of a time step indexed by $n$.
IMC is made semi-implicit and linear within a time
step by setting $\beta=\beta_{n}$,
$\sigma_{a,g}=\sigma_{a,g,n}$, and $\sigma_{P}=\sigma_{P,n}$
\citep{fleck1971,fleck1984}.
Additionally, setting $\Delta t_{n}\bar{I}_{0,g}=
\int_{t_{n}}^{t_{n+1}}I_{0,g}dt$,
$\Delta t_{n}[\alpha T_{n+1}^{4}+(1-\alpha)T_{n}^{4}]=
\int_{t_{n}}^{t_{n+1}}T^{4}$, and
$\Delta t_{n}\bar{g}_{0,s}^{(0)}=\int_{t_{n}}^{t_{n+1}}
g_{0,s}^{(0)}dt$ gives
\begin{multline}
\label{eq44}
aT_{n+1}^{4}-aT_{n}^{4}=
\beta_{n}\Delta t_{n}\sum_{g=1}^{G}\int_{4\pi}\sigma_{a,g,n}
\bar{I}_{0,g}d\Omega_{0}\\
- c\Delta t_{n}\beta_{n}\sigma_{P,n}[\alpha 
aT_{n+1}^{4}+(1-\alpha)aT_{n}^{4}]
-\Delta t_{n}\beta_{n}\bar{g}_{0,s}^{(0)}
\end{multline}
where $\Delta t_{n}=t_{n+1}-t_{n}$ and $\alpha\in[0,1]$
is the standard IMC time centering parameter.
With Eq.~\eqref{eq44}, an expression may be
found for $\alpha aT_{n+1}^{4}+(1-\alpha)aT_{n}^{4}$
that excludes $T_{n+1}$.
Introducing the Fleck factor,
\begin{equation}
\label{eq45}
f_{n}=\frac{1}{1+\alpha\beta_{n}c\Delta t_{n}\sigma_{P,n}}\;\;,
\end{equation}
the time centered $aT^{4}$ is~\citep{abdikamalov2012}
\begin{multline}
\label{eq46}
\alpha aT_{n+1}^{4}+(1-\alpha)aT_{n}^{4} = \\
\frac{1}{c\sigma_{P,n}}
(1-f_{n})\sum_{g=1}^{G}\int_{4\pi}\sigma_{a,g,n}\bar{I}_{0,g}d\Omega_{0}
+f_{n}aT_{n}^{4}\\
-\frac{1}{c\sigma_{P,n}}(1-f_{n})\bar{g}_{0,s}^{(0)}
\;\;.
\end{multline}
By replacing $\bar{I}_{0,g}$ with $I_{0,g}$, the thermal
emission source term for a group $g$ in the comoving
transport equation may be approximated as
\begin{multline}
\label{eq47}
\sigma_{a,g,n}B_{0,g}=\frac{1}{4\pi}caT^{4}\sigma_{a,g,n}
b_{g,n}=\\
(1-f_{n})\frac{\sigma_{a,g,n}b_{g}}{4\pi\sigma_{P}}
\sum_{g'=1}^{G}\int_{4\pi}\sigma_{a,g',n}\bar{I}_{0,g'}d\Omega_{0}
+\frac{\sigma_{a,g,n}b_{g,n}}{4\pi\sigma_{P,n}}f_{n}acT_{n}^{4}\\
-(1-f_{n})\frac{\sigma_{a,g,n}b_{g,n}}{4\pi\sigma_{P,n}}
\bar{g}_{0,s}^{(0)}\;\;.
\end{multline}
Equations~\eqref{eq43}-\eqref{eq47} are not the only way
to semi-implicitly discretize the temperature equation in
time.
Moreover, in certain circumstances it may be appropriate
to apply different approximations in order to avoid problematic
IMC errors.
In particular,~\cite{larsen1987} derive a ``Maximum Principle''
for IMC that supplies a sufficient but not necessary upper
bound on time step sizes.
It follows from their analysis that IMC is not guaranteed to
give a physical result for any possible numerical setup.
If IMC numerical parameters are ill-conditioned, spurious
temperature oscillations and overheating may occur
\citep{mcclarren2012}.
\cite{gentile2011} performs a similar discretization but
linearly expands opacity and $aT^{4}$ from their values
at $n$ to values at $n+1$.
Despite severe approximations~\citep{gentile2011}, the result
is a modified Fleck factor that adapts to the state of the
radiation field.
Instead of expanding material quantities in $T$, an alternative
approach to obtaining the result of~\cite{gentile2011} is to
make a change of variables in the time derivative similar to
that of~\cite{fleck1971}.
Defining
\begin{equation}
\label{eq48}
E_{*}=\frac{1}{c\Delta t_{n}\bar{\sigma}_{P}}\int_{t_{n}}^{t_{n+1}}
\sum_{g=1}^{G}\int_{4\pi}\sigma_{a,g}I_{0,g}d\Omega_{0}dt
\;\;,
\end{equation}
where $\bar{\sigma}_{P}$ is time centered, Equation~\eqref{eq42}
may be stated as
\begin{multline}
\label{eq49}
\frac{1}{\sigma_{P}\tilde{\beta}}
\frac{D}{Dt}[\sigma_{P}(aT^{4}-E_{*})] \\
= \sum_{g=1}^{G}\int_{4\pi}\sigma_{a,g}
I_{0,g}d\Omega_{0} - c\sigma_{P}aT^{4}-g_{0,s}^{(0)}
\;\;,
\end{multline}
where
\begin{equation}
\label{eq50}
\tilde{\beta}=\frac{1}{C_{v}}\left[4aT^{3}+(aT^{4}-E_{*})
\frac{1}{\sigma_{P}}\frac{\partial\sigma_{P}}{\partial T}
\right] \;\;.
\end{equation}
Evaluating $\sigma_{P}\tilde{\beta}$ on the left hand side
of Eq.~\eqref{eq49} at the beginning of a time step,
integrating Eq.~\eqref{eq49} with respect to time,
setting $\int_{t_{n}}^{t_{n+1}}\sigma_{P}aT^{4}=\Delta t_{n}
[\alpha\sigma_{P,n+1}aT_{n+1}^{4}+(1-\alpha)\sigma_{P,n}aT_{n}^{4}]$,
setting $\bar{\sigma}_{P}=\alpha\sigma_{P,n+1}+(1-\alpha)
\sigma_{P,n}$, and setting $\Lambda_{a,n}=\sigma_{P,n}
(aT_{n}^{4}-E_{*})$ give
\begin{multline}
\label{eq51}
\Lambda_{a,n+1}-\Lambda_{a,n}=\\
c\Delta t_{n}\sigma_{P,n}\tilde{\beta}_{n}
\left(-\alpha\Lambda_{a,n+1}-(1-\alpha)\Lambda_{a,n}-\bar{g}_{0,s}^{(0)}
\right) \;\;.
\end{multline}
Defining the Gentile-Fleck factor as
\begin{equation}
\label{eq52}
\tilde{f}_{n}=\frac{1}{1+\alpha\tilde{\beta}_{n}c\Delta_{t}\sigma_{P,n}}
\;\;,
\end{equation}
The time centered emission term is found to be
\begin{multline}
\label{eq53}
\alpha\sigma_{P,n+1}aT_{n+1}^{4}+(1-\alpha)\sigma_{P,n}aT_{n}^{4}=\\
\tilde{f}_{n}\sigma_{P,n}aT_{n}^{4}-(1-\tilde{f}_{n})\bar{g}_{0,s}^{(0)}+
\bar{\sigma}_{P}\left(1-\frac{\sigma_{P,n}}{\bar{\sigma}_{P}}
\tilde{f}_{n}\right)E_{*}\;\;.
\end{multline}
The next simplification is $\sigma_{P,n}/\bar{\sigma}_{P}$ in the
last term on the right hand side of Eq.~\eqref{eq53}.
By incorporating Eq.~\eqref{eq48} for $E_{*}$, Eq.~\eqref{eq48}
may be a substitute for the emission term in the comoving
thermal transport equation.
The value $\tilde{f}_{n}$ may be interpreted in the same manner
as $f_{n}$ to control the amount of effective scattering and
absorption in IMC.
Unfortunately, the form of $\tilde{\beta}_{n}$ allows
$\tilde{f}_{n}$ to be negative.
\cite{gentile2011} constrains $\tilde{f}_{n}\in[0,1]$ by
setting
\begin{multline}
\label{eq54}
\tilde{\beta}_{n}=\\
\frac{1}{C_{v}}\left[4aT_{n}^{3}+
\max\left((aT_{n}^{4}-E_{*})\frac{1}{\sigma_{P,n}}
\left.\frac{\partial\sigma_{P}}{\partial T}\right|_{T_{n}},
0\right)\right]
\end{multline}
Additionally, $E_{*}$ is estimated with the tallied
radiation energy density from time step $n-1$.
Equations~\eqref{eq52} and~\eqref{eq54} are the exact same
equations for the modified Fleck factor derived by~\cite{gentile2011}.
If the Planck opacity decreases with temperature
and the radiation temperature is higher than the
material temperature, then $\tilde{\beta}_{n}>\beta_{n}$
and $\tilde{f}_{n}<f_{n}$.
From Eq.~\eqref{eq54}, it is evident that
$\tilde{f}_{n}\leq f_{n}$ and the Gentile-Fleck factor
always increases effective scattering over the standard
Fleck factor~\citep{gentile2011}.
Unfortunately, the cost of more stability in IMC
temperature update is a decrease in IMC efficiency.
However, hybridizing IMC with a diffusion scheme
mitigates the added cost~\citep{gentile2011}.

It remains to assess whether or not such a modification
to IMC is needed for problems like the W7 SN Ia described
by~\cite{nomoto1984}.
The grey form of the Maximum Principle of~\cite{larsen1987}
is
\begin{equation}
\label{eq55}
\Delta t_{n}\left[ac\sup_{T_{L}<T<T_{U}}\left\{\frac{\sigma_{P}}
{C_{v}}\left(\frac{T_{U}^{4}-T^{4}}{T_{U}-T}-4\alpha T^{3}\right)
\right\}\right]\leq 1 \;\;,
\end{equation}
where $T_{L}$ and $T_{U}$ are physical lower and upper
bounds on temperature.
To reiterate the grey Maximum Principle, Eq.~\eqref{eq55},
provides a sufficient time step limit but is not
necessary~\citep{larsen1987}.
\cite{larsen1987} prove the general form of the IMC Maximum
Principle by induction over the grid of time steps $n$.
If $T_{L}\leq T_{n}\leq T_{U}$ and
$B_{0,\nu_{0}}(T_{L})\leq I_{0,\nu_{0},n}\leq B_{0,\nu_{0}}(T_{U})$
then $T_{L}\leq T_{n+1}\leq T_{U}$ and
$B_{0,\nu_{0}}(T_{L})\leq I_{0,\nu_{0},n+1}\leq B_{0,\nu_{0}}(T_{U})$
if there is no external source of radiation or material
energy.
For $\sigma_{P}/\rho=0.13$ cm$^{2}$/g, $C_{v}/\rho=2.0\times10^{7}$,
$T_{U}=100000$ K, and $T_{L}=10000$ K, the grey Maximum
Principle gives $\Delta t_{n}\leq 0.6$ milliseconds.
The nominal opacity and heat capacity are from the
analytic SN Ia analysis performed by~\cite{pinto2000}.
W7 results in Section~\ref{sec:NumRes} indicate the
modified Fleck factor derived by~\cite{gentile2011} mitigates
temperature instabilities in outer spatial cells at late
time in the SN evolution.

For the remainder of this section (Section~\ref{sec:mceqs}),
we will write down the IMC-DDMC equations with $f_{n}$ but
note that modified IMC-DDMC merely replaces $f_{n}$ with
$\tilde{f}_{n}$.
The multigroup, semi-relativistic IMC equations
in differential form are
\begin{equation}
\label{eq56}
C_{v}\frac{DT}{Dt}=f_{n}\sum_{g=1}^{G}\int_{4\pi}\sigma_{a,g}
I_{0,g}d\Omega_{0}-f_{n}\sigma_{P}caT^{4}-g_{0,s}^{(0)}
\;\;,
\end{equation}
and~\citep[p.~112]{castor2004}
\begin{multline}
\label{eq57}
\left(1+\hat{\Omega}_{0}\cdot\frac{\vec{U}}{c}\right)\frac{1}{c}
\frac{DI_{0,g}}{Dt}+
\hat{\Omega}_{0}\cdot\nabla I_{0,g}\\
+\frac{4}{c}\hat{\Omega}_{0}\cdot\nabla\vec{U}\cdot\hat{\Omega}_{0}
I_{0,g}
-\frac{1}{c}\hat{\Omega}_{0}\cdot\nabla\vec{U}\cdot(\mathbf{I}
-\hat{\Omega}_{0}\hat{\Omega}_{0})\cdot\nabla_{\hat{\Omega}_{0}}I_{0,g}\\
-\frac{1}{c}\hat{\Omega}_{0}\cdot\nabla\vec{U}\cdot\hat{\Omega}_{0}
\left(\nu_{g-1/2}I_{0,\nu_{g-1/2}}-\nu_{g+1/2}I_{0,\nu_{g+1/2}}\right)\\
+(\sigma_{s,g,n}+\sigma_{a,g,n})I_{0,g}=
\frac{f_{n}}{4\pi}\sigma_{a,g,n}b_{0,g,n}acT_{n}^{4}\\+
\frac{b_{0,g,n}\sigma_{a,g,n}}{4\pi\sigma_{P,n}}(1-f_{n})
\sum_{g'=1}^{G}\int_{4\pi}\sigma_{a,g',n}I_{0,g'}d\Omega_{0}'\\+
\int_{\nu_{g+1/2}}^{\nu_{g-1/2}}\int_{4\pi}\int_{0}^{\infty}\frac{\nu_{0}}{\nu_{0}'}\sigma_{s,n}
(\vec{r},\nu_{0}'\rightarrow\nu_{0},
\hat{\Omega}_{0}'\cdot\hat{\Omega}_{0})\times\\
I_{0,\nu_{0}'}d\nu_{0}'d\Omega_{0}'d\nu_{0}
\;\;,
\end{multline}
where $g_{0,s}^{(0)}$ has been grouped back into the
material equation, Eq.~\eqref{eq57}.
Following~\cite{abdikamalov2012}, Eq.~\eqref{eq57} may be
integrated in $\Omega_{0}$ and operator split into a transport
component, a Doppler shift component, and an advection-expansion
component.
Fick's Law may be applied to the transport component to
obtain a diffusion equation.
To obtain a DDMC equation, the diffusion component is
discretized in space to obtain ``leakage opacities''
\citep{densmore2007} which determine the likelihood of a DDMC
particle moving to an adjacent cell.
The DDMC equation is hybridized with solutions to the IMC
equation in space and frequency through an asymptotic
diffusion limit boundary condition and effective scattering,
respectively
\citep{densmore2007,densmore2012,abdikamalov2012,wollaeger2013}.
The operator-split Doppler-shift and advection-expansion
equations are
\begin{multline}
\label{eq58}
\left(\frac{\partial\phi_{0,g}}{\partial t}\right)_{\text{Doppler}}+
\frac{\nabla\cdot\vec{U}}{3}
\phi_{0,g}=\\\frac{\nabla\cdot\vec{U}}{3}
\left(\nu_{g-1/2}\phi_{0,\nu_{g-1/2}}-
\nu_{g+1/2}\phi_{0,\nu_{g+1/2}}\right) \;\;,
\end{multline}
and
\begin{equation}
\label{eq59}
\left(\frac{\partial\phi_{0,g}}{\partial t}\right)_{\text{Adv/Exp}}
+\nabla\cdot(\vec{U}\phi_{0,g}) = 0 \;\;,
\end{equation}
respectively, where $\phi_{0,g}=\int_{4\pi}I_{0,g}d\Omega_{0}$.
Neglecting physical inelastic scattering, on a spatial domain indexed by $j\in\{1\ldots J\}$, the hybrid DDMC
component of the operator split is~\citep{densmore2012,wollaeger2013}
\begin{multline}
\label{eq60}
\frac{1}{c}\frac{\partial\phi_{0,j,g}}{\partial t}+\bigg(
\sum_{j'}\tilde{\sigma}_{j\rightarrow j',g}+(1-\gamma_{j,g,n})(1-f_{j,n})\sigma_{a,j,g,n}\\
+f_{j,n}\sigma_{a,j,g,n}\bigg)\phi_{0,j,g}=f_{j,n}\gamma_{j,g,n}\sigma_{P,j,n}acT_{j,n}^{4}\\
+\frac{1}{V_{j}}\sum_{j'}V_{j'}\sum_{g_{D}'}
\frac{b_{j',g\leftrightarrow g_{D}',n}}{b_{j',g,n}}\sigma_{j'\rightarrow j,g_{D}'}\phi_{0,j',g_{D}'}\\
+\frac{1}{V_{j}}\sum_{j'}\sum_{g_{T}'}\int_{A_{b(j,j')}}\int_{\hat{\Omega}_{0}\cdot\vec{n}<0}
\int_{g\leftrightarrow g_{T}'}G_{\vec{U},b(j,j')}(|\hat{\Omega}_{0}\cdot\vec{n}|)\times\\
P_{b(j,j')}(|\hat{\Omega}_{0}\cdot\vec{n}|)|\hat{\Omega}_{0}\cdot
\vec{n}|I_{0,\nu_{0}}d\nu_{0}d\Omega_{0}d^{2}\vec{r}\\
+\frac{\gamma_{j,g,n}(1-f_{j,n})}{V_{j}}\times\\
\sum_{g_{T}}\int_{V_{j}}\int_{4\pi}\int_{\nu_{g_{T}+1/2}}^{\nu_{g_{T}-1/2}}
\sigma_{0,\nu_{0},a,j,n}I_{0,\nu_{0}}d\nu_{0}d\Omega_{0}d^{3}\vec{r}\\
+\gamma_{j,g,n}(1-f_{j,n})\sum_{g_{D}}\sigma_{a,j,g_{D},n}\phi_{0,j,g_{D}}
\end{multline}
where the subscript $j$ indicates a finite volume or spatially
piecewise-constant evaluation, $\tilde{\sigma}_{j\rightarrow j',g}$
is the leakage opacity for particle transition from cell $j$ to
$j'$, $\gamma_{j,g,n}=b_{j,g,n}\sigma_{a,j,g,n}/\sigma_{P,j,n}$,
$(1-\gamma_{j,g,n})(1-f_{j,n})\sigma_{a,j,g,n}$ is the effective
scattering opacity for scattering out of group $g$, $V_{j}$ is the
volume of cell $j$, $g_{D}$ ($g_{T}$) are group indexes in cell $j$
that are DDMC (IMC), $b_{j',g\leftrightarrow g_{D}',n}$ is the integral
of the normalized Planck function evaluated at $T_{j'}$ and integrated
over the intersection in frequency of the current group, $g$, and a
diffusion group in cell $j'$, $g_{D}'$.
Furthermore, $A_{b(j,j')}$ indicates the area of spatial interface
between an IMC cell $j'$ and the current cell $j$, $\vec{n}$ is a
unit vector normal to surface $A_{b(j,j')}$ pointing from the interior
of cell $j$, $G_{\vec{U},b(j,j')}(\mu)\approx 1+(2/c)\vec{n}\cdot\vec{U}
(\vec{r}_{b},t)(0.55/\mu-1.25\mu)$ is a particle weight modification
factor for semi-relativistic boundaries~\citep{wollaeger2013}, and
$P_{b(j,j')}$ is the probability of IMC to DDMC particle transition
corresponding to the asymptotic diffusion limit boundary condition
\citep{densmore2008,malvagi1991}.
The $\sim$ notation over the leakage opacity indicates it may be a
composite of leakage opacities for DDMC to IMC transitions and
DDMC to DDMC transitions.
The form of the leakage opacity is~\citep{densmore2012}
\begin{multline}
\label{eq61}
\tilde{\sigma}_{j\rightarrow j',g} = \\
\left(\frac{\sum_{g_{D}'}b_{j,g\leftrightarrow g_{D}',n}}{b_{j,g,n}}\right)
\sigma_{j\rightarrow j',g}
+\left(\frac{\sum_{g_{T}'}b_{j,g\leftrightarrow g_{T}',n}}{b_{j,g,n}}\right)
\sigma_{b(j,j'),g}
\end{multline}
where $\sigma_{j\rightarrow j',g}$ is the leakage opacity to
DDMC groups and $\sigma_{b(j,j'),g}$ is the leakage opacity to IMC
groups in cell $j'$.
The pure leakage opacities may themselves be weighted averages
of leakage opacities corresponding to $(j,g)\rightarrow(j',g_{D}')$
and $(j,g)\rightarrow(j',g_{T}')$ transitions.
A resolved form of Eq.~\eqref{eq61} is
\begin{multline}
\label{eq62}
\tilde{\sigma}_{j\rightarrow j',g} =
\left(\frac{\sum_{g_{D}'}
b_{j,g\leftrightarrow g_{D}',n}\sigma_{j\rightarrow j',g\rightarrow g_{D}'}}
{b_{j,g,n}}\right)\\
+\left(\frac{\sum_{g_{T}'}b_{j,g\leftrightarrow g_{T}',n}
\sigma_{b(j,j'),g\rightarrow g_{T}'}}{b_{j,g,n}}\right)
\end{multline}
where the form of $\sigma_{j\rightarrow j',g}$ and
$\sigma_{b(j,j'),g}$ may be solved for in Eq.~\eqref{eq62} from
Eq.~\eqref{eq61}.

\section{Opacity Regrouping}
\label{sec:gregroup}

Opacity regrouping is an optimization of DDMC that may be incorporated
into Eq.~\eqref{eq60} without having to modify the form of the
equation.
The process involves combining DDMC frequency intervals and
properties corresponding to DDMC frequency intervals to make
larger groups.
This scheme was devised by~\cite{densmore2012} as an approximation
of an adaptive threshold frequency between grey DDMC and multigroup
IMC.
Since the set of groups is divided into a DDMC set and an IMC
set, the DDMC groups corresponding to a set of frequency intervals
do not have to match the set of IMC groups corresponding to the
same set of frequency intervals.
Equation~\eqref{eq60} accommodates adaptive grouping, unaligned
groups at spatial boundaries, and opacity regrouping.

To illustrate the opacity regrouping process, we consider a subset
with subindex $l\in\{1\ldots L\}$ of a resolved group structure.
Groups that satisfy given regrouping criteria belong to the subset
and form a group denoted $\cup_{l=1}^{L}g_{l}$.
The union $\cup_{l=1}^{L}$ implies a union of the frequency intervals
for each group index $g_{l}$.
The regrouped absorption opacity is set to
\begin{equation}
\label{eq63}
\sigma_{a,j,\cup_{l}g_{l},n}=\frac{\sum_{l=1}^{L}b_{j,g_{l},n}\sigma_{a,j,g_{l},n}}
{\sum_{l=1}^{L}b_{j,g_{l},n}} \;\;.
\end{equation}
Similarly, the regrouped leakage opacity is
\begin{equation}
\label{eq64}
\tilde{\sigma}_{j\rightarrow j',\cup_{l}g_{l}}=\frac{\sum_{l=1}^{L}
b_{j,g_{l},n}\tilde{\sigma}_{j\rightarrow j',g_{l}}}{\sum_{l=1}^{L}b_{j,g_{l},n}} \;\;.
\end{equation}
Incorporating Eq.~\eqref{eq62} into Eq.~\eqref{eq64} yields
\begin{multline}
\label{eq65}
\tilde{\sigma}_{j\rightarrow j',\cup_{l}g_{l}}=
\left(\sum_{l=1}^{L}b_{j,g_{l},n}\right)^{-1}\times\\
\sum_{l=1}^{L}\left[\sum_{g_{D}'}b_{j,g_{l}\leftrightarrow g_{D}'}
\sigma_{j\rightarrow j',g_{l}\rightarrow g_{D}'}+
\right.\\
\left.
\sum_{g_{T}'}b_{j,g_{l}\leftrightarrow g_{T}',n}
\sigma_{b(j,j'),g_{l}\rightarrow g_{T}'}\right] \;\;.
\end{multline}
If a leakage event from $\cup_{l=1}^{L}g_{l}$ is sampled, the
probability of leaking to an interfacing group $g_{D}'$
is $(\tilde{\sigma}_{j\rightarrow j',\cup_{l}g_{l}}
\sum_{l=1}^{L}b_{j,g_{l},n})^{-1}\sum_{l=1}^{L}b_{j,g_{l}\leftrightarrow g_{D}'}
\sigma_{j\rightarrow j',g_{l}\rightarrow g_{D}'}$.
The regrouped term responsible for the increase in efficiency
over DDMC without regrouping is
\begin{equation}
\label{eq66}
\gamma_{j,\cup_{l}g_{l},n} = \sum_{l=1}^{L}\gamma_{j,g_{l},n}\;\;,
\end{equation}
which reduces overall effective scattering since a
DDMC particle in $g_{l}$ may no longer scatter to $g_{l'}$ if
these groups are in $\cup_{l=1}^{L}g_{l}$.
Equations~\eqref{eq63}-\eqref{eq66} may be used in place of the
non-opacity-regrouped (non-OR) counterparts in Eq.~\eqref{eq60} to solve
Eq.~\eqref{eq60} for a regrouped intensity, $\phi_{0,j,\cup_{l}g_{l}}$.
The values indexed by $g_{D}$ in the last term on the
right hand side of Eq.~\eqref{eq60} correspond
to DDMC groups not used to construct $\cup_{l=1}^{L}g_{l}$.

The cost of regrouping opacities is a loss in accuracy of the
distribution of the radiation field over the groups.
However, the use of the Planck function in weighting
the group quantities for regrouping may suffice when
effective scattering is a dominant interaction.

\section{IMC and DDMC Processes}
\label{sec:procs}

We now summarize the MC implementation of the equations
from Section~\ref{sec:mceqs} for a homologous outflow.
Following~\cite{lucy2005} and~\cite{abdikamalov2012},
IMC particles are streamed in a lab frame and converted to
the fluid frame when a collision is sampled.
To first order in $\vec{U}/c$, IMC particle lab-frame frequency
and direction may be expressed in terms of their comoving counterparts
as~\cite[p.~104]{castor2004}
\begin{equation}
\label{eq67}
\nu=\nu_{0}\left(1+\frac{\hat{\Omega}_{0}\cdot\vec{U}}{c}\right)\;\;,
\end{equation}
and
\begin{equation}
\label{eq68}
\hat{\Omega}=\frac{\hat{\Omega}_{0}+\vec{U}/c}
{1+\hat{\Omega}_{0}\cdot\vec{U}/c} \;\;.
\end{equation}
Equations~\eqref{eq67} and~\eqref{eq68} account for Doppler shift
and aberration, respectively~\citep{lucy2005}.
An opacity $\sigma_{0}$ transforms to a lab frame value, $\sigma$,
with $\sigma=\nu_{0}\sigma_{0}/\nu$~\citep[p.~106]{castor2004}.
Equation~\eqref{eq67} may be used to express opacity in terms of
direction.

Despite occurring in a moving spatial grid, MC processes may be
tracked over an unchanging ``velocity grid''~\citep{kasen2006,wollaeger2013}.
The collision and census IMC velocity distances computed tracking a
particle, labeled $p$, with coordinate $(t_{p},\vec{U}_{p})$ in cell $j$,
in time step $n$, and group $g$ are~\citep{wollaeger2013}
\begin{equation}
\label{eq69}
u_{\text{col}}=\frac{-\ln(\xi)}
{t_{n}(1-\hat{\Omega}_{p}\cdot\vec{U}/c)
((1-f_{n})\sigma_{a,j,g,n}+\sigma_{s,j,g,n})}\;\;,
\end{equation}
\begin{equation}
\label{eq70}
u_{\text{cen}}=c\frac{1}{t_{n}}(t_{n}+\Delta t_{n}-t_{p})
\;\;,
\end{equation}
respectively, where $\xi\in(0,1]$ is a uniformly sampled
random variable.
Eq.~\eqref{eq69} assumes effective absorption
is treated exactly during streaming.
The velocity distance to the boundary of cell $j$ is geometry
dependent.
For one dimensional spherical geometry the velocity distance
to a boundary is
\begin{equation}
\label{eq71}
u_{b}=\begin{cases}
|(U_{j-1/2}^{2}-(1-\mu_{p}^{2})U_{p}^{2})^{1/2}+\mu_{p}U_{p}| \\
\;\;\;\;\text{if }\mu_{p}<-\sqrt{1-(U_{j-1/2}/U_{p})^{2}}\\
\\
(U_{j+1/2}^{2}-(1-\mu_{p}^{2})U_{p}^{2})^{1/2}-\mu_{p}U_{p}\\
\;\;\;\;\text{otherwise}
\end{cases}
\end{equation}
where $\mu_{p}=\hat{\Omega}_{p}\cdot\vec{U}_{p}/|\vec{U}_{p}|$.
A distance required for an IMC particle to stream into
another group through Doppler shift may be incorporated.
In spherical coordinates, the distance to redshift
between groups is~\citep{wollaeger2013}
\begin{equation}
\label{eq72}
u_{\text{Dop}}=
c\left(1-\frac{\nu_{g+1/2}}{\nu_{p}}\right)-
\vec{U}_{p}\cdot\hat{\Omega}_{p}
\end{equation}
for continuous frequency transport.
Converting $\nu_{p}$ from the lab frame to the fluid frame,
$\nu_{0,p}$, in Eq.~\eqref{eq72} yields
$u_{\text{Dop}}=c(1-\nu_{g+1/2}/\nu_{0,p})
(1-\vec{U}_{p}\cdot\hat{\Omega}_{p}/c)$.
Since $\nu_{0,p}\geq\nu_{g+1/2}$ and
$\vec{U}_{p}\cdot\hat{\Omega}_{p}/c<1$, $u_{\text{Dop}}\geq0$.

Each IMC particle has its spatial coordinate stored after
transport.
Thus, the velocity coordinate of each IMC particle must be
updated before or after a transport step~\citep{wollaeger2013}.
If a DDMC region advects into an IMC particle, the IMC particle
is placed on the cell surface so that the IMC-DDMC interface
condition may be applied in the subsequent transport phase.

In DDMC, Eqs.~\eqref{eq58},~\eqref{eq59} and
\eqref{eq60} determine appropriate modifications to DDMC
particle properties.
Eq.~\eqref{eq60} has no velocity terms and may be solved
with static material DDMC~\citep{abdikamalov2012}.
Equation~\eqref{eq58} determines the Doppler correction
to a particle energy weight and frequency.
Our Doppler shift group coupling scheme is:
\begin{enumerate}
\item For each particle: solve Eq.~\eqref{eq58} to
  modify particle energy weight.
  For a homologous expansion, the energy weight is
  multiplied by $e^{-\Delta t_{n}/t_{n}}$.
\item For the particle's current cell and group, $(j,g)$,
  determine the inelastic opacity.
  If only absorption, then $\sigma_{a,j,g,n}$
  is the inelastic opacity.
\item Make a uniformly random sample, $\xi\in[0,1]$.
\item If $\xi\leq\frac{\nu_{g+1/2}}{ct\Delta\nu_{g}}/
  (\frac{\nu_{g+1/2}}{ct\Delta\nu_{g}}+\sigma_{a,j,g,n})$,
  sample comoving frequency in the group then multiply
  comoving frequency by $e^{-\Delta t_{n}/t_{n}}$.
  Otherwise, do not sample or redshift comoving
  frequency.
\end{enumerate}
In the above list, the first step ensures grey outflow
radiation diffusion problems are solved correctly
\citep[p.~474]{mihalas1984}.
If $(j,g)$ only has elastic scattering, then $\nu_{0,p}$
is updated in the same manner as particle energy weight
in IMC and DDMC.
We constrain source particle frequency to be uniform
at the subgroup level; for pure elastic scattering
problems, the fourth step above (with uniformly sampled
frequency) then emulates the cumulative progression
of redshift from elastic scattering in IMC.
In the last portion of Section~\ref{sec:dopge}, it is
found that uniform redistribution in frequency furnishes
a grouped transport equation that can be solved without
coupling groups with Doppler corrections
(see Eqs.~\eqref{eq41d} and~\eqref{eq41e}).
The fourth step heuristically mitigates frequency shift
when redistribution is a strong effect.
In terms of Section~\ref{sec:dopge}, the condition in
the fourth step is similar to $\xi\leq\varepsilon$,
where $\varepsilon$ is the asymptotic parameter that
makes scattering large.

Keeping all terms associated with Doppler shift operator
split from the MC solution of Eq.~\eqref{eq60} makes opacity
regrouping simpler.
Moreover, Doppler shifting for non-OR groups in the operator
split fashion described is permissible despite use of
regrouped groups in Eq.~\eqref{eq60}.
We ensure DDMC particles have a definite non-OR group
before and after the MC solution of Eq.~\eqref{eq60};
this is accomplished by resampling a non-OR group after
a leakage or effective scattering event.
Equation~\eqref{eq59} is solved by advecting DDMC particles
with their velocity cells; cell expansion naturally dilutes
radiation energy density.

Following~\cite{densmore2012} and~\cite{abdikamalov2012},
DDMC is determined to be applicable to a cell-group by a
mean free path threshold, $\tau_{D}$.
Specifically, if the number of mean free paths in a
cell-group is greater than $\tau_{D}$, then the cell-group
may apply DDMC.
Typical values of $\tau_{D}$ are around 3 to 6 mean free
paths per some characteristic cell length (e.g., the
minimum length of a rectangular cell).
For spherical spatial grids we use the radial length,
$\Delta r = t_{n}\Delta U$.
For a three dimensional Cartesian spatial grid, a conservative
value might be the minimum of three orthogonal cell lengths.
In addition to $\tau_{D}$, we introduce a mean free path
threshold, $\tau_{L}$, for regrouping groups.
This parameter is primarily used for testing solution quality
versus degree of opacity regrouping in DDMC.
Elastic scattering is not included in computing the mean free
paths to check against $\tau_{L}$ since it does not couple
DDMC groups.
For a DDMC particle, the opacity regrouping algorithm may
be delineated as:
\begin{enumerate}
\item For each particle: find current cell and group, $(j,g)$, and
  measure the inelastic collision mean free paths.
  For absorption, $t_{n}\Delta U_{j}\sigma_{a,j,g,n}$
  is a measure of effective scattering and effective absorption mean
  free paths.
\item If $t_{n}\Delta U_{j}\sigma_{a,j,g,n}>\tau_{L}$,
  then search about $g$ for neighboring groups
  $g_{l}$ in cell $j$ satisfying 
  $t_{n}\Delta U_{j}\sigma_{a,j,g_{l},n}>\tau_{L}$.
\item For the set of frequencies corresponding to $\cup_{l}g_{l}$
  where $g\in\cup_{l}g_{l}$,
  apply Eqs.~\eqref{eq63},~\eqref{eq65}, and~\eqref{eq66}.
\item Perform a DDMC step for each particle to leak into adjacent
  cell, effectively scatter out of group $\cup_{l}g_{l}$, get
  absorbed, reach census.
\item If not censused, return to first step.
\end{enumerate}

The material temperature field may be updated upon completion
of all particle processes.
The temperature is updated with Eq.~\eqref{eq56} where
$f_{n}\sum_{g=1}^{G}\int_{4\pi}\sigma_{a,g}I_{0,g}$ is estimated
with the tallied particle energy deposition.

We obtain luminosity and spectra in the lab frame
directly from tallying particles~\citep{lucy2005}.
To do so, either a lab frame wavelength grid can be introduced
or the comoving wavelength grid can be repurposed as an
observational grid in the lab frame.
In our scheme, particles are tracked with a lab frame
wavelength in IMC; thus determining the group of the IMC
particle with a comoving group structure requires a
frame transformation.
For IMC, a lab frame spectral tally is unambiguous
since particle direction, $\hat{\Omega}$, is known.
For escaping DDMC particles, we sample direction isotropically
at the surface and use the sampled direction to determine
the lab frame group of the particle.

\section{Numerical Results}
\label{sec:NumRes}

In the following calculations, we
consider one dimensional spherical problems that test
the Gentile-Fleck factor and opacity regrouping in high-velocity
outflow.
Additionally, Section~\ref{sec:W7} explores mixed weighting in
computing group opacities.
In the plot legends, ``HMC'' denotes hybrid
Monte Carlo with opacity regrouping (opacity-regrouped IMC-DDMC);
``Non-OR HMC'' denotes hybrid Monte
Carlo without opacity regrouping (non-OR IMC-DDMC).
The labels ``Standard IMC'' and ``Standard DDMC'' indicate
IMC and DDMC solutions that do not apply the modified Fleck factor.
For all results shown, source particles and particles undergoing
effective scattering have their frequencies uniformly sampled at
the subgroup level.

\subsection{Quasi-Manufactured Verification}
\label{sec:manu}

Our first problem is a test of the Gentile-Fleck factor
using a quasi-manufactured solution~\citep{oberkampf2010} for
grey transport in a high-velocity outflow.
Here, a quasi-manufactured radiation transport solution has an
assumed, or manufactured, radiation energy density profile and, in contrast,
a material temperature that is solved for using the manufactured radiation
energy density and the material equation.
The manufactured source term is incorporated in the radiation transport
equation to counter redshift and preserve the constancy of the manufactured
radiation energy density.
For the numerical regime considered, we obtain a positive definite source
that is simple to implement.
The quasi-manufactured solution provides a benchmark demonstrating
that the Gentile-Fleck factor (or modified Fleck factor) provides
better accuracy relative to the standard Fleck factor.
Specifically, the Gentile-Fleck factor decreases effective absorption,
which mitigates potential violations of the IMC Maximum Principle~\citep{larsen1987}.

Equation~\eqref{eq53} is implemented approximately~\citep{gentile2011}
in an optimized form since computing the derivative of opacity with
respect to temperature may be computationally expensive.
We use $\partial\sigma_{P,j,n}/\partial T\approx
(\sigma_{P,j,n}-(\rho_{j,n}/\rho_{j,n-1})\sigma_{P,j,n-1})/(T_{j,n}-T_{j,n-1})$
for $n\geq 2$, and $\partial\sigma_{P,j,1}/\partial T\approx
(\sigma_{P,j}((1+\varepsilon)T_{j,1})-\sigma_{P,j,1})/(\varepsilon T_{j,1})$
where $\varepsilon$ is a user defined parameter.
The source term from the manufacturing is positive-definite
and yields a solution with non-trivial time dependence.
\cite{gentile2001} provides an analytic solution to a spatially
independent problem that is used as a benchmark for modified
IMC in static material.
The opacity is proportional to $T^{-5}$, implying that increasing
temperature reduces emission.
The manufacturing and outflow are an extension of the solution,
but we find our analytic result somewhat simpler in form.
Assuming pure absorption, integrating the comoving transport
equation (Eq.~\eqref{eq2}) over frequency, and assuming no
spatial dependence yields
\begin{equation}
\label{eq74}
\frac{\partial E}{\partial t}+\frac{4}{t}E = c\sigma(T)
(aT^{4}-E)+S_{m}\;\;,
\end{equation}
and
\begin{equation}
\label{eq75}
C_{v}\frac{\partial T}{\partial t} = c\sigma(T)(E-aT^{4})\;\;,
\end{equation}
where $E$ is radiation energy density and $S_{m}$ is
the manufactured source.
The heat capacity $C_{v}=\rho c_{v}$ and the opacity is
\begin{equation}
\label{eq76}
\sigma(T)=\frac{\kappa\rho}{T^{5}}\;\;,
\end{equation}
where $c_{v}$ and $\kappa$ are constants.
We manufacture the radiation field as constant and
solve Eq.~\eqref{eq75} to obtain a transcendental expression
for temperature and time.
The manufactured source may then be found from
\begin{equation}
\label{eq77}
S_{m}=\frac{4}{t}E+C_{v}\frac{\partial T}{\partial t}
\end{equation}
by adding Eqs.~\eqref{eq74} and~\eqref{eq75}.
It is clear from Eq.~\eqref{eq77} that a monotonically
increasing temperature over all time ensures a positive
definite source.
This should be the case when $T$ is initialized lower
than $(E/a)^{1/4}$.
Fortunately a low initial temperature and high initial
radiation field is the setup that induces the overheating
pathology in standard IMC.
Following the approach of~\cite{gentile2011}, Eq.~\eqref{eq75}
may be re-expressed as
\begin{equation}
\label{eq78}
\left(\frac{(E/a)T}{E/a-T^{4}}-T\right)\frac{\partial T}
{\partial t} = \frac{ac\kappa}{c_{v}}
\end{equation}
where conveniently, $\rho$ cancels through division of
$\sigma(T)$ by $C_{v}$.
Equation~\eqref{eq78} yields
\begin{multline}
\label{eq79}
\frac{1}{4}\sqrt{\frac{E}{a}}\ln\left(\frac{[\sqrt{E/a}+T^{2}]
[\sqrt{E/a}-T_{1}^{2}]}{[\sqrt{E/a}-T^{2}][\sqrt{E/a}+T_{1}^{2}]}
\right) \\
- \frac{1}{2}(T^{2}-T_{1}^{2}) = \frac{ac\kappa}{c_{v}}(t-t_{1}) \;\;,
\end{multline}
where $t_{1}$ and $T_{1}$ are the initial time and material
temperature, respectively.
For material and radiation properties of interest, Eq.~\eqref{eq79},
indicates long equilibration time between the fields.
Specifically, for an initial radiation temperature of
1.70$\times10^{7}$ K, an initial material temperature of 1.16$\times10^{5}$ K,
a specific heat capacity of 9.3$\times10^{17}$ erg/K/g, and
$\kappa=1.42\times10^{35}$ cm$^{2}$K$^{5}$/g, the characteristic
equilibrium time is on the order of 10$^{91}$ seconds.
These numbers are borrowed or adapted from~\cite{gentile2011}.
If the scope of simulation time is much smaller, it may safely
be assumed that $T^{2},T_{1}^{2}\ll (E/a)^{1/2}=T_{r}^{2}$ for
the numbers given.
When the material temperature and initial temperature are much
smaller than the radiation temperature, Eq.~\eqref{eq79} may
be approximated by
\begin{equation}
\label{eq80}
T(t)=T_{r}\left[6\frac{ac\kappa}{c_{v}T_{r}^{2}}(t-t_{1})+
\left(\frac{T_{1}}{T_{r}}\right)^{6}\right]^{1/6}\;\;.
\end{equation}
From Eq.~\eqref{eq77}, the time integrated manufactured radiation
source is approximately
\begin{equation}
\label{eq81}
\Delta t_{n}S_{m,n} = \frac{4}{t_{n}}E\Delta t_{n}+C_{v,n}(T_{n+1}-T_{n})
\;\;,
\end{equation}
for small time steps.
Equation~\eqref{eq81} is positive definite when Eq.~\eqref{eq80}
is used ($T(t_{n})=T_{n}$).

We construct a problem that induces a ``temperature flip''
pathology in standard IMC or DDMC.
In the first time step, standard IMC-DDMC causes over deposition;
this results in the radiation energy density and material temperature
respectively dropping and increasing abruptly despite the more
gradual nature of the actual solution.
Given the strong inverse dependence of opacity on temperature,
emission abruptly becomes low, causing the material temperature to
remain too high for time spans of interest.
\cite{gentile2011} demonstrates this IMC pathology in the context
of static material.
Our problem consists of a homologous outflow over 10 spatial cells
with a maximum speed of 10$^{9}$ cm/s.
The material temperature is uniformly initialized to 1.16$\times10^{5}$ K
and the radiation temperature is initialized to the manufactured
value of 1.70$\times10^{7}$ K.
Starting from an expansion time of 2 days, we compute the MC
results over a 10th of a millisecond, or 
$t\in[2,2+1.1574\times10^{-9}]$ days.
We test both 100 and 1000 time steps in the time span given.
The source, Eq~\eqref{eq81}, is applied uniformly across the
10 spatial cells.
The density is uniform over the spatial domain with a total
constant mass of $M=1\times10^{33}$ g.
Additionally, $\kappa=1.42\times10^{35}$ cm$^{2}$K$^{5}$/g,
$c_{v}=9.3\times10^{17}$ erg/K/g.

\begin{figure}
\subfloat[]{\includegraphics[height=70mm]{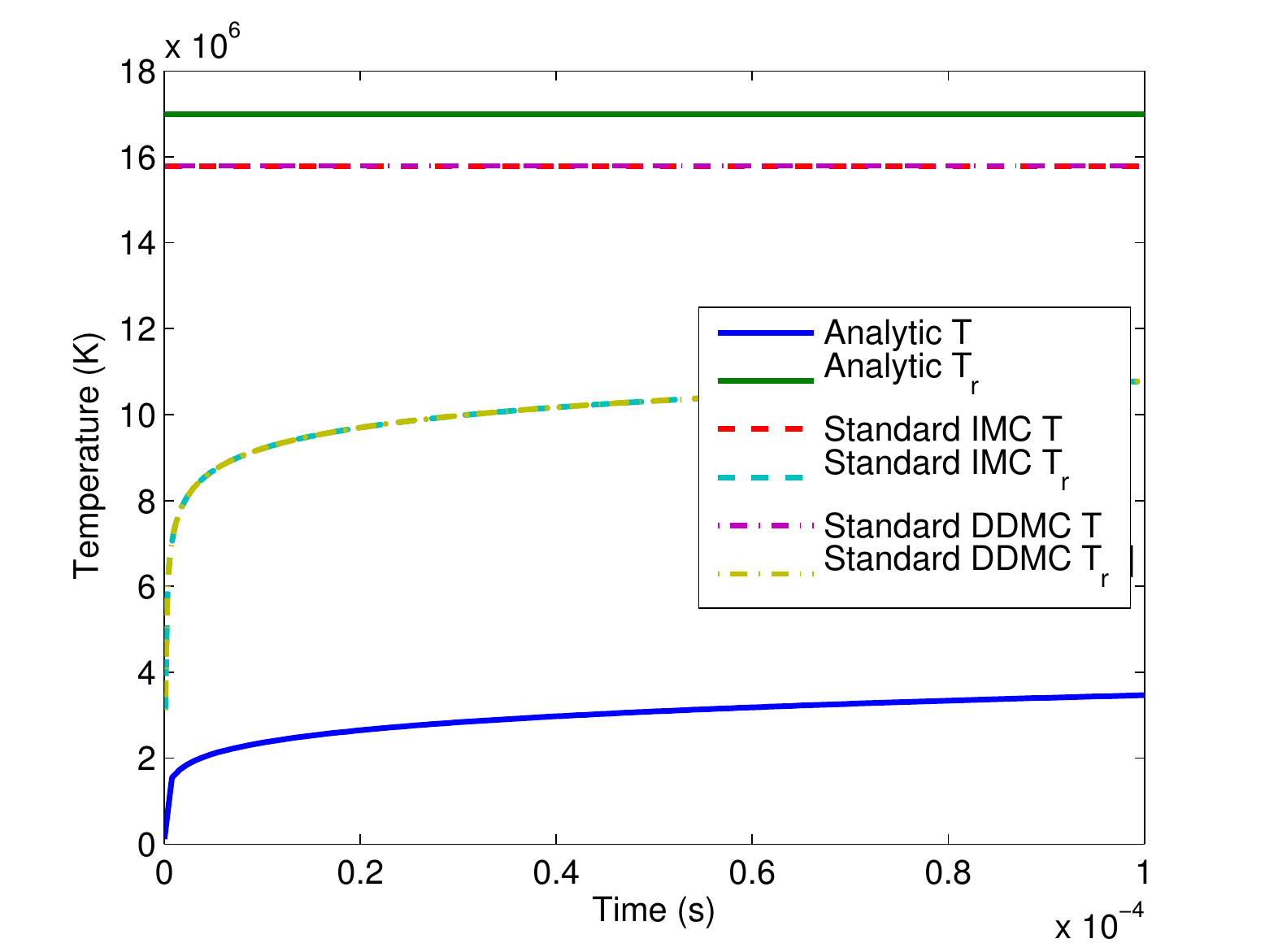}\label{fgA}}\\
\subfloat[]{\includegraphics[height=70mm]{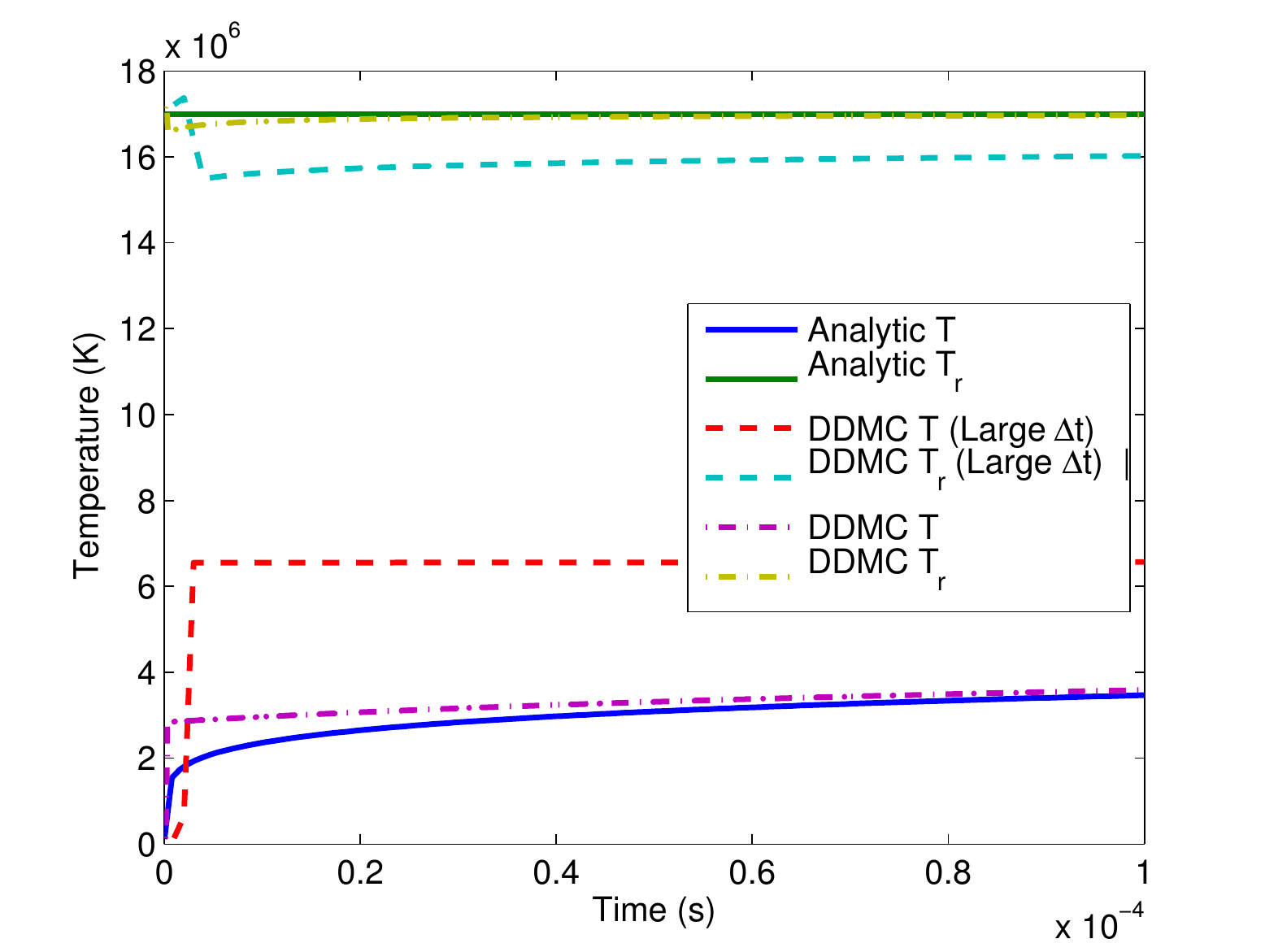}\label{fgB}}
\caption{
For the quasi-manufactured problem described in Section~\ref{sec:manu},
we compare standard IMC, standard DDMC, and DDMC with the modified Fleck factor against
analytic solutions.
In Fig.~\ref{fgA}: analytic (solid), standard IMC (dashed),
and standard DDMC (dash-dotted) material (T)
and radiation (T$_{r}$) temperatures for the 1000 time step
case.
The IMC and DDMC results agree very closely but are both
wrong.
The IMC (dashed light blue) and DDMC (dash-dotted yellow) radiation
temperatures are closer to the analytic material temperature
(solid blue) than the analytic radiation temperature (solid green).
Inversely, the IMC (dashed red) and DDMC (dash-dotted purple)
material temperatures are closer to the analytic radiation
temperature than the analytic material temperature.
This pathology indicates the standard Fleck factor is
insufficient for the time step sizes used.
In Fig.~\ref{fgB}: material (T) and radiation (T$_{r}$)
temperatures from the analytic solution (solid), DDMC with the modified Fleck factor and 100 time
steps (dashed, ``Large $\Delta$t''), and
DDMC with the modified Fleck factor and 1000 time steps (dash-dotted).
The modified Fleck factor prevents the radiation and material
temperatures from ``flipping'' (see Fig.~\ref{fgA}).
Moreover, a decrease in time step size causes further correction
of the MC solutions towards the analytic solution.
}
\label{fgAB}
\end{figure}

Similar to findings of~\cite{gentile2011}, for this test problem
it is found that modified pure IMC is very inefficient;
the Gentile-Fleck factor increases effective scattering in IMC
to a large extent relative to the standard Fleck factor in IMC.
Since grey DDMC does not model effective scattering
\textit{explicitly}, we test the
Gentile-Fleck factor in DDMC; this approach is similar to the use
of RW by~\cite{gentile2011} to accelerate a test calculation.
In Figure~\ref{fgAB}, analytic material temperature is calculated with Eq.~\eqref{eq80}.
The MC temperatures are obtained by implementing the manufactured source, Eq.~\eqref{eq81},
with Eq.~\eqref{eq80} used to evaluate $T_{n}$ and $T_{n+1}$.
For the MC results, the average of the temperature profiles are taken over the 10
spatial cells (temperature change from cell to cell is insignificant, however).
Figure~\ref{fgA} has material and radiation temperature results of
IMC and DDMC with the standard Fleck factor, and the quasi-manufactured solution versus time.
In Fig.~\ref{fgA}, both the IMC and DDMC solutions suffer the ``temperature flip'' error, in which
material temperature becomes non-physically higher than radiation temperature in the first time step.
Figure~\ref{fgB} has material and radiation temperature results
for DDMC with the modified Fleck factor using 100 (denoted ``Large $\Delta t$'') and 1000
time steps.
Results demonstrate the ``temperature flip'' error is avoided for
DDMC modified with the Gentile-Fleck factor.
Increasing the number of time steps from 100 to 1000 further improves
agreement towards the quasi-manufactured solution.
We conclude that the overheating pathology in IMC
and DDMC can occur in high-velocity flows and that
the Gentile-Fleck factor mitigates the overheating error
in high-velocity outflow.
However, the ability of the Gentile-Fleck factor to correct the error
is apparently limited, since in the early time steps the material
temperature becomes too high while the radiation temperature drops too
low relative to the analytic solutions.

\subsection{Ten Group Outflow Test}
\label{sec:heav}

With 10 group, spherical Heaviside source, outflow problems
described by~\cite{wollaeger2013}, we test the effect of opacity
regrouping in IMC-DDMC for simple yet highly structured opacities.
Specifically, we demonstrate the utility of regrouping
non-contiguous groups for radiation transport in a high-velocity
fluid with astrophysical properties.
The approach is described in Section~\ref{sec:gregroup} for LTE
transport.
The form of the opacities is meant to only allow for significant
code speed-up when opacities for non-adjacent frequency intervals can
be regrouped.
With opacity regrouping allowed for non-contiguous group intervals,
a DDMC particle has a probability of being in any group that
satisfies the regrouping criteria; this generalization improves
speed without significant detriment to accuracy relative to
the non-opacity-regrouped (non-OR) results for the numerical
specifications considered.

The problems consist of a homologous outflow with a maximum
outer speed of $U_{\max}=10^{9}$ cm/s.
The time domain of the problem is $t\in [2,5]$ days.
The temperature of the domain is uniformly initialized to
1.16$\times10^{7}$ K.
There is a uniform radiation source density of
$4\times10^{24}/t_{n}^{3}$ erg/cm$^{3}$/s for
$|\vec{U}|\in [0,0.8U_{\max}]$.
The source is uniform in frequency as well.
The total mass is set to $1\times10^{33}$ g equally divided
amongst spatial cells.
The heat capacity is $C_{v}=2\times10^{7}\rho$ erg/cm$^{3}$/K.
The groups are spaced logarithmically from $1.2398\times10^{-9}$
cm to $1.2398\times10^{-3}$ cm in wavelength with $g=1$ being
the lowest wavelength group.
The opacity in cm$^{-1}$ (with $\rho$ in g/cm$^{3}$) is
\begin{equation}
\label{eq82}
\sigma_{a,g}=\begin{cases}
0.13\rho \;\;,\;\;g=2k-1\\
0.13\times 10^{-m}\rho \;\;,\;\;g=2k \;\;,
\end{cases}
\end{equation}
where $k\in\{1\ldots 5\}$ and $m$ is set to 4 or 7~\citep{wollaeger2013}.
For both values of $m$, we use 50 uniform spatial cells, 128 uniform
time steps, 0 initial particles, and 100,000 source particles per
time step.
For all the IMC-DDMC calculations presented, $\tau_{L}=\tau_{D}=3$
mean free paths.

\begin{figure}
\subfloat[]{\includegraphics[height=65mm]{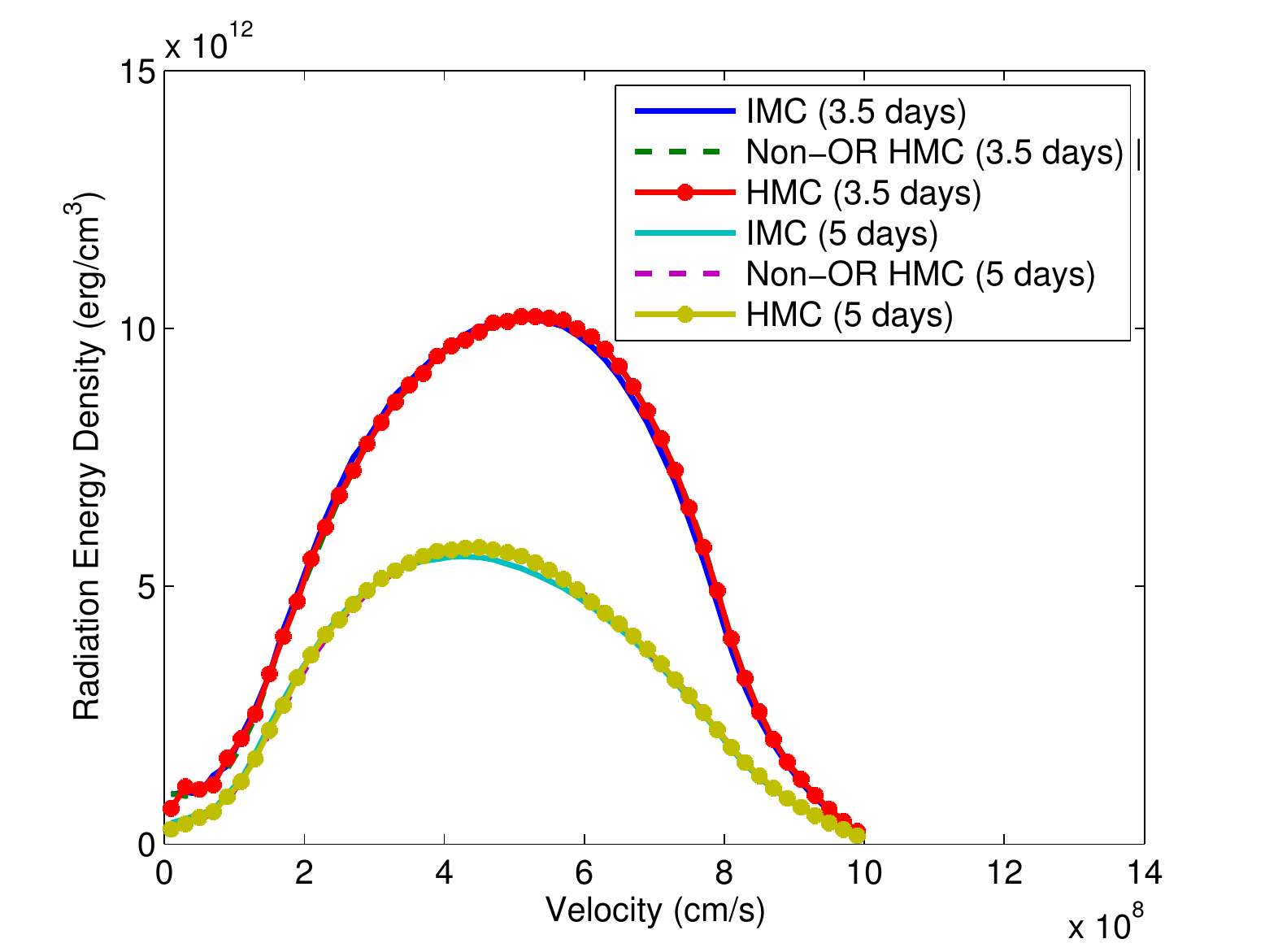}\label{fg1Ha}}\\
\subfloat[]{\includegraphics[height=65mm]{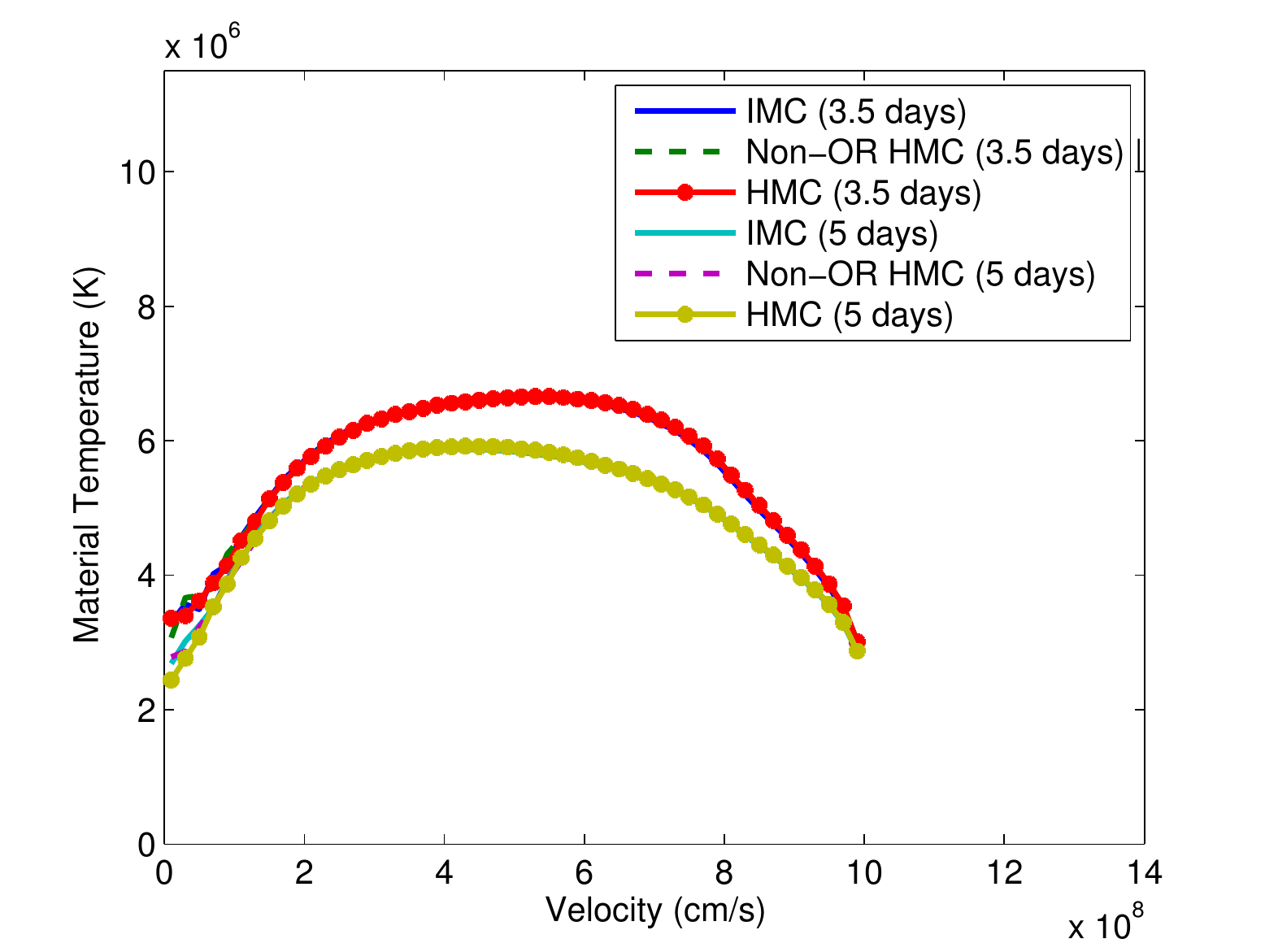}\label{fg1Hb}}\\
\subfloat[]{\includegraphics[height=65mm]{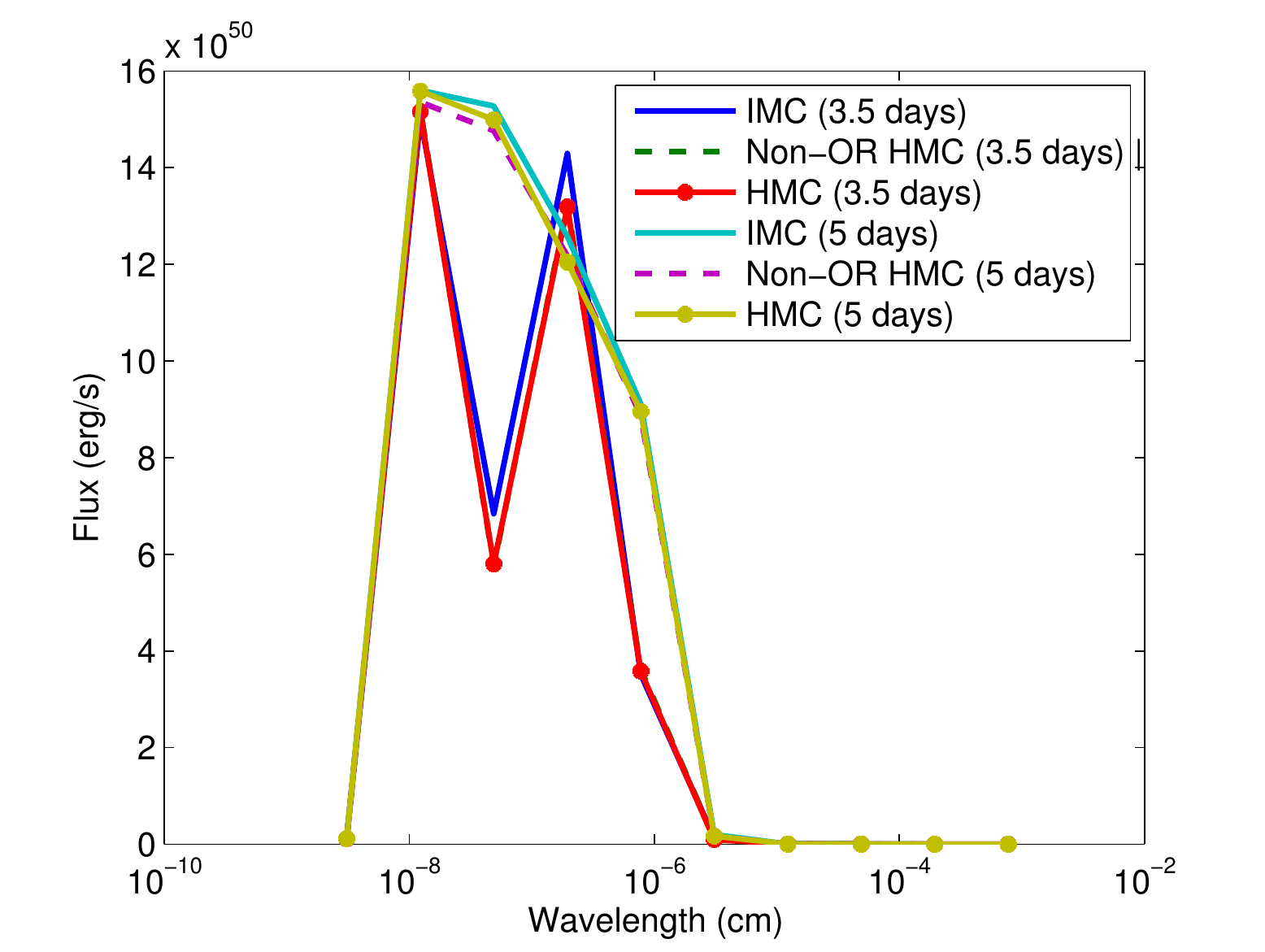}\label{fg1Hc}}
\caption{
Radiation energy density, material temperature, and grouped spectra
of IMC (solid), non-opacity-regrouped (non-OR) IMC-DDMC (dashed, $g_{c}=0$),
and opacity-regrouped IMC-DDMC (dot-solid, $g_{c}=10$) at 3.5 and 5 days for the 10
group, outflow problem with a spherical Heaviside source described
in Section~\ref{sec:heav}.
Radiation energy density and material temperature are plotted versus
fluid velocity and spectra are plotted at group wavelength centers.
The opacity is described by Eq.~\eqref{eq82} with $m=4$.
In Figs.~\ref{fg1Ha},~\ref{fg1Hb}, and~\ref{fg1Hc}, radiation energy
density, material temperature, group spectra, respectively.
For this problem, the IMC-DDMC results with opacity regrouping show good
agreement with the non-OR results.
}
\label{fg1H}
\end{figure}

\begin{figure}
\includegraphics[height=70mm]{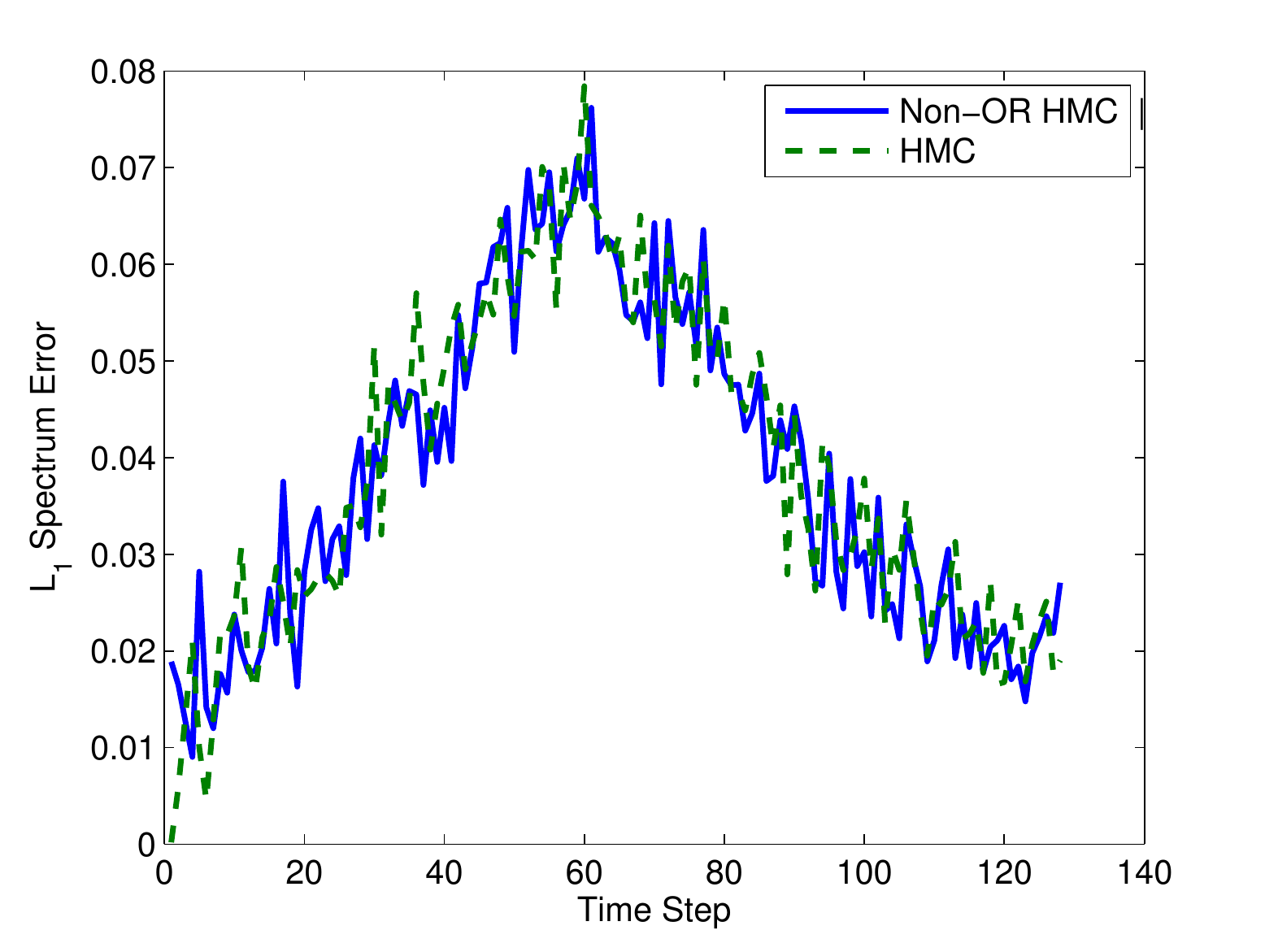}
\caption{
Non-opacity-regrouped (non-OR) IMC-DDMC (solid), and opacity regrouped IMC-DDMC
(dashed) L$_{1}$ error versus time step of group spectra relative to pure IMC
for radiation escaping the outermost cell of the 10 group, Heaviside
source problem described in Section~\ref{sec:heav}.
The opacity is described by Eq.~\eqref{eq82} with $m=4$.
The spectral error for both non-OR and opacity-regrouped IMC-DDMC progressively
increases relative to IMC until pure IMC is applied in the outer cells.
Error from opacity regrouping appears insignificant relative to error from
DDMC.
}
\label{fg2H}
\end{figure}

Considering the $m=4$ disparity, Fig.~\ref{fg1H} has radiation energy
densities and material temperatures for IMC, non-OR IMC-DDMC, and
opacity-regrouped IMC-DDMC; opacity regrouping is not apparently a significant detriment
to these solutions.
In Fig.~\ref{fg2H}, the L$_{1}$ error for the spectra (in erg/s) of
non-OR and opacity-regrouped IMC-DDMC relative to IMC increase while DDMC is
dominant and subsequently decrease as outer cells transition to IMC.
The DDMC approximation for the lab frame spectral tally becomes steadily less
accurate relative to the IMC tally as the cells become optically thin.
The influence of opacity regrouping in the $m=7$ case is similar to
that of the $m=4$ case.
In other words the conclusions from Figs.~\ref{fg1H} and~\ref{fg2H} hold
for the $m=7$ case.

We also incorporate a regrouping cutoff index, $g_{c}$, as an
experimental parameter.
For a group $g$ that meets the regrouping criteria,
only groups in the neighborhood $g\pm g_{c}$ with a number
of mean free paths for inelastic collisions greater than $\tau_{L}$
may have their properties used to accelerate the diffusion of
particles in $g$.
For $t\in[2,3.5]$, we test solution speed versus the cutoff group
displacement for different regrouping cutoffs, $g_{c}\in\{0\ldots10\}$.
Table~\ref{tb1} has times of IMC-DDMC for each $g_{c}$ value
along with the time for IMC.
All times presented are for simulation on one core.
\begin{table}[H]
\caption{Run Times for First 64 Time Steps of Heaviside
Problem with $g_{c}\in\{0\ldots10\}$ with 1 Core (minutes)}
\label{tb1}
\centering
\resizebox{50 mm}{!}{
\begin{tabular}{|l|c|c|c|}
\hline
Method & $g_{c}$ & $m=4$ & $m=7$\\
\hline
IMC & - & 202.23 & 505.71 \\
\hline
HMC & 0 & 23.11 & 45.09 \\
\hline
HMC & 1 & 19.60 & 37.62 \\
\hline
HMC & 2 & 5.80 & 6.74 \\
\hline
HMC & 3 & 5.74 & 6.71 \\
\hline
HMC & 4 & 5.72 & 6.31 \\
\hline
HMC & 5 & 5.79 & 6.44 \\
\hline
HMC & 6 & 5.80 & 6.41 \\
\hline
HMC & 7 & 5.79 & 6.47 \\
\hline
HMC & 8 & 5.83 & 6.42 \\
\hline
HMC & 9 & 5.76 & 6.54 \\
\hline
HMC & 10 & 5.81 & 6.64 \\
\hline
\end{tabular}
}
\end{table}
From Table~\ref{tb1}, it is evident that regrouping only
adjacent groups provides no significant speed up in computation
due to the highly non-monotonic structuring of opacity
versus group.
However, when the regrouping cutoff parameter, $g_{c}$, is
set to 2, there is a significant reduction in computational
cost.

For the problems considered in this section, opacity regrouping
in IMC-DDMC is seen to be a large computational advantage
without large cost of accuracy to important quantities
(spectra and temperatures).
For different problems, the control parameters for opacity
regrouping may need to be adjusted to maintain good agreement
with IMC.
To balance efficiency with solution accuracy, adaptive
regrouping parameters might be considered.
However, for the calculations in the following section,
opacity regrouping is constrained to $\tau_{D}=\tau_{L}=3$
with $g_{c}$ set to the number of groups.

\subsection{W7 Tests}
\label{sec:W7}

We now turn to the W7 problem described by~\cite{nomoto1984} and solved
by several authors [see, e.g.~\citet{kasen2006,kromer2009,vanrossum2012}].
The W7 problem consists of simulating radiative transfer in a one
dimensional model of Type Ia supernovae.
The W7 specifications include density and mass fractions for elements
up to Ni on a velocity grid.
The radial outflow speed at the outer boundary is $\sim$7\% of the speed of light.
In the free-expansion phase of the supernova
radioactive decay of $^{56}$Ni heats the fluid and causes it to radiate
in the UV, visible and infrared ranges of the spectrum.
For this problem, we apply the modified Fleck factor, tested in Section
\ref{sec:manu}, and opacity regrouping, tested in Section~\ref{sec:heav}.
Additionally, we test different calculations of the grouped opacity
by introducing uniform subgroups for each group.
Despite the physical and algorithmic complexities of the opacity,
IMC-DDMC yields light curves and spectra that are in good agreement with those of
{\tt PHOENIX} for the numerical specifications considered.
Moreover, the total computation times are on the order of hours
(see Table~\ref{tb2}).

For IMC-DDMC, a method that in our formulation requires a group structure,
the W7 problem has the difficulty of requiring many groups for accurate
spectra.
Specifically, we find that the number of groups required to achieve a
resolved light curve is on the order of thousands.
While IMC-DDMC is easily extensible to 2 and 3 spatial dimensions in theory,
storing $\sim$10,000 groups per spatial cell is expensive in memory.
Apart from memory overhead, there is the difficult question of spectral
accuracy.
In particular, it may be advantageous to implement adaptive group bounds
so that important portions of the spectrum are properly resolved;
no part of the theory presented precludes adaptive wavelength bounds or
even non-uniform group number per cell.
In this section, we focus mainly on the performance of IMC-DDMC with
opacity regrouping.
We test the effect of mixing reciprocal (Rosseland) and
arithmetic (Planck) computations of the opacity on light curves
and spectra.
Additionally, we show that spikes in the temperature profile at late
time are mitigated with the Gentile-Fleck factor.
However, the application of the Gentile-Fleck factor reveals uncertainty
in the spectra around day 6 post-explosion for the numerical set-up
presented.
For the following simulations, the code \supernu\ is run on
192 cores on the Cray XE6 supercomputer Beagle at the Computation
Institute of the University of Chicago.

In each time step, the opacity per group is computed using a subgroup
structure to allow for non-trivial opacity profile weighting.
Opacity contributions to each group include bound-bound (bb),
bound-free (bf), and free-free (ff) transitions.
Unless otherwise specified, groups are spaced logarithmically while
subgroups are treated uniformly.
Additionally, there is a grey scattering opacity that is isotropic
in the comoving frame calculated as~\citep[p.~161]{castor2004}
\begin{equation}
\label{eq88}
\sigma_{s}=\frac{8\pi}{3}n_{e^{-}}
\left(\frac{e^{-}}{m_{e^{-}}c^{2}}\right)^{2}\;\;,
\end{equation}
where $e^{-}$ is electron charge, $n_{e^{-}}$ is electron number density,
and $m_{e^{-}}$ is electron mass in cgs units.
With mass fractions known \textit{a priori} and given the assumption
of LTE, the Saha-Boltzmann equations are used to obtain the excitation
densities for each atom in the W7 model~\citep[p.~49]{mihalas1984}.
To calculate opacity, we introduce a subgrid for each group $g$ with
index $g_{g}\in\{1\ldots G_{g}\}$.
Values for bb opacities are calculated from oscillator strength data
for each atomic species~\citep{kurucz1994}.
Furthermore, it is assumed that a line is entirely included in the
subgroup its line center is located.
So~\citep[pp.~329-332]{mihalas1984},
\begin{multline}
\label{eq90}
\sigma_{a,g_{g},bb} = \\
\frac{1}{\Delta\lambda_{g_{g}}}\sum_{s}\sum_{i}\sum_{i'>i}
\left(\frac{\pi(e^{-})^{2}}{m_{e^{-}}c}\right)f_{i,i',s}
\frac{\lambda_{i,i',s}^{2}}{c}\times\\
[\Theta(\lambda_{i,i',s}-\lambda_{g_{g}-1/2})-
\Theta(\lambda_{i,i',s}-\lambda_{g_{g}+1/2})]\times\\n_{i,s}
(1-e^{\frac{hc}{kT\lambda_{i,i',s}}}) \;\;,
\end{multline}
where $\sigma_{a,g_{g},bb}$ is the bb contribution to subgroup $g_{g}$,
$f_{i,i',s}$ is the non-dimensional oscillator strength from state $i$ to
$i'$ of species $s$, $\lambda_{i,i',s}$ is the wavelength center of the
line corresponding to the $i\rightarrow i'$ transition, $n_{i,s}$ is the
total density of species $s$ occupying state $i$, and the $\Theta$ are
Heaviside step functions constraining the sum to opacity profiles centered
in the subgroup.
The bound-free opacities are tabulated according to the analytic fit
prescription of~\cite{verner1996}.
We approximate the bf opacity, $\sigma_{a,g_{g},bf}$, of the subgroup as the
value of the fit at the center wavelength in the subgroup.
The ff opacities, $\sigma_{a,g_{g},ff}$, are computed with tabulated Gaunt
factors based on the work of~\cite{sutherland1998} and are similarly
evaluated in the subgroup.
The total absorption opacity for subgroup $g_{g}$ is
$\sigma_{a,g_{g}}=\sigma_{a,g_{g},bb}+\sigma_{a,g_{g},bf}+\sigma_{a,g_{g},ff}$
\citep[p.~332]{mihalas1984}.
The total group opacity may then be averaged in some manner over the
sub group contributions.
We introduce an opacity mixing control parameter $\alpha_{\sigma}\in[0,1]$
to linearly combine reciprocal (``Rosseland type'') and direct
averages of opacity.
Averages of reciprocal opacity may preferentially weight lower
opacity.
For instance, Rosseland opacity is lower than Planck opacity.
For some weight function, $w(\lambda)$, the group absorption opacity
is calculated as
\begin{multline}
\label{eq91}
\sigma_{a,g} = (1-\alpha_{\sigma})\frac{1}{w_{g}}\sum_{g_{g}}^{G_{g}}\sigma_{a,g_{g}}
\int_{\lambda_{g-1/2}}^{\lambda_{g+1/2}}w(\lambda)d\lambda + \\
\frac{\alpha_{\sigma}w_{g}}{\sum_{g_{g}}^{G_{g}}
\sigma_{a,g_{g}}^{-1}\int_{\lambda_{g-1/2}}^{\lambda_{g+1/2}}w(\lambda)d\lambda}\;\;,
\end{multline}
where $w_{g}=\int_{\lambda_{g-1/2}}^{\lambda_{g+1/2}}w(\lambda)d\lambda$.
For a uniform weight function, Eq.~\eqref{eq91} simplifies to
\begin{equation}
\label{eq92}
\sigma_{a,g} = (1-\alpha_{\sigma})\frac{1}{G_{g}}\sum_{g_{g}}^{G_{g}}\sigma_{a,g_{g}}
+
\frac{\alpha_{\sigma}G_{g}}{\sum_{g_{g}}^{G_{g}}1/\sigma_{a,g_{g}}} \;\;.
\end{equation}
If LTE is considered, the weight function might be set to the normalized
Planck function; in this case Eq.~\eqref{eq91} is a mix of grouped Planck
and Rosseland opacities.

For the W7 tests discussed, gamma ray energy deposition profiles and the initial
material and radiation temperatures are borrowed from the \phoenix\ code
\citep{hauschildt1992,hauschildt1999,hauschildt2004,vanrossum2012}.
We estimate and apply a nominal value of heat capacity of $C_{v}=2.0\times10^{7}\rho$
erg/K/cm$^{3}$ from~\cite{pinto2000} to compute the Fleck factor and update the
material temperature.
It has been found that changing $C_{v}$ by a factor of 3 does not change
temperatures and spectra; the insignificance of $C_{v}$ is attributable to the
disparity of energy storage between the radiation and material fields.
In the W7 problem, the Fleck factor is found to be very small in IMC and IMC-DDMC.
Consequently, even a modest group resolution in IMC causes effective scattering
to dominate particle processes.
For the W7 tests attempted, it is apparently unfeasible to use pure IMC, non-OR
IMC-DDMC, or even IMC-DDMC where opacity regrouping is limited to adjacent groups.
For a 100 group W7 simulation with groups logarithmically spaced from $1\times10^{-6}$
cm to $3.2\times10^{-4}$ cm, 64 velocity cells spaced uniformly from 0 cm/s to
$2.2027\times10^{9}$ cm/s, a time domain of $t\in[40,64]$ days post explosion with
0.25 day time steps, 250,000 initial particles, and 250,000 source particles per time step,
neither IMC nor non-OR IMC-DDMC completed the simulation with 192 cores and a wall time
of 40 hours each.
In contrast, fully opacity-regrouped IMC-DDMC ($g_{c}=100$) completed the same problem with 24 cores
in 1018.9 seconds.
For the scope of this paper, we focus our attention to opacity-regrouped IMC-DDMC simulations.

\begin{figure*}
\subfloat[]{\includegraphics[height=65mm]{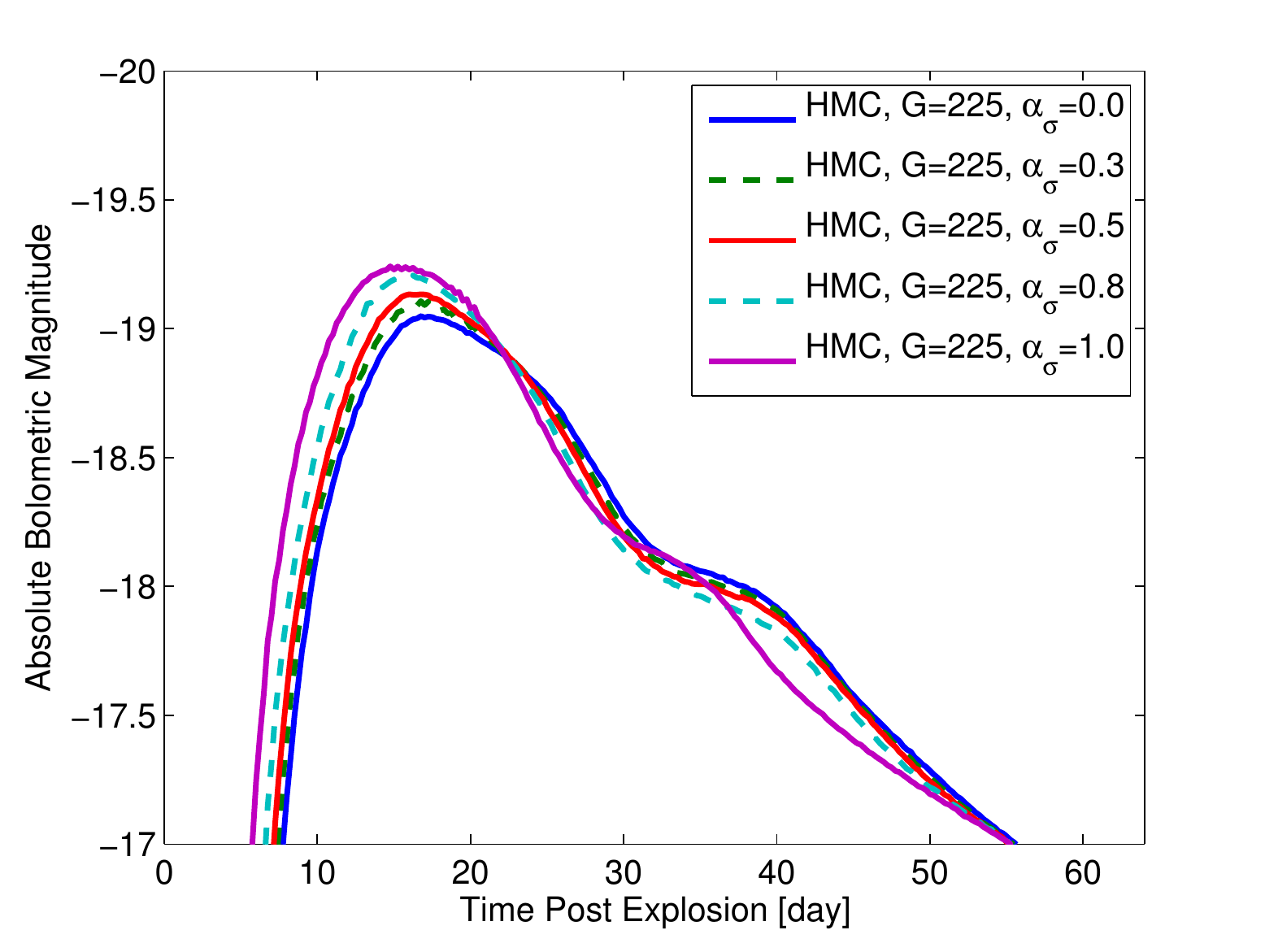}\label{fg4a}}
\subfloat[]{\includegraphics[height=65mm]{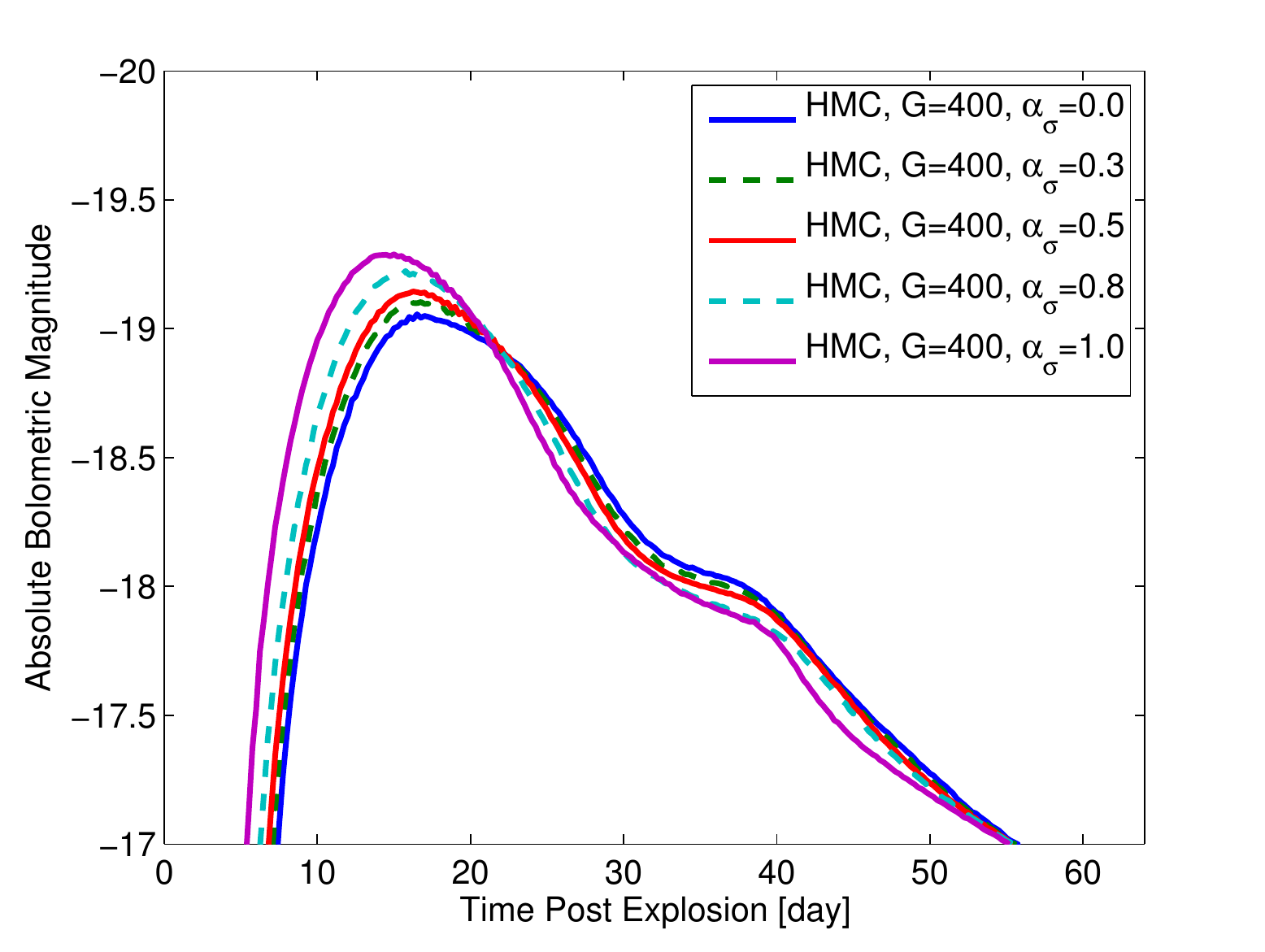}\label{fg4b}}\\
\subfloat[]{\includegraphics[height=65mm]{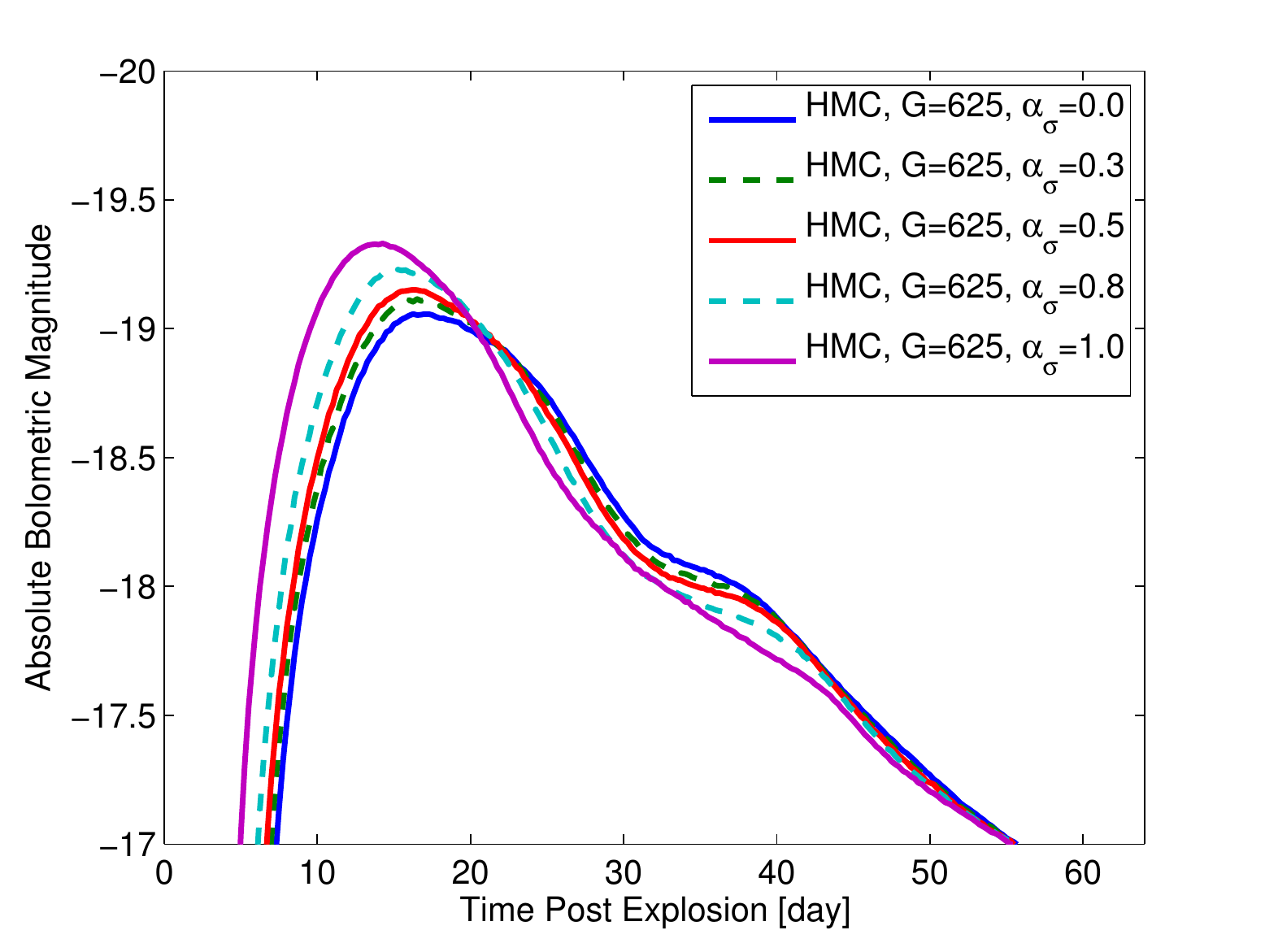}\label{fg4c}}
\subfloat[]{\includegraphics[height=65mm]{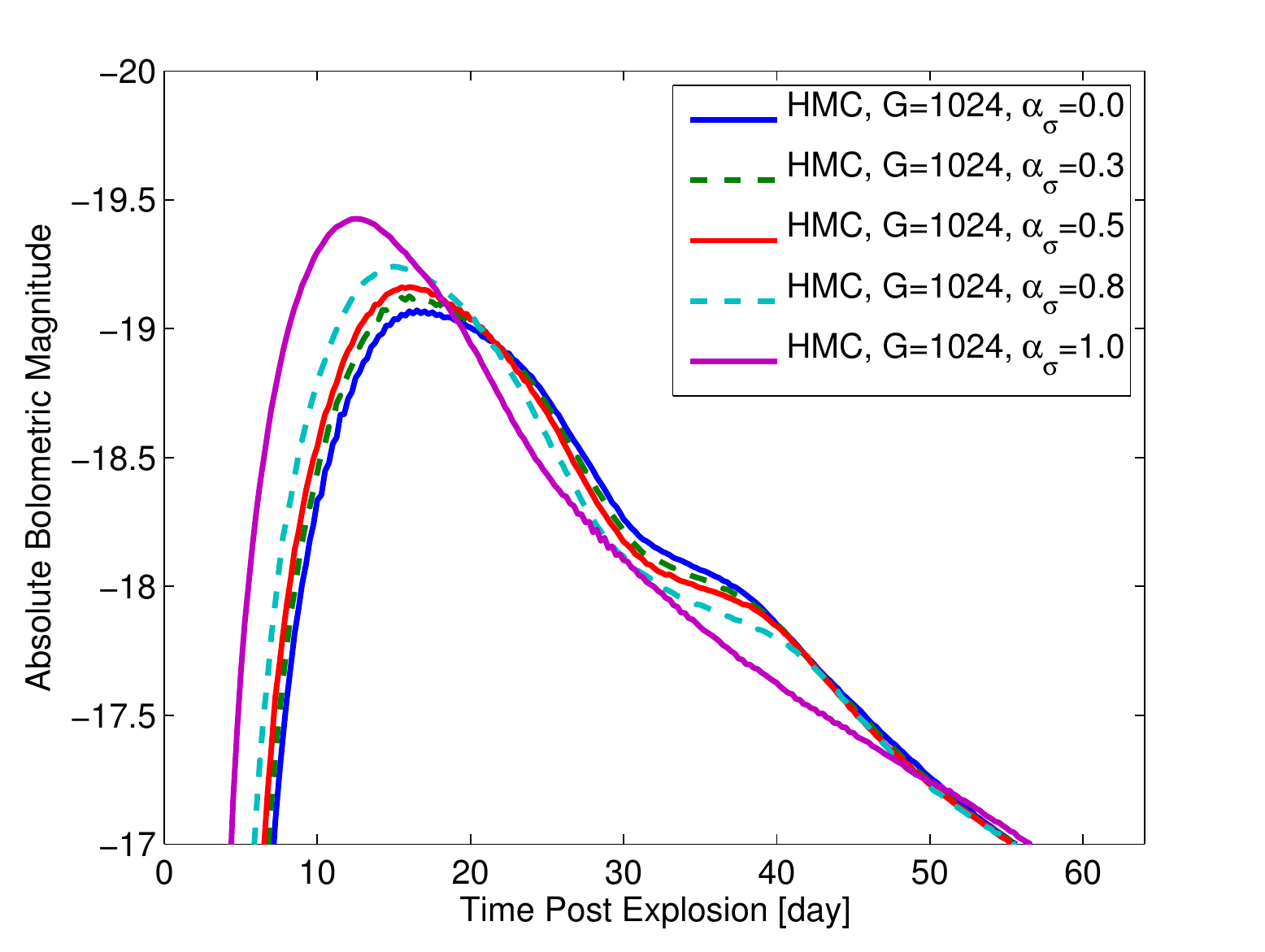}\label{fg4d}}
\caption{
Opacity-regrouped IMC-DDMC W7 bolometric light curves for opacity mixing
$\alpha_{\sigma}\in\{0.0,0.5,1.0\}$ (solid) and $\alpha_{\sigma}\in\{0.3,0.8\}$
(dashed; so solid and dashed curves alternate versus $\alpha_{\sigma}$)
and a fixed number of subgroups, $G_{g}=20$.
Equation~\eqref{eq92} has been applied for opacity mixing.
Light curves are calculated by tallying particles that have escaped the
spatial (velocity) domain per time step and applying Eq.~\eqref{eq93}.
In Figs.~\ref{fg4a},~\ref{fg4b},~\ref{fg4c}, and~\ref{fg4d} group
resolutions are $G=225$, $G=400$, $G=625$, and $G=1024$, respectively.
As expected, peak luminosity is earlier and brighter for opacity mixing
that favors reciprocal averaging since smaller subgroup opacity values
are favored.
Values of $\alpha_\sigma$ close to 1 are not realistic as opacities of
strong absorption lines are more and more neglected.
The opacity mixing parameter can be used to calibrate simulations with
modest group resolution to emulate the diffusion characteristics of
equivalent high-resolution simulations.
}
\label{fg4}
\end{figure*}

\begin{figure*}
\subfloat[]{\includegraphics[height=65mm]{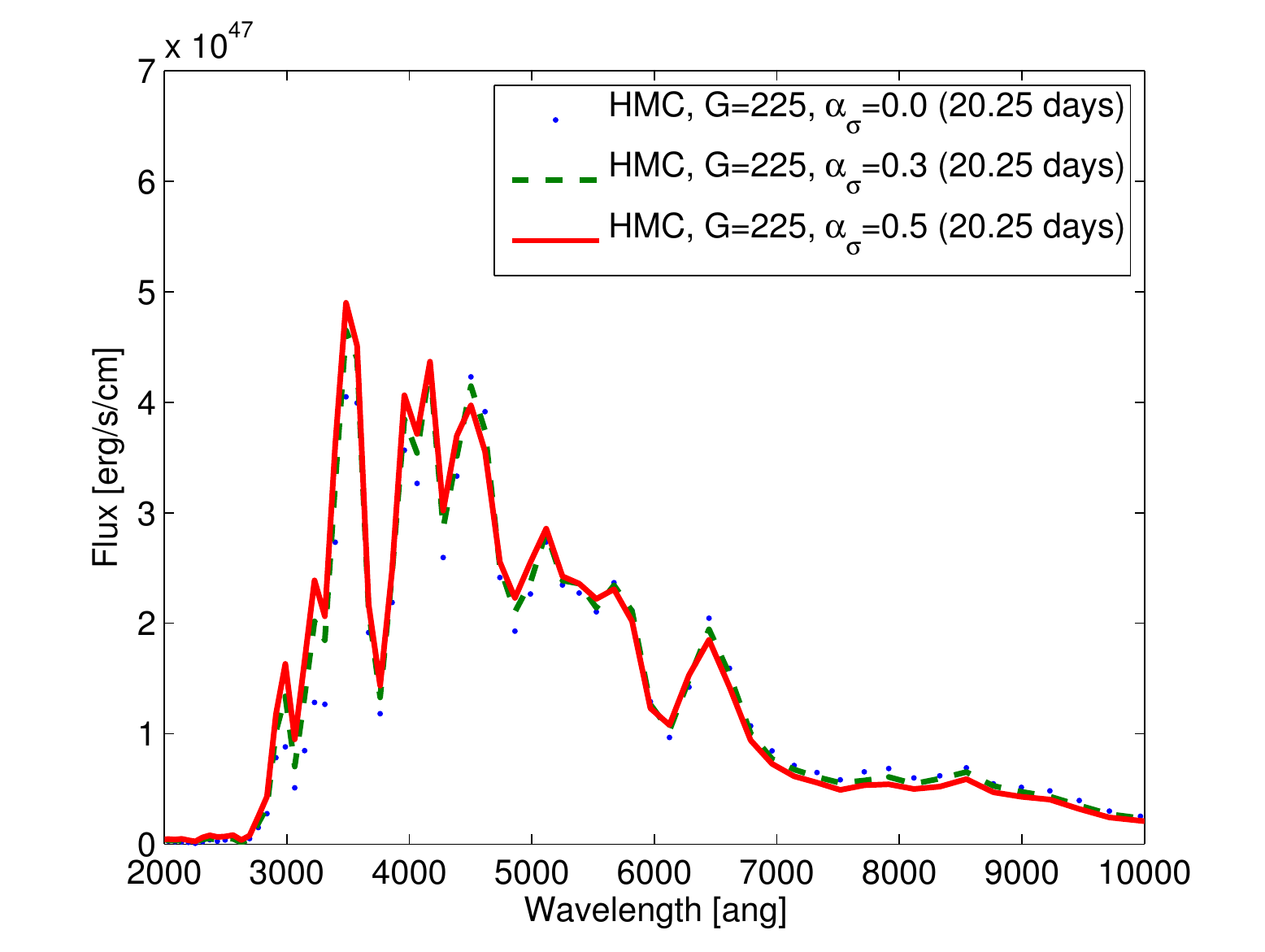}\label{fg5a}}
\subfloat[]{\includegraphics[height=65mm]{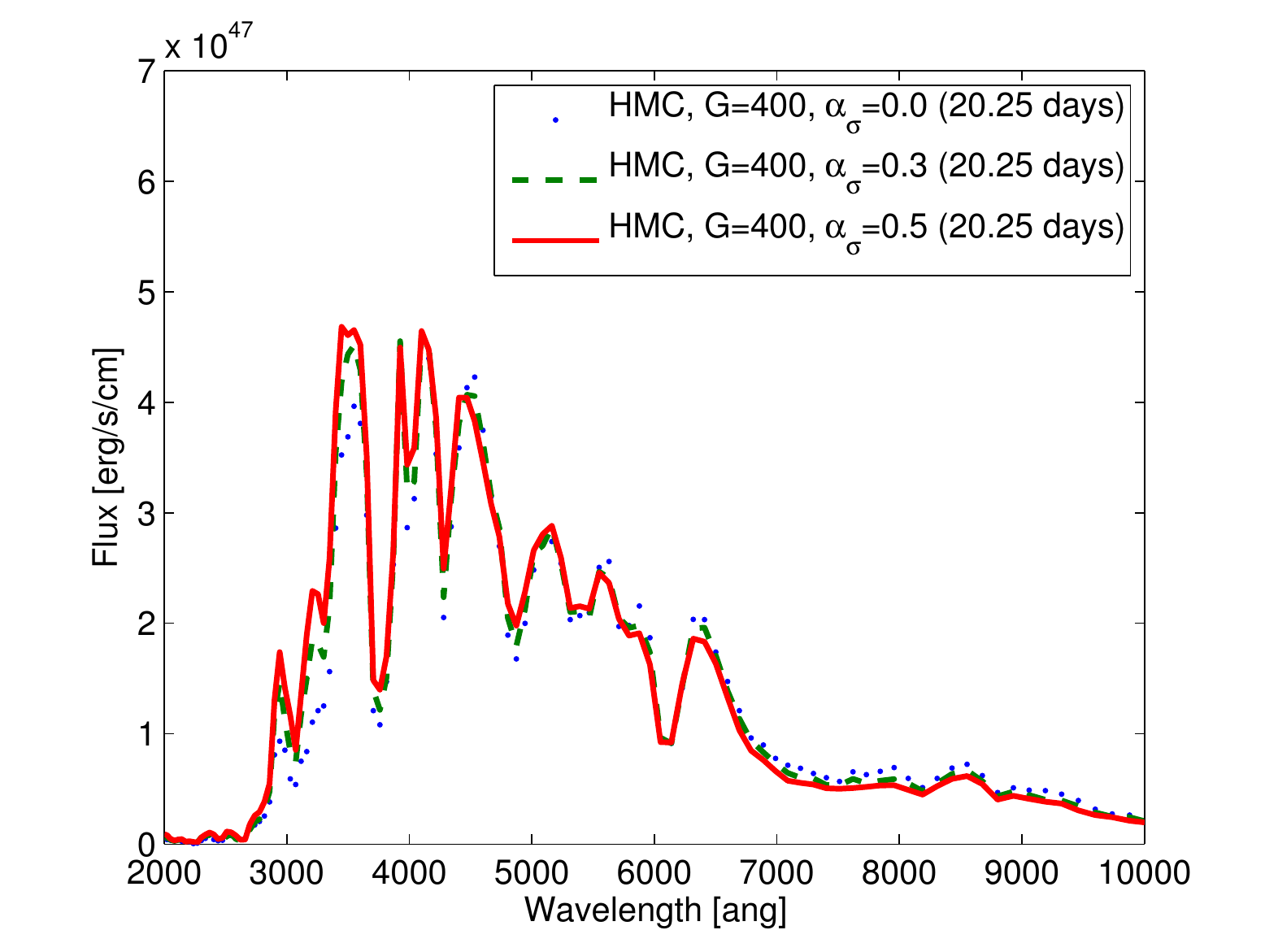}\label{fg5b}}\\
\subfloat[]{\includegraphics[height=65mm]{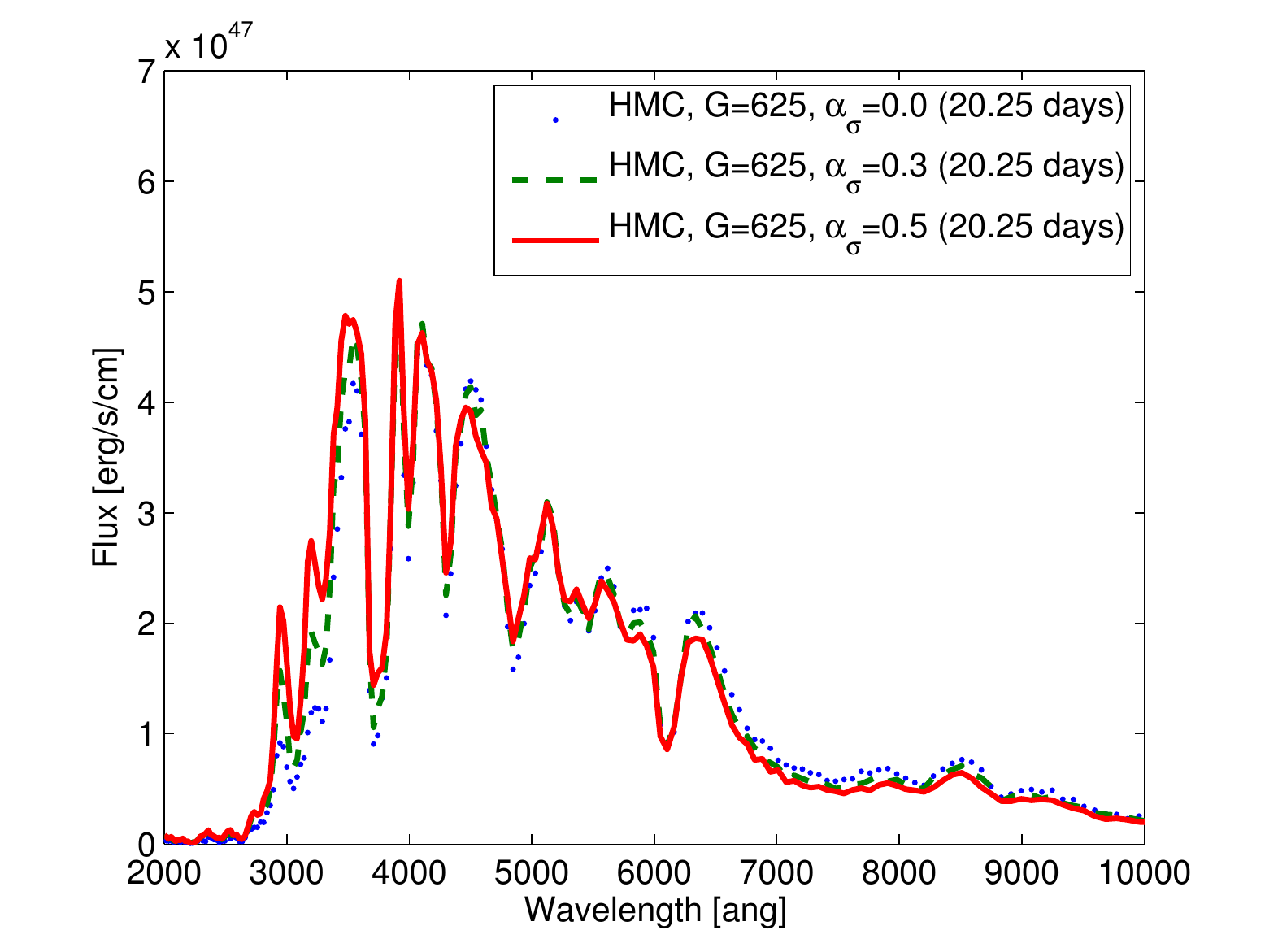}\label{fg5c}}
\subfloat[]{\includegraphics[height=65mm]{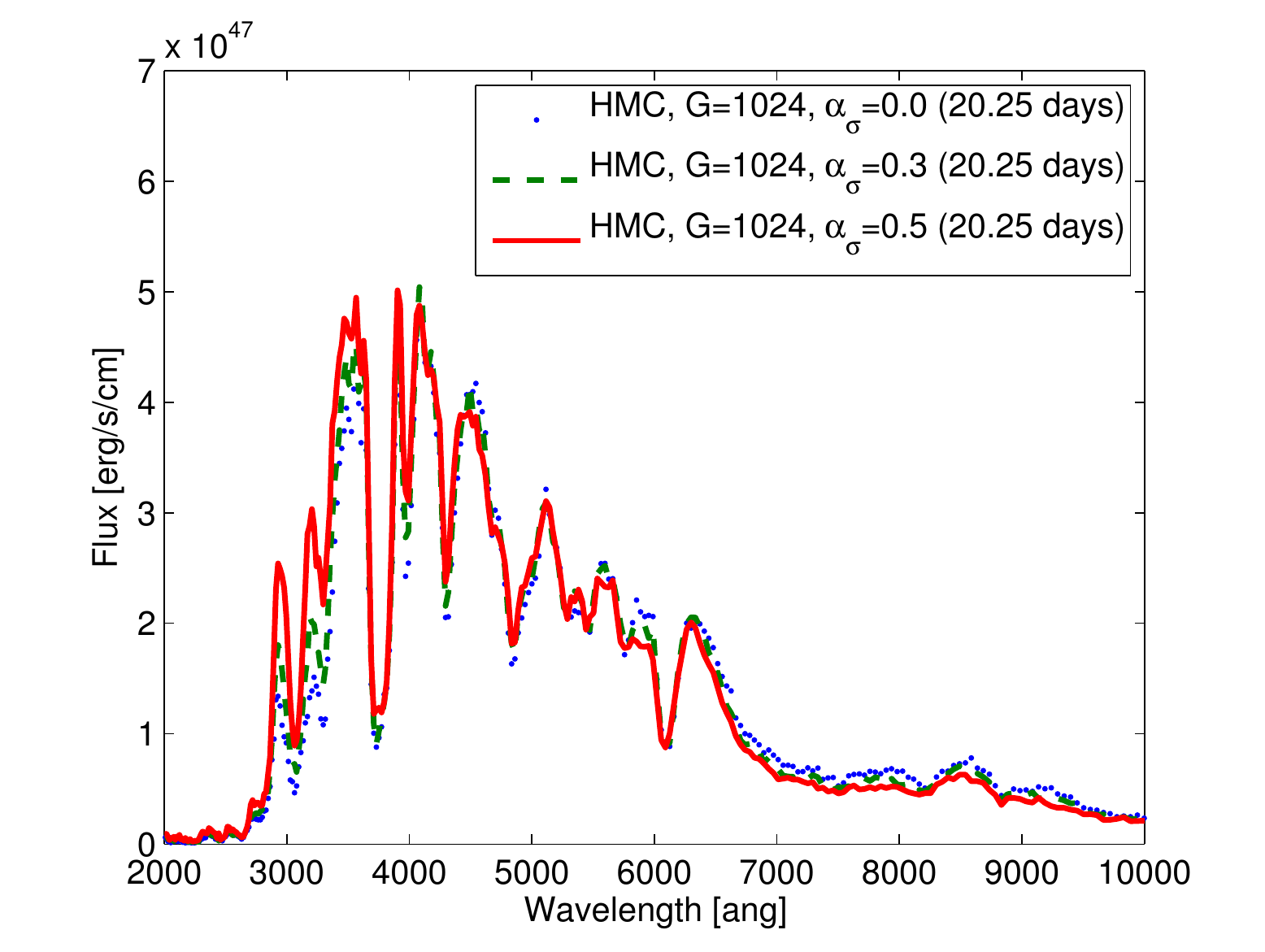}\label{fg5d}}
\caption{
Opacity-regrouped IMC-DDMC W7 spectra for opacity mixing
$\alpha_{\sigma}=0.0,0.3,0.5$ (dotted, dashed, and solid, respectively)
and a fixed number of subgroups, $G_{g}=20$.
Equation~\eqref{eq92} has been applied for opacity mixing.
Spectra are calculated by tallying escaping particles energies per group per
time and dividing by group wavelength range.
Data are plotted at group centers.
In Figs.~\ref{fg5a},~\ref{fg5b},~\ref{fg5c}, and~\ref{fg5d} group
resolutions are $G=225$, $G=400$, $G=625$, and $G=1024$, respectively.
Locations of peaks and troughs amongst the different opacity mixings presented
appear consistent.
For $\lambda\in[2000,4000]$, radiation transmission is larger for larger values
of $\alpha_{\sigma}$.
}
\label{fg5}
\end{figure*}

Our first W7 test problems explore the effect of different group opacity averaging and
group resolution.
Specifically, Eq.~\eqref{eq92} is implemented.
The problems considered have 225, 400, 625, and 1024 groups, 20 subgroups per
group, and an opacity mixing parameter $\alpha_{\sigma}\in\{0.0,0.3,0.5,0.8,1.0\}$.
Each calculation has 64 velocity cells uniformly spaced from 0 cm/s to $2.2027\times10^{9}$
cm/s, 248 uniform time steps for $t\in[2,64]$ days, 250,000 initial radiation particles,
250,000 source particles generated per time step, $\tau_{D}=\tau_{L}=3$ mean free paths,
and the opacity-regrouped neighborhoods span the entire set of groups ($g_{c}=G$).
Absolute bolometric magnitudes are calculated with
\begin{equation}
\label{eq93}
M_{\text{bol}}=4.74-2.5\log_{10}\left(\frac{L}{3.84\times10^{33}}\right) \;\;,
\end{equation}
where $L$ is luminosity in erg/s.
The luminosities are computed by tallying lab frame particle energies escaping the domain
and dividing by time step size.
Figures~\ref{fg4a},~\ref{fg4b},~\ref{fg4c}, and~\ref{fg4d} have light curves
calculated with Eq.~\eqref{eq93} for $G=225$, $G=400$, $G=625$, and $G=1024$,
respectively, and a fixed number of subgroups, $G_{g}=20$.
Similarly Figs.~\ref{fg5a},~\ref{fg5b},~\ref{fg5c}, and~\ref{fg5d} have spectra at
20 days post explosion calculated with Eq.~\eqref{eq93} for $G=225$, $G=400$,
$G=625$, and $G=1024$, respectively, and $G_{g}=20$.
For the group resolutions presented, the $\alpha_{\sigma}=1.0$ case does not appear
to converge at the same rate as the other results.
In other words, the $\alpha_{\sigma}=1.0$ case for Eq.~\eqref{eq92} produces
more sensitivity in brightness and spectrum versus course group resolutions.
As the mixing parameter is increased towards 1, the opacity calculation
applies more reciprocal averaging.
Since reciprocal averaging favors smaller subgroup opacity values, it is
expected that larger $\alpha_{\sigma}$ yield earlier and brighter light curves.
Despite producing unrealistic light curves for $\alpha_{\sigma}\approx1$,
$\alpha_{\sigma}$ may be calibrated between 0 and $\sim0.3$ to make
simulations with low or modest group numbers emulate high-resolution
simulations.

Table~\ref{tb2} has computation times for each curve.
Timing results for the problem described are for 24 cores.
With source particle numbers kept constant, simulation time scales sub-linearly
with increasing group number.
\begin{table}[H]
\caption{Total Run Times for Opacity-Regrouped HMC W7 with 24 Cores (hours)}
\label{tb2}
\centering
\begin{tabular}{|l|c|c|c|c|c|}
\hline
$G\,\backslash\,\alpha_{\sigma}$ & 0.0 & 0.3 & 0.5 & 0.8 & 1.0 \\
\hline
225 & 0.92 & 0.92 & 0.91 & 0.89 & 0.83 \\
\hline
400 & 1.33 & 1.32 & 1.32 & 1.28 & 1.21 \\
\hline
625 & 1.92 & 1.88 & 1.91 & 1.87 & 1.89 \\
\hline
1024 & 2.73 & 2.71 & 2.70 & 2.70 & 3.32 \\
\hline
\end{tabular}
\end{table}

\begin{figure*}
\subfloat[]{\includegraphics[height=65mm]{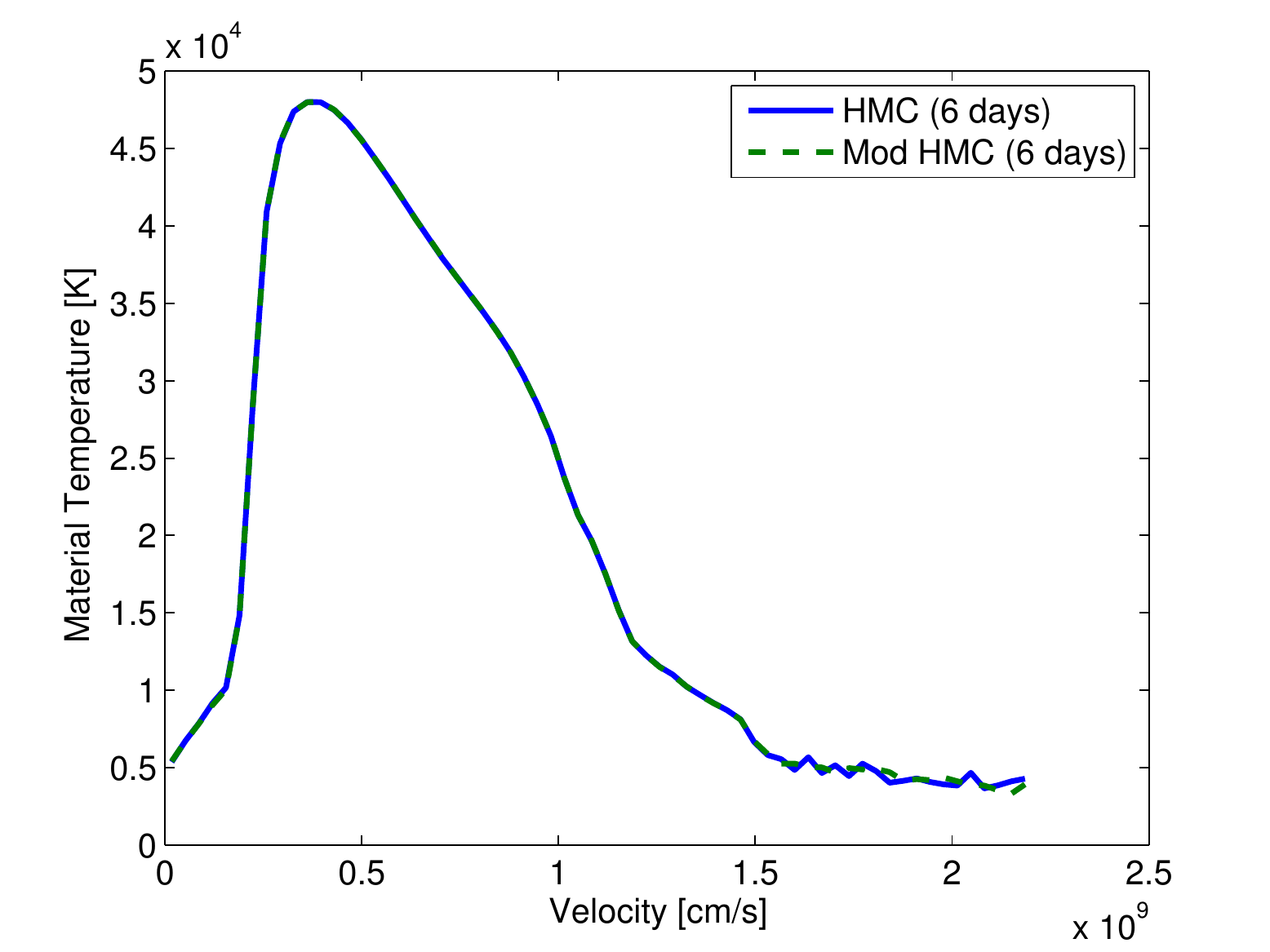}\label{fg6a}}
\subfloat[]{\includegraphics[height=65mm]{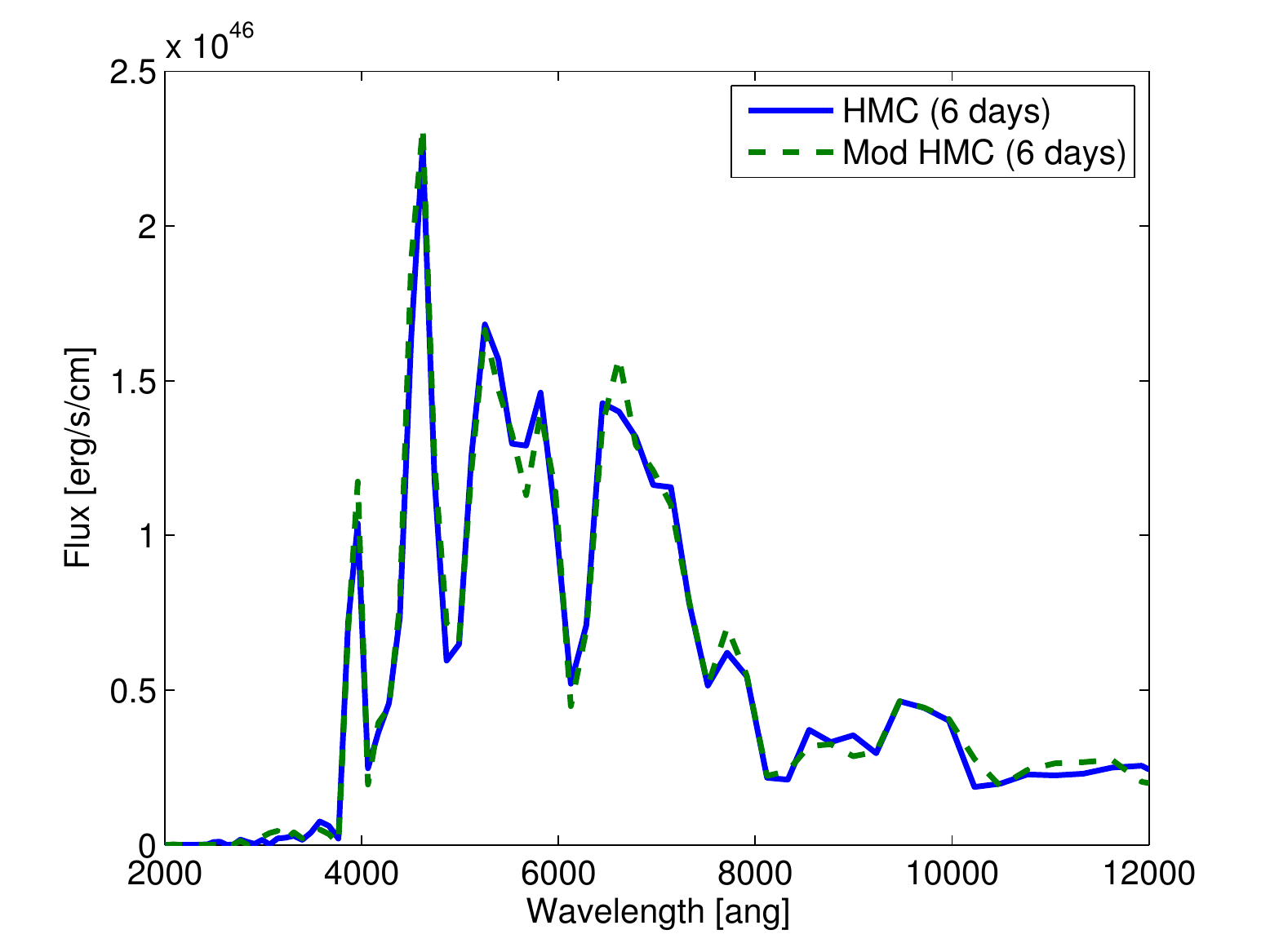}\label{fg6b}}\\
\subfloat[]{\includegraphics[height=65mm]{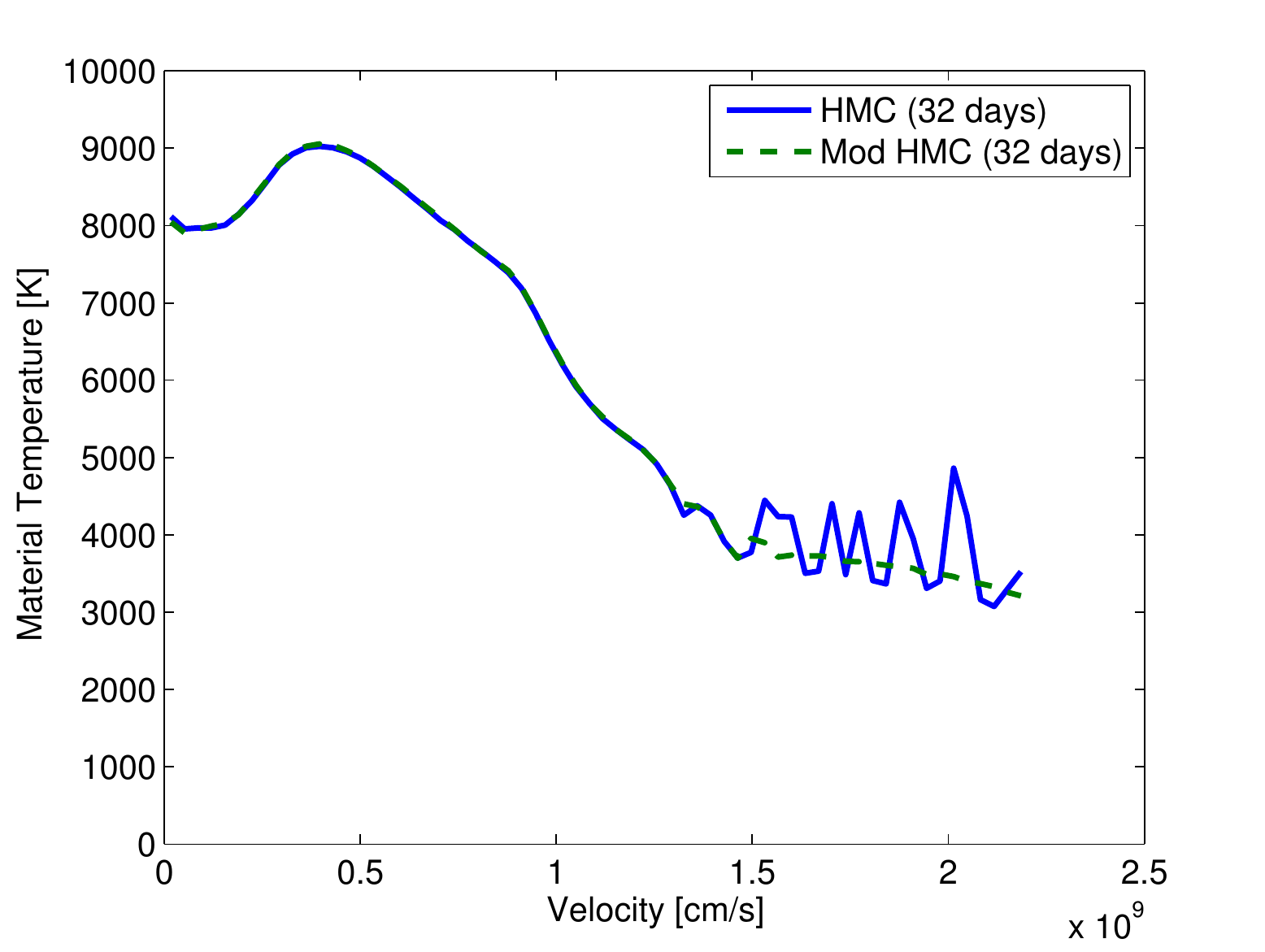}\label{fg6c}}
\subfloat[]{\includegraphics[height=65mm]{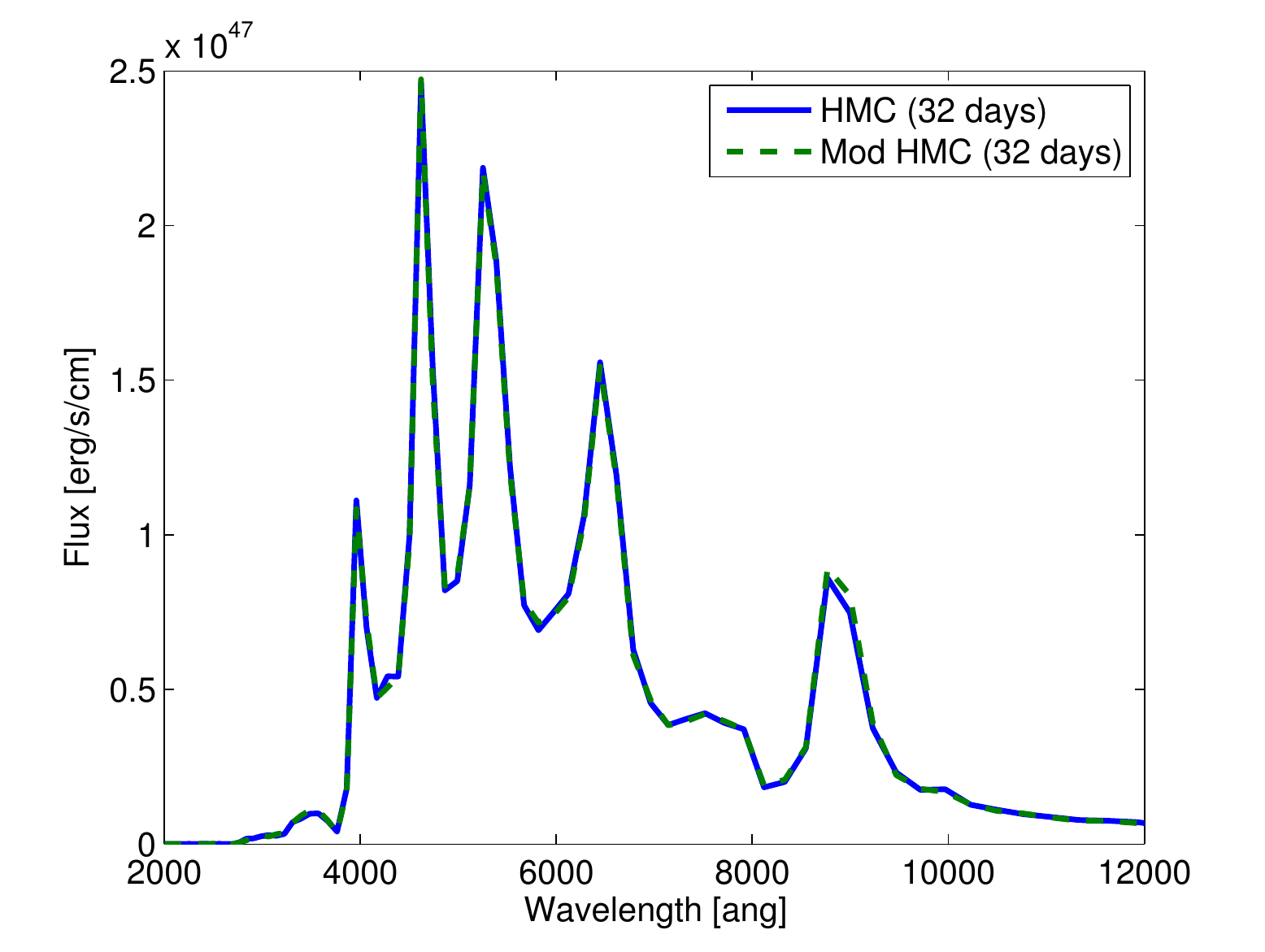}\label{fg6d}}
\caption{
IMC-DDMC (solid) and Gentile-Fleck factor modified IMC-DDMC (dashed) material
temperatures (left) and spectra (right) at days 6 and 32 post-explosion for the
W7 problem described in Section~\ref{sec:W7} with $G=225$ and $\alpha_{\sigma}=0.5$.
At early time ($t\lesssim10$ days), the Gentile-Fleck factor slightly modifies the spectrum which is sensitive to its effect on the fluctuations in the outer regions of the ejecta.
At late times in the W7 expansion ($t\gtrsim25$ days), the Gentile-Fleck factor
consistently mitigates temperature fluctuations in the outer cells.
Despite the continued temperature fluctuations in the outer cells for standard
IMC-DDMC at later times, the difference in spectra at late times is no longer
significant.
}
\label{fg6}
\end{figure*}

We now examine the effect of the Gentile-Fleck factor, or Eqs.~\eqref{eq50} and
\eqref{eq54} along with the optimization described in the last paragraph of
Section~\ref{sec:manu}, on W7 temperatures.
Figure~\ref{fg6} has spectra and material temperature profiles shown at day
3 and 32 post-explosion for the W7 problem described with $\alpha_{\sigma}=0.5$
and $G=225$.
At early times ($t\lesssim10$ days), both IMC-DDMC and modified IMC-DDMC
yield outer-cell temperature fluctuations for the numerical specifications
considered.
The fluctuations are different between the standard and modified methods.
Consequently, the application of the Gentile-Fleck factor in IMC-DDMC
uncovers some uncertainty in early spectra.
At later times ($t\gtrsim25$ days), the Gentile-Fleck factor yields
consistently smoother material temperature profiles than the standard
Fleck factor.
However, the spectra at later times are not significantly affected by
the fluctuations in the outer-cell temperatures
because that region is optically thin at that point.

\begin{figure*}
\subfloat[]{\includegraphics[height=65mm]{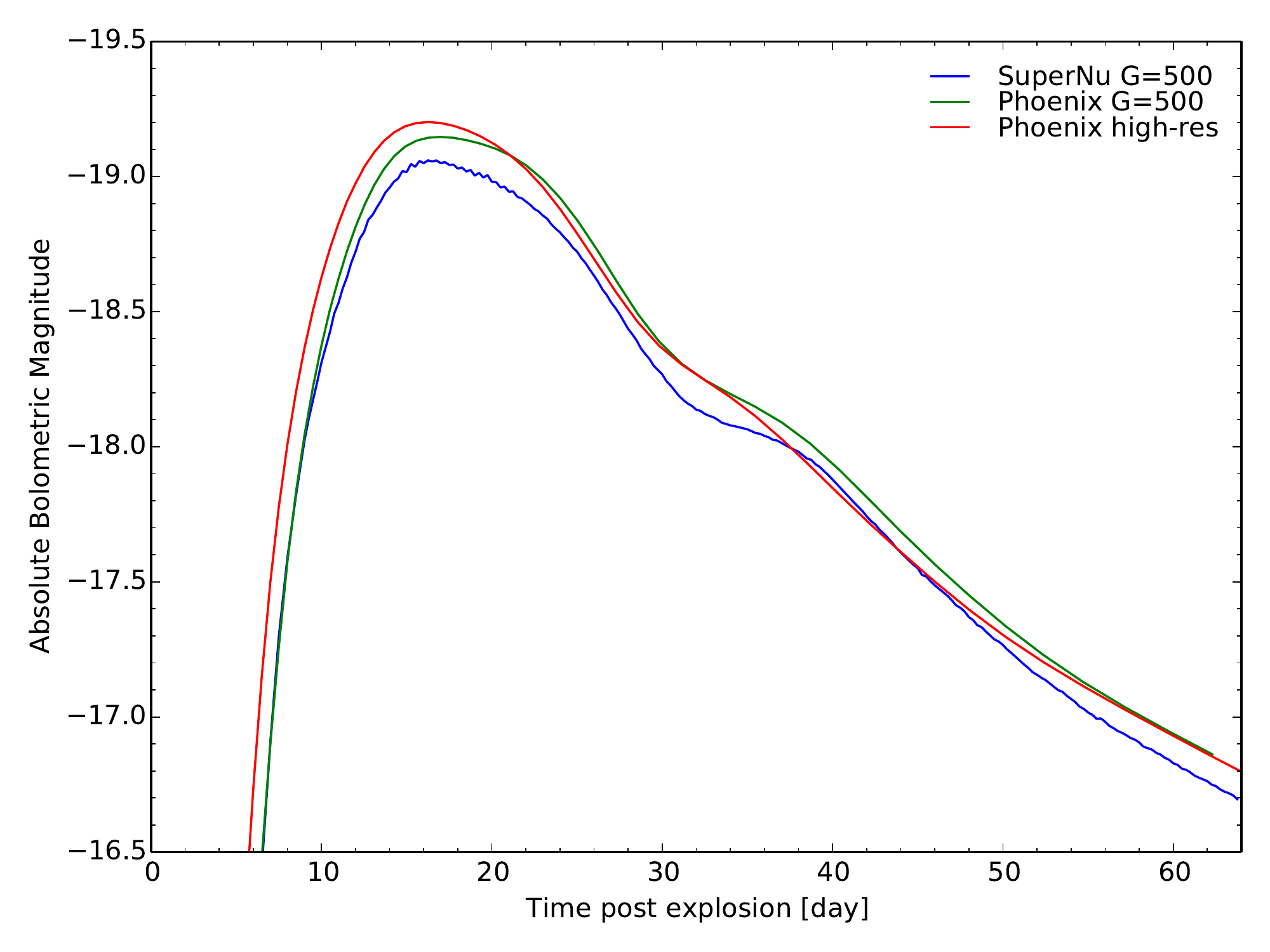}\label{fg7a}}
\subfloat[]{\includegraphics[height=65mm]{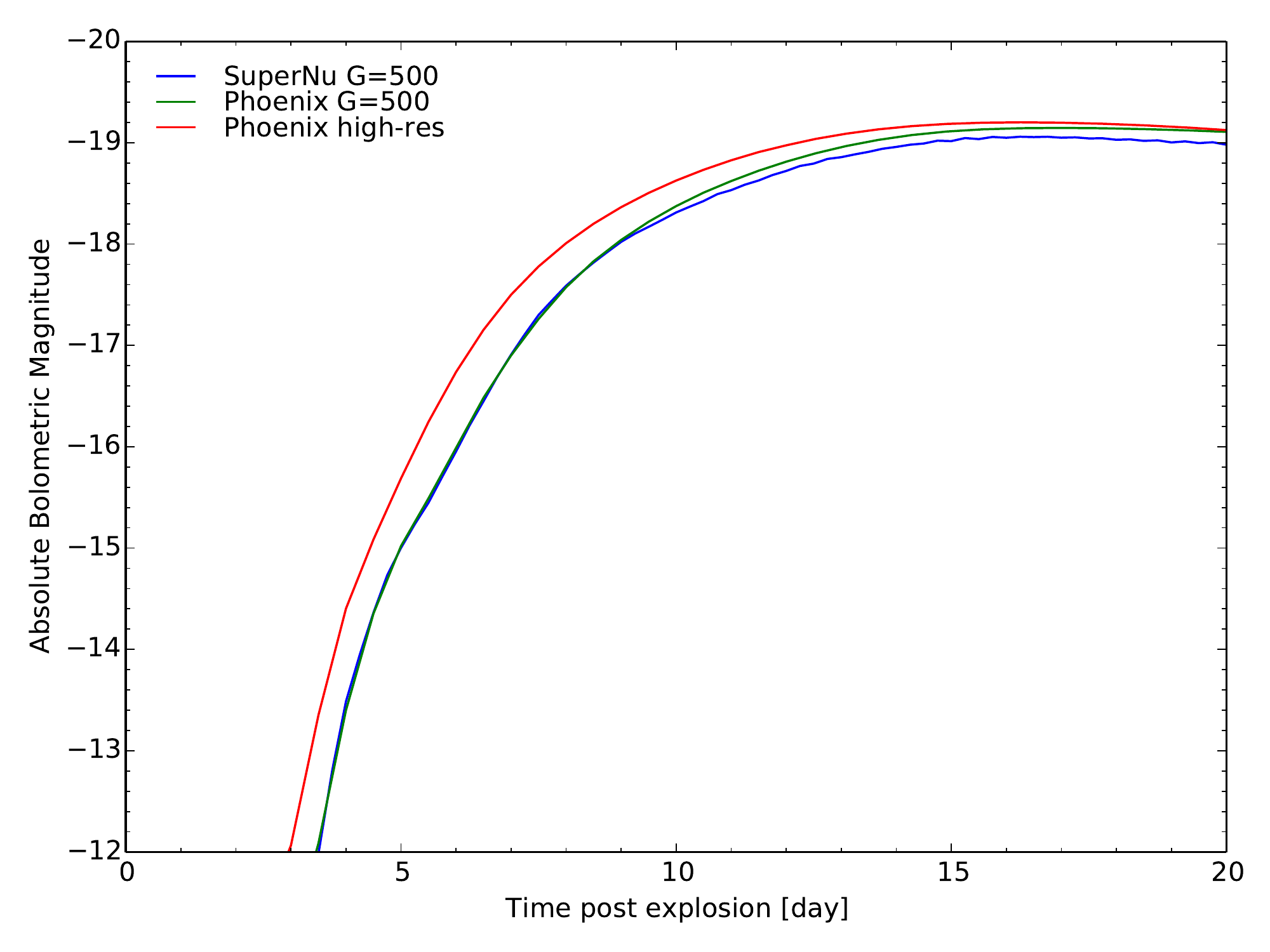}\label{fg7b}}
\caption{
\supernu\ (blue), with multigroup \phoenix\ (green),
and standard \phoenix\ (red) light curves.
\supernu\ and multigroup \phoenix\ apply 500 groups and directly
averaged group opacity, or an $\alpha_{\sigma}=0$ mix.
\phoenix\ is run in LTE for consistency with \supernu.
There exists a systematic difference of $\sim$10-15\% in luminosity for
much of the W7 evolution between the multigroup results.
Differences in transport, EOS, or opacity routines along with
spatial grid resolution may account for some of the discrepancy.
In Fig.~\ref{fg7b}, it is notable that the multigroup results give
very similar early rising light curves,
meaning that the different diffusion treatments in the two codes are in good agreement.
The standard \phoenix\ light curve rises earlier than multigroup \phoenix.
This is due to the high resolution that enables windows of lower opacity
through which diffusion is enhanced.
Diffusion at low group resolutions can be simulated and calibrated
using the opacity mixing parameter $\alpha_\sigma$ (see Figure \ref{fg4}).
}
\label{fg7}
\end{figure*}

\begin{figure*}
\subfloat[]{\includegraphics[height=65mm]{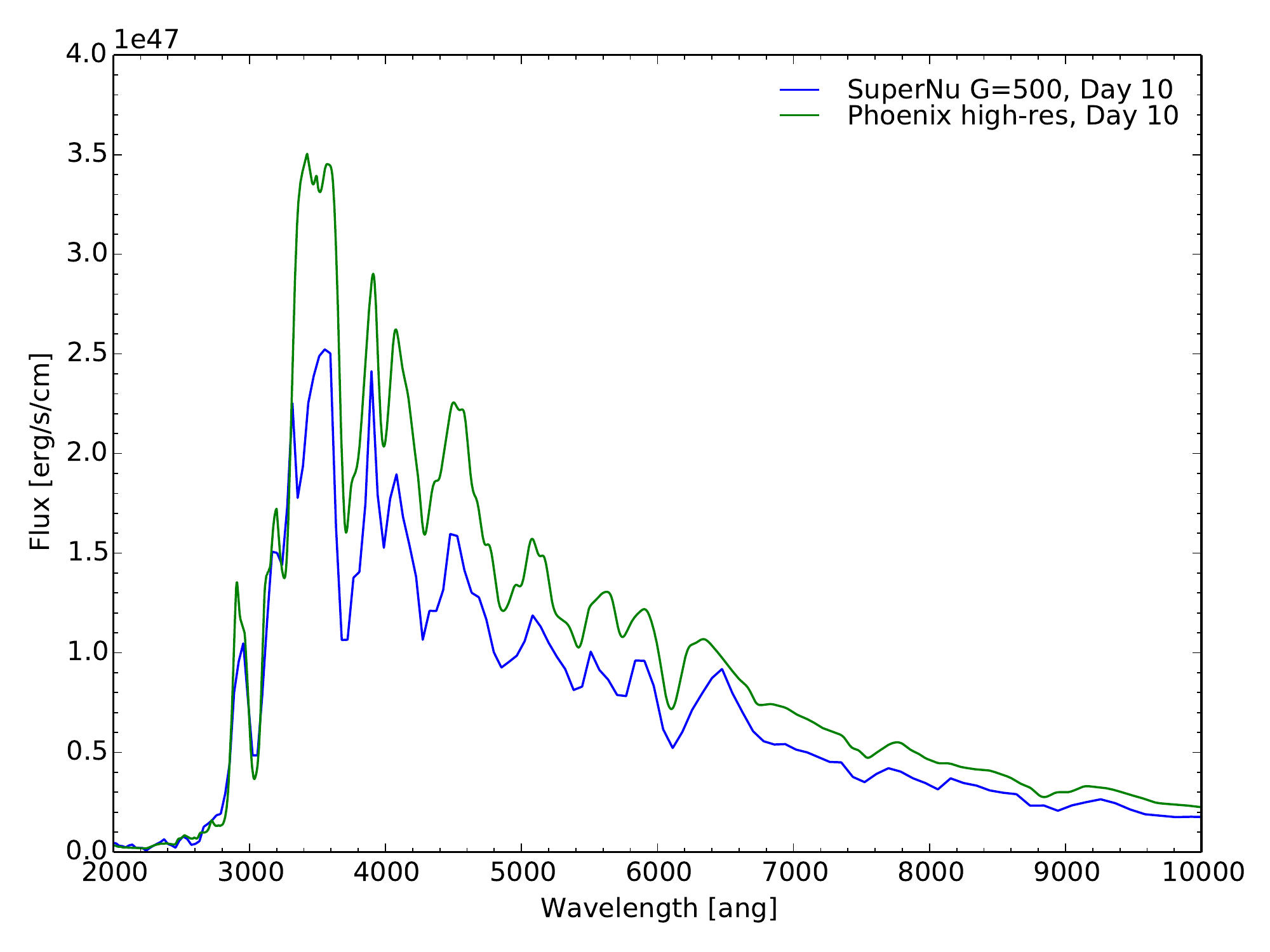}\label{fg8a}}
\subfloat[]{\includegraphics[height=65mm]{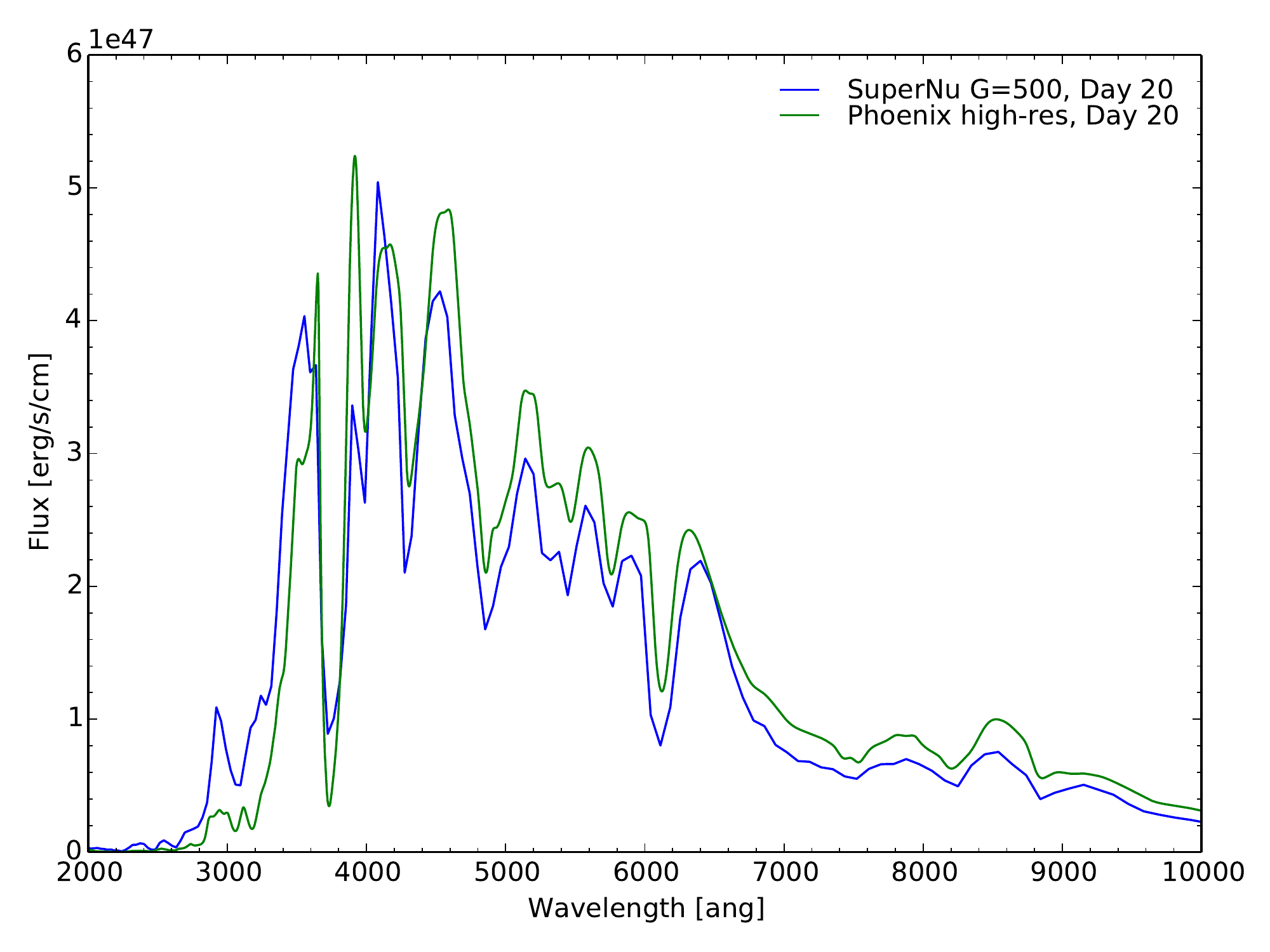}\label{fg8b}}\\
\subfloat[]{\includegraphics[height=65mm]{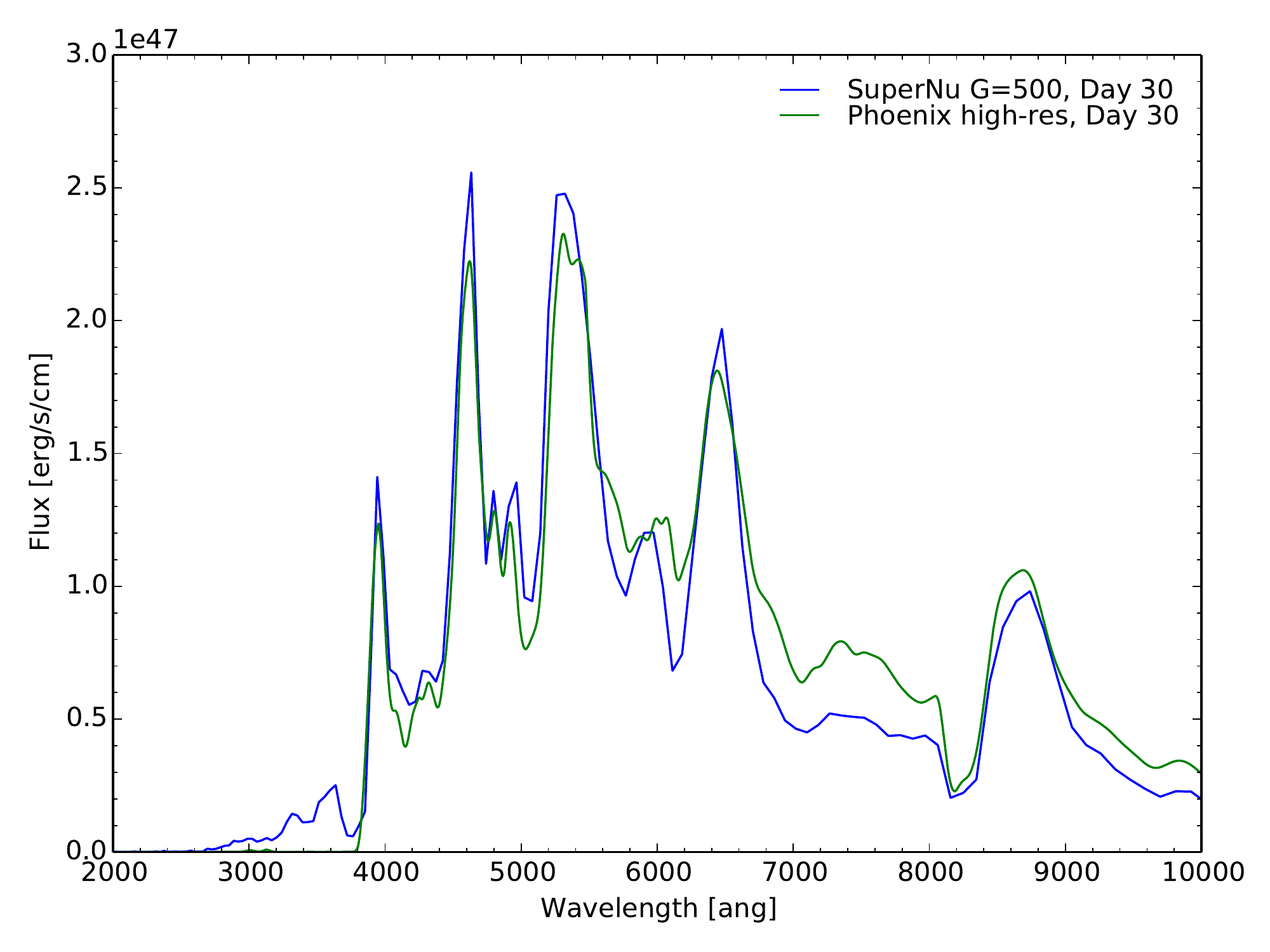}\label{fg8c}}
\subfloat[]{\includegraphics[height=65mm]{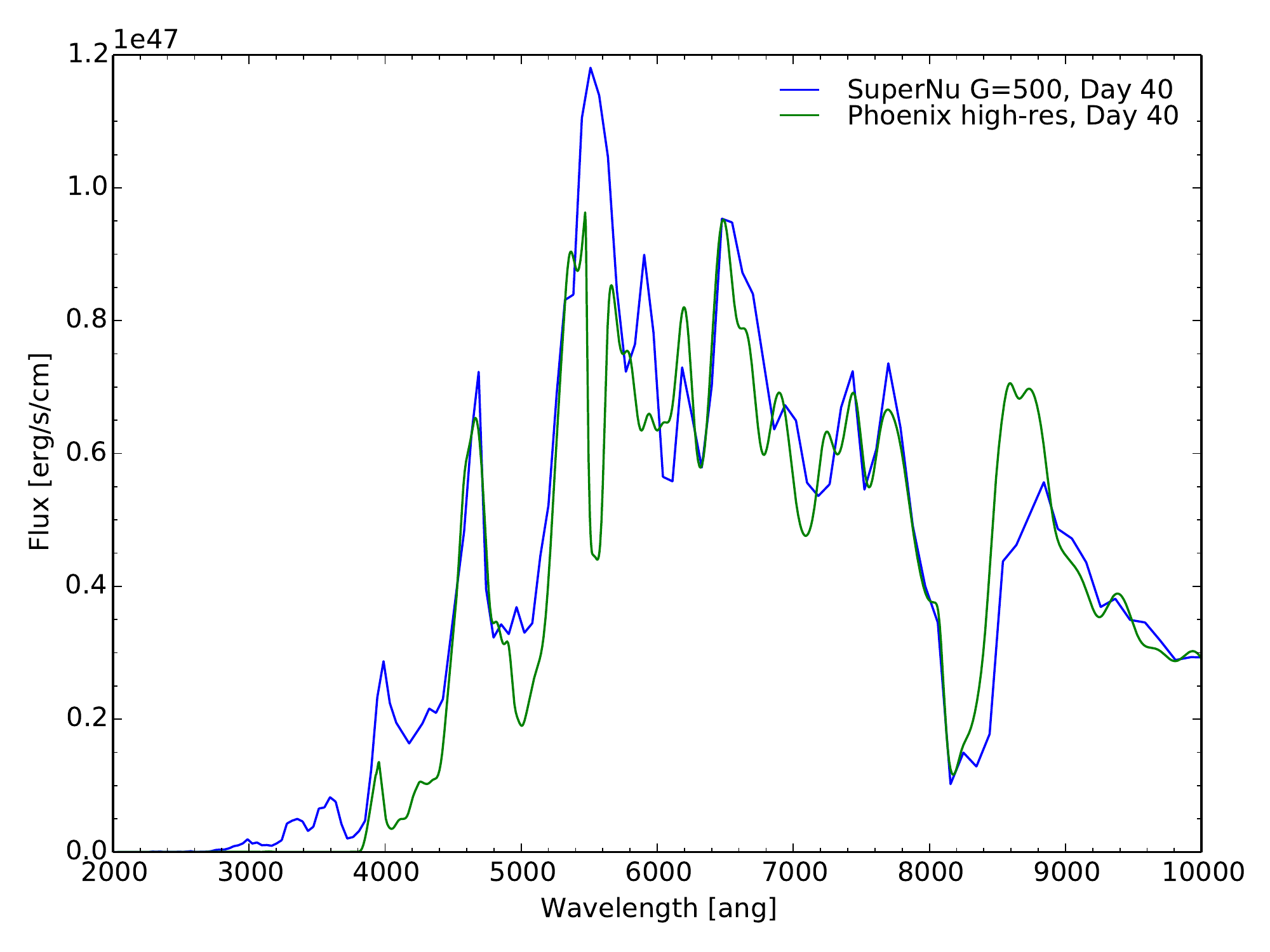}\label{fg8d}}
\caption{
\supernu\ (blue) with 500 groups and standard \phoenix\ (green)
spectra for the W7 problem at 10, 20, 30,
and 40 days post-explosion.
In Fig.~\ref{fg8a}, the difference in flux is partly
attributable to the earlier rise of the \phoenix\ high-resolution
luminosity (see Fig.~\ref{fg7}).
In Fig.~\ref{fg8b}, the W7 supernova is near peak luminosity; resolving
the discrepancy in flux requires further code-to-code comparison.
In Fig.~\ref{fg8d}, the flux of \phoenix\ is not
systematically larger than \supernu.
Around day 40, the high-resolution \phoenix\ light curve is
at a lower luminosity than the 500 group \supernu\ light curve.
Given the considerable differences in computational methods between
the codes, the temporal behavior and shape of the spectra are in
good agreement.
}
\label{fg8}
\end{figure*}

Finally, we compare the results of \supernu\ and \phoenix\ for the
W7 problem in LTE.
We find that the light curve generated by \supernu\ is systematically
$\sim$10\% dimmer at peak than the light curve generated by \phoenix\ for
various time step and group resolutions.
For controlled testing, grouped opacities have been introduced
into \phoenix.
The multigroup computations have no opacity mixing, or
$\alpha_{\sigma}=0$.
Figure~\ref{fg7} has 500 group light curve results from \phoenix\ and
\supernu\ along with a standard, high-resolution
(30,000 wavelength points) \phoenix\ light curve.
From inspection of Fig.~\ref{fg7b}, it is worth noting that the
luminosities of \emph{multigroup} \phoenix\ and \supernu\ have
similar early rising light curves.
This means that the different diffusion treatments in the two codes are in good agreement.
The standard \phoenix\ light curve rises earlier than the
multigroup \phoenix\ light curve, as expected.
This effect can be emulated in low group resolution simulations
using the opacity mixing parameter (see Figure \ref{fg4}).
Increasing $\alpha_\sigma$ from 0 to $\sim 0.3$ has a similar effect on the
light curve shape as increasing the resolution to convergence.
Figure~\ref{fg8} has spectra at 10, 20, and 40 days post-explosion
for the 500 group \supernu\ and high-resolution \phoenix\
simulations.
Despite differences in magnitudes, the time evolution of the light curves and the shapes
of the spectra are in good agreement.
The codes use the same atomic data but the EOS and opacity routines are
different; these factors may account for some differences in the
luminosities and spectra.

Resolving the sources of the 10-15\% discrepancy will require more in-depth
code-to-code comparisons which is work in progress but beyond the scope of this paper.
Having performed time step and group resolution tests, we also plan to
perform resolution tests on the spatial grid.
It is possible the codes have different convergence properties
with grid resolution.
In particular, the standard leakage opacity at IMC-DDMC spatial
method interfaces may underpredict particle transmission across
cell surfaces when DDMC interface cells are optically thick
\citep{densmore2007}.
\cite{densmore2006} performs an emissivity based derivation
to generalize the standard IMC-DDMC boundary condition and improve
the emission from DDMC to IMC at spatial interfaces.
If increased grid resolution in \supernu\ increases the
luminosity, then the alternate boundary condition
presented by~\cite{densmore2006} may increase the absolute
bolometric magnitude of the light curve at the current 64 cell resolution.
We have performed preliminary tests with an emissivity based
boundary condition and find a $\sim$2\% increase in the absolute
bolometric magnitude at peak; despite this modest change, exploring the
effects of increasing the spatial resolution may be revealing.
Apart from grid resolution, EOS, opacities, and transport
methods, there may be other important reasons for the observed
differences.

\section{Conclusions and Future Work}
\label{sec:Conc}

We have incorporated techniques to mitigate overheating errors
and combine DDMC groups with high opacity in the IMC-DDMC code,
\supernu.
In Section~\ref{sec:procs}, we described an approach to Doppler
shift DDMC particles.
The Doppler shift scheme accounts for the effect of inelastic
collisions with uniform subgroup redistribution.
Following~\cite{abdikamalov2012}, the Doppler shift scheme is
operator split from the diffusion scheme; it does not conflict
with the opacity regrouping process.

We found that opacity regrouping is needed in IMC-DDMC to make the
W7 problem feasible; the optimization mitigates computational
cost in performing the multidimensional calculation.
Additionally, we have described and tested an approach to treating
the opacity that involves refining the wavelength grid to subgroups.

In Section~\ref{sec:manu} we used the Gentile-Fleck factor
to mitigate an overheating pathology in the presence of strong
outflow.
The MC results are benchmarked against a quasi-manufactured
solution.
In Section~\ref{sec:heav}, we treated structured multigroup
problems with IMC-DDMC to test the effect of non-contiguous
opacity regrouping.
For the problem presented, opacity regrouping significantly improves
efficiency without a significant cost of accuracy in the
temperatures and spectra.
In Section~\ref{sec:W7}, we tested IMC-DDMC with opacity regrouping and
subgrouping on the W7 problem.
We also compared light curves and spectra for the W7 test
problem calculated using \supernu\ and \phoenix\ for
a similar set-up.
We modified \phoenix\ to be able to use multigroup opacities,
which enabled us to do more controlled code-to-code comparisons.
The light-curve rise times given by multigroup \phoenix\ and
\supernu\ are in good agreement for the same group resolution.
We find satisfactory agreement in the shape of the spectra.
However, there exists a $\sim$10-15\% discrepancy between \supernu\ 
and \phoenix\ in the luminosity of the light curve around and after peak
that is currently not fully understood.
Time step resolution tests indicate the light curves
compared between codes are converged in time.
For certain spatial grid resolutions, DDMC may underpredict spatial
leakage of diffusion particles to IMC
\citep{densmore2006,densmore2007}.
Consequently, spatial grid resolution tests of
\supernu\ may be informative.

We plan to extend our code to multiple dimensions.
The IMC-DDMC method is simple to extend to two and three
dimensions for simple grid geometries.
The challenges in performing multidimensional
simulations of SN Ia light curves and spectra with IMC-DDMC
lies in optimization and memory requirements.
In addition to spatial geometry, we plan to investigate
methods and algorithms that further mitigate spurious
temperature spikes due to the Maximum Principle or MC
noise.

\section{Acknowledgements}

We would like to thank Donald Lamb, Gregory Moses, and Carlo Graziani
for supporting and guiding this work.
We would like to thank Donald Lamb for the constructive recommendations
and suggestions.
We especially thank our referee, Ernazar Abdikamalov, for
the valuable recommendations that improved this paper.
This research was supported in part by the NSF under grant AST-0909132,
and by NIH through resources provided by the Computation Institute and the
Biological Sciences Division of the University of Chicago and Argonne
National Laboratory, under grant S10 RR029030-01.
This work is supported in part at the University of Chicago
by the National Science Foundation under grant
PHY-0822648 for the Physics Frontier Center "Joint     
Institute for Nuclear Astrophysics" (JINA).            

\nocite{*}
\bibliography{Bibliography}

\begin{thebibliography}{67}
\expandafter\ifx\csname natexlab\endcsname\relax\def\natexlab#1{#1}\fi

\bibitem[{Abdikamalov {et~al.}(2012)Abdikamalov, Burrows, Ott, Loffler,
  O'Connor, Dolence, \& Schnetter}]{abdikamalov2012}
Abdikamalov, E., Burrows, A., Ott, C.~D., Loffler, F., O'Connor, E., Dolence,
  J.~C., \& Schnetter, E. 2012, ApJ, 755, 111

\bibitem[{{Adams}(2001)}]{adams2001}
{Adams}, M.~L. 2001, Nucl. Sci. Eng., 137

\bibitem[{Atzeni \& ter Vehn(2004)}]{atzeni2004}
Atzeni, S., \& ter Vehn, J.~M. 2004, The Physics of Inertial Fusion (Oxford
  University Press)

\bibitem[{{Baron} \& {Hauschildt}(2007)}]{baron2007}
{Baron}, E., \& {Hauschildt}, P.~H. 2007, \aap, 468, 255

\bibitem[{Branch \& Khokhlov(1995)}]{branch1995}
Branch, D., \& Khokhlov, A. 1995, Physics Reports, 256, 53

\bibitem[{Brooks(1989)}]{brooks1989}
Brooks, E.~D. 1989, J. Comput. Phys., 83

\bibitem[{Buchler(1983)}]{buchler1983}
Buchler, J.~R. 1983, JQSRT, 30, 395

\bibitem[{Calder {et~al.}(2004)Calder, Plewa, Vladimirova, Lamb, \&
  Truran}]{calder2004}
Calder, A.~C., Plewa, T., Vladimirova, N., Lamb, D.~Q., \& Truran, J.~W. 2004,
  Astrophysical Journal, Letters

\bibitem[{Calder {et~al.}(2002)Calder, Fryxell, Plewa, Rosner, Dursi, Weirs,
  Dupont, Robey, Kane, Remington, Drake, Dimonte, Zingale, Timmes, Olson,
  Ricker, MacNeice, \& Tufo}]{calder2002}
Calder, A.~C., {et~al.} 2002, Astrophysical Journal, Supplement, 143, 201

\bibitem[{Carter \& Forest(1973)}]{carter1973}
Carter, L.~L., \& Forest, C.~A. 1973, lA-5038, Los Alamos National Laboratory

\bibitem[{Castor(2004)}]{castor2004}
Castor, J.~I. 2004, Radiation Hydrodynamics (Cambridge University Press)

\bibitem[{Cleveland \& Gentile(2014)}]{cleveland2014}
Cleveland, M.~A., \& Gentile, N. 2014, Transport Theory and Statistical
  Physics, 1

\bibitem[{Cleveland {et~al.}(2010)Cleveland, Gentile, \&
  Palmer}]{cleveland2010}
Cleveland, M.~A., Gentile, N.~A., \& Palmer, T.~S. 2010, J. Comput. Phys., 229,
  5707

\bibitem[{Densmore(2011)}]{densmore2011}
Densmore, J.~D. 2011, J. Comput. Phys., 230, 1116

\bibitem[{Densmore {et~al.}(2006)Densmore, Davidson, \&
  Carrington}]{densmore2006}
Densmore, J.~D., Davidson, G., \& Carrington, D.~B. 2006, Ann. Nucl. Energy,
  33, 583

\bibitem[{Densmore {et~al.}(2008)Densmore, Evans, \& Buksas}]{densmore2008}
Densmore, J.~D., Evans, T.~M., \& Buksas, M.~W. 2008, Nucl. Sci. Eng., 159, 1

\bibitem[{Densmore \& Larsen(2004)}]{densmore2004}
Densmore, J.~D., \& Larsen, E.~W. 2004, J. Comput. Phys., 199, 175

\bibitem[{Densmore {et~al.}(2012)Densmore, Thompson, \&
  Urbatsch}]{densmore2012}
Densmore, J.~D., Thompson, K.~G., \& Urbatsch, T.~J. 2012, J. Comput. Phys.,
  231, 6925

\bibitem[{Densmore {et~al.}(2007)Densmore, Urbatsch, Evans, \&
  Buksas}]{densmore2007}
Densmore, J.~D., Urbatsch, T.~J., Evans, T.~M., \& Buksas, M.~W. 2007, J.
  Comput. Phys., 222, 485

\bibitem[{Fleck \& Canfield(1984)}]{fleck1984}
Fleck, Jr., J.~A., \& Canfield, E.~H. 1984, J. Comput. Phys., 54, 508

\bibitem[{Fleck \& Cummings(1971)}]{fleck1971}
Fleck, Jr., J.~A., \& Cummings, J.~D. 1971, J. Comput. Phys., 8, 313

\bibitem[{{Fryxell} {et~al.}(2000){Fryxell}, {Olson}, {Ricker}, {Timmes},
  {Zingale}, {Lamb}, {MacNeice}, {Rosner}, {Truran}, \& {Tufo}}]{fryxell2000}
{Fryxell}, B., {et~al.} 2000, \apjs, 131, 273

\bibitem[{Gamezo {et~al.}(2003)Gamezo, Khokhlov, Oran, Chtchelkanova, \&
  Rosenberg}]{gamezo2003}
Gamezo, V.~N., Khokhlov, A.~M., Oran, E.~S., Chtchelkanova, A.~Y., \&
  Rosenberg, R.~O. 2003, Science, 299, 77

\bibitem[{Gentile(2001)}]{gentile2001}
Gentile, N.~A. 2001, J. Comput. Phys., 172, 543

\bibitem[{Gentile(2011)}]{gentile2011}
---. 2011, J. Comput. Phys., 230

\bibitem[{Habetler \& Matkowsky(1975)}]{habetler1975}
Habetler, G.~J., \& Matkowsky, B.~J. 1975, J. Math. Phys., 16, 846

\bibitem[{Hauschildt(1992)}]{hauschildt1992}
Hauschildt, P.~H. 1992, JQSRT, 47, 433

\bibitem[{Hauschildt \& Baron(1999)}]{hauschildt1999}
Hauschildt, P.~H., \& Baron, E. 1999, Journal of Computational and Applied
  Mathematics, 109

\bibitem[{{Hauschildt} \& {Baron}(2004)}]{hauschildt2004}
{Hauschildt}, P.~H., \& {Baron}, E. 2004, \aap, 417, 317

\bibitem[{Hauschildt \& Wehrse(1991)}]{hauschildt1991}
Hauschildt, P.~H., \& Wehrse, R. 1991, JQSRT, 46

\bibitem[{Hillebrandt \& Niemeyer(2000)}]{hillebrandt2000}
Hillebrandt, W., \& Niemeyer, J. 2000, {\araa}, 38, 191

\bibitem[{Kasen {et~al.}(2006)Kasen, Thomas, \& Nugent}]{kasen2006}
Kasen, D., Thomas, R.~C., \& Nugent, P. 2006, ApJ, 651, 366

\bibitem[{Kromer \& Sim(2009)}]{kromer2009}
Kromer, M., \& Sim, S.~A. 2009, Mon. Not. R. Astron. Soc., 398

\bibitem[{Kurucz(1994)}]{kurucz1994}
Kurucz, R.~L. 1994

\bibitem[{Larsen \& Mercier(1987)}]{larsen1987}
Larsen, E.~W., \& Mercier, B. 1987, J. Comput. Phys., 71

\bibitem[{Lewis \& Miller(1993)}]{lewis1993}
Lewis, E.~E., \& Miller, Jr., W.~F. 1993, Computational Methods of Neutron
  Transport (American Nuclear Society)

\bibitem[{{Long} {et~al.}(2013){Long}, {Jordan}, {van Rossum}, {Diemer},
  {Graziani}, {Kessler}, {Meyer}, {Rich}, \& {Lamb}}]{long2014}
{Long}, M., {et~al.} 2013, ArXiv e-prints

\bibitem[{{Lowrie} {et~al.}(2001){Lowrie}, {Mihalas}, \& {Morel}}]{lowrie2001}
{Lowrie}, R.~B., {Mihalas}, D., \& {Morel}, J.~E. 2001, JQRST, 69, 291

\bibitem[{Lucy(2005)}]{lucy2005}
Lucy, L.~B. 2005, A\&A, 429, 19

\bibitem[{{Malvagi} \& {Pomraning}(1991)}]{malvagi1991}
{Malvagi}, F., \& {Pomraning}, G.~C. 1991, J. Math. Phys., 32, 805

\bibitem[{McClarren {et~al.}(2008{\natexlab{a}})McClarren, Holloway, \&
  Brunner}]{mcclarren2008}
McClarren, R.~G., Holloway, J.~P., \& Brunner, T.~A. 2008{\natexlab{a}}, JQSRT,
  109, 389

\bibitem[{McClarren {et~al.}(2008{\natexlab{b}})McClarren, Lowrie, Prinja, \&
  Morel}]{mcclarren2008b}
McClarren, R.~G., Lowrie, R.~B., Prinja, A.~K., \& Morel, J.~E.
  2008{\natexlab{b}}, JQSRT, 109, 2590

\bibitem[{McClarren \& Urbatsch(2009)}]{mcclarren2009}
McClarren, R.~G., \& Urbatsch, T.~J. 2009, J. Comput. Phys., 228, 5669

\bibitem[{{McClarren} \& {Urbatsch}(2012)}]{mcclarren2012}
{McClarren}, R.~G., \& {Urbatsch}, T.~J. 2012, in Transactions of the American
  Nuclear Society

\bibitem[{{McKinley} {et~al.}(2003){McKinley}, {Brooks}, \&
  {Sz\H{o}ke}}]{mckinley2003}
{McKinley}, M.~S., {Brooks}, E.~D., \& {Sz\H{o}ke}, A. 2003, J. Comput. Phys.,
  189, 330

\bibitem[{Mihalas \& Mihalas(1984)}]{mihalas1984}
Mihalas, D., \& Mihalas, B.~W. 1984, Foundations of Radiation Hydrodynamics
  (Oxford University Press)

\bibitem[{N'Kaoua(1991)}]{nkaoua1991}
N'Kaoua, T. 1991, SIAM J. Stat. Comput., 12, 505

\bibitem[{Nomoto {et~al.}(1984)Nomoto, Thielemann, \& Yokoi}]{nomoto1984}
Nomoto, K., Thielemann, F., \& Yokoi, K. 1984, ApJ, 286, 644

\bibitem[{Oberkampf \& Roy(2010)}]{oberkampf2010}
Oberkampf, W.~L., \& Roy, C.~J. 2010, Verification and Validation in Scientific
  Computing (Cambridge University Press)

\bibitem[{Olson \& Kunasz(1987)}]{olson1987}
Olson, G.~L., \& Kunasz, P.~B. 1987, JQSRT, 38

\bibitem[{Perlmutter(2003)}]{perlmutter2003}
Perlmutter, S. 2003, Physics Today, 53

\bibitem[{{Perlmutter} {et~al.}(1999){Perlmutter}, {Aldering}, {Goldhaber},
  {Knop}, {Nugent}, {Castro}, {Deustua}, {Fabbro}, {Goobar}, {Groom}, {Hook},
  {Kim}, {Kim}, {Lee}, {Nunes}, {Pain}, {Pennypacker}, {Quimby}, {Lidman},
  {Ellis}, {Irwin}, {McMahon}, {Ruiz-Lapuente}, {Walton}, {Schaefer}, {Boyle},
  {Filippenko}, {Matheson}, {Fruchter}, {Panagia}, {Newberg}, {Couch}, \&
  {Supernova Cosmology Project}}]{perlmutter1999}
{Perlmutter}, S., {et~al.} 1999, \apj, 517, 565

\bibitem[{Petschek(1990)}]{petschek1990}
Petschek, A. 1990, Supernovae (Springer-Verlag)

\bibitem[{{Phillips}(1993)}]{phillips1993}
{Phillips}, M.~M. 1993, \apjl, 413, L105

\bibitem[{Pinto \& Eastman(2000)}]{pinto2000}
Pinto, P.~A., \& Eastman, R.~G. 2000, ApJ, 530, 744

\bibitem[{Pomraning(1973)}]{pomraning1973}
Pomraning, G.~C. 1973, The Equations of Radiation Hydrodynamics (Pergamon
  Press)

\bibitem[{{Riess} {et~al.}(1998){Riess}, {Filippenko}, {Challis},
  {Clocchiatti}, {Diercks}, {Garnavich}, {Gilliland}, {Hogan}, {Jha},
  {Kirshner}, {Leibundgut}, {Phillips}, {Reiss}, {Schmidt}, {Schommer},
  {Smith}, {Spyromilio}, {Stubbs}, {Suntzeff}, \& {Tonry}}]{riess1998}
{Riess}, A.~G., {et~al.} 1998, \aj, 116, 1009

\bibitem[{{Scannapieco} {et~al.}(2008){Scannapieco}, {Tissera}, {White}, \&
  {Springel}}]{scannapieco2008}
{Scannapieco}, C., {Tissera}, P.~B., {White}, S.~D.~M., \& {Springel}, V. 2008,
  MNRAS, 389

\bibitem[{Seitenzahl {et~al.}(2013)Seitenzahl, Ciaraldi-Schoolmann, Ropke,
  Fink, Hillebrandt, Kromer, Pakmor, Ruiter, Sim, \&
  Taubenberger}]{seitenzahl2012}
Seitenzahl, I.~R., {et~al.} 2013, MNRAS, 429, 1156

\bibitem[{Su \& Olson(1999)}]{su1999}
Su, B., \& Olson, G.~L. 1999, JQSRT, 62, 279

\bibitem[{Sutherland(1998)}]{sutherland1998}
Sutherland, R.~S. 1998, \mnras, 300

\bibitem[{{Sz\H{o}ke} \& {Brooks}(2005)}]{szoke2005}
{Sz\H{o}ke}, A., \& {Brooks}, E.~D. 2005, JQSRT, 91, 95

\bibitem[{{Urbatsch} {et~al.}(1999){Urbatsch}, {Morel}, \&
  {Gulick}}]{urbatsch1999}
{Urbatsch}, T.~J., {Morel}, J.~E., \& {Gulick}, J.~C. 1999, in Proc. Int. Conf.
  Mathematics and Computation, Reactor Physics, and Environment Analysis in
  Nuclear Applications

\bibitem[{{van Rossum}(2012)}]{vanrossum2012}
{van Rossum}, D.~R. 2012, \apj, 756, 31

\bibitem[{{Verner} {et~al.}(1996){Verner}, {Ferland}, {Korista}, \&
  {Yakovlev}}]{verner1996}
{Verner}, D.~A., {Ferland}, G.~J., {Korista}, K.~T., \& {Yakovlev}, D.~G. 1996,
  \apj, 465

\bibitem[{Warsa \& Densmore(2010)}]{warsa2010}
Warsa, J.~S., \& Densmore, J.~D. 2010, Nucl. Sci. Eng., 166, 36

\bibitem[{{Wollaeger} {et~al.}(2013){Wollaeger}, {van Rossum}, {Graziani},
  {Couch}, {Jordan}, {Lamb}, \& {Moses}}]{wollaeger2013}
{Wollaeger}, R.~T., {van Rossum}, D.~R., {Graziani}, C., {Couch}, S.~M.,
  {Jordan}, G.~C., {Lamb}, D.~Q., \& {Moses}, G.~A. 2013, \apjs, 209

\end{thebibliography}

\end{document}